\documentclass[fleqn,usenatbib]{mnras}

\usepackage[T1]{fontenc}

\DeclareRobustCommand{\VAN}[3]{#2}
\let\VANthebibliography\thebibliography
\def\thebibliography{\DeclareRobustCommand{\VAN}[3]{##3}\VANthebibliography}

\usepackage{graphicx}	
\usepackage{amsmath}	
\usepackage{amssymb}	
\usepackage[capitalise, noabbrev]{cleveref}
\crefname{section}{\S}{\SS}
\Crefname{section}{\S}{\SS}

\graphicspath{{figures/}}
\newcommand{\HII}{\ion{H}{ii}}
\newcommand{\Hii}{\ion{H}{ii}}
\newcommand{\Ha}{H$\alpha$}
\newcommand{\Hb}{H$\beta$}
\newcommand{\EWHa}{EW(H$\alpha$)}
\newcommand{\age}{$\mathcal{A}_{\star,L}$}
\newcommand{\met}{$\mathcal{Z}_{\star,L}$}
\newcommand{\SII}{[\ion{S}{II}]$\lambda\lambda 6717,31$}
\newcommand{\OII}{[\ion{O}{II}]$\lambda\lambda 3727,29$}

\title[HII regions in CALIFA survey]{HII regions in CALIFA survey: II. The relation between their physical properties and  galaxy evolution}

\author[C. Espinosa-Ponce et al.]{C. Espinosa-Ponce$^{1}$\thanks{E-mail: cespinosa@astro.unam.mx, carlcespinosa@gmail.com},
S. F. S\'anchez$^{1}$,
C. Morisset$^{2}$,
J. K. Barrera-Ballesteros$^{1}$,
\newauthor{L. Galbany$^{3,4}$, R. Garc\'ia-Benito$^{5}$, E. A. D. Lacerda$^{1}$, D. Mast$^{6,7}$}
\\
$^{1}$Universidad Nacional Aut\'onoma de M\'exico, Instituto de Astronom\'ia, AP 70-264, CDMX  04510, M\'exico\\
$^{2}$Universidad Nacional Aut\'onoma de M\'exico, Instituto de Astronom\'ia, AP 106,  Ensenada 22800, BC, M\'exico\\
$^{3}$Institute of Space Sciences (ICE, CSIC), Campus UAB, Carrer de Can Magrans, s/n, E-08193 Barcelona, Spain.\\
$^{4}$Institut d’Estudis Espacials de Catalunya (IEEC), E-08034 Barcelona, Spain.\\
$^{5}$Instituto de Astrof\'isica de Andaluc\'ia, CSIC, Apartado de correos 3004, E-18080 Granada, Spain\\
$^{6}$Universidad Nacional C\'ordoba. Observatorio Astron\'omico de C\'ordoba. C\'ordoba, Argentina.\\
$^{7}$Consejo de Investigaciones Cient\'{i}ficas y T\'ecnicas de la Rep\'ublica Argentina, Avda. Rivadavia 1917, C1033AAJ, CABA, Argentina
}

\date{Accepted XXX. Received YYY; in original form ZZZ}

\pubyear{2020}

\usepackage{newtxtext,newtxmath}
\begin{document}
\label{firstpage}
\pagerange{\pageref{firstpage}--\pageref{lastpage}}
\maketitle

\begin{abstract}
We present in here the exploration of the physical properties of the sample of \HII\ regions and aggregations of the last \HII\ regions catalog of the CALIFA survey. This sample comprises the optical spectroscopic properties of more than $\sim$26,000 ionized regions corresponding to 924 galaxies from the Integral Field Spectroscopy data, including the flux intensity and equivalent widths and the properties of their underlying stellar population. In the current study we derive a set of physical quantities for all these regions based on those properties, including (i) the fraction of young stars; (ii) the ionization strength (using six different estimations);  (iii) the oxygen abundance (using 25 different calibrators); (iv) the nitrogen and nitrogen-to-oxygen abundance; (v) the dust extinction and (vi) the electron density. Using this dataset we explore how the loci in the classical diagnostic diagrams are connected with those quantities, the radial distributions of these parameters, and the inter-relations between themselves and with the properties of the underlying stellar populations. We conclude that many properties of the \HII\ regions are tightly related to the galactic stellar evolution at the location where those regions are observed. Those properties are modulated only as a second-order effect by the properties of the ionizing stars and the ionized nebulae that do not depend on the astrophysical context in which they are formed. Our results highlight the importance of \HII\ regions to explore the chemical evolution in galaxies, clarifying which of their properties can be used as proxies of that evolution.
\end{abstract}



\section{Introduction}
\label{sec:intro}

\HII\ regions are gas nebulae ionized by young and massive, short-lived, OB stars \citep[e.g.,][]{strom39}. Due to their nature, they are tracers of the star-formation processes in galaxies, frequently found in late-type/disk galaxies \citep[e.g.,][]{kennicutt80}, although they have been recently observed in some early-type ones \citep[e.g.,][]{gomes16b}. Their
nebular emission lines make possible to directly measure the gas-phase abundance at discrete spatial positions in galaxies
\citep[e.g.,][]{sear71,peim78,1996Kennicutt_ApJ456, 2003Kennicutt_ApJ591}. Hence, they are one of the primary tools for constraining galactic chemical evolution theories \citep[e.g.,][]{carigi19}, as the present abundances are the consequence of the enrichment history \citep{2019Maiolino_A&ARv27, 2019Kewley_ARA&A57, 2020Sanchez_ARA&A58}. This connection induces relations between their observational properties, the chemical composition, the properties of the underlying stellar populations, and the global properties of their host galaxies \citep[e.g.][]{2015Sanchez_AA574,2020EspinosaPonce_MNRAS494}.

Previous studies have determined essential relationships, galaxy wide patterns, and scaling laws between the chemical abundances in galaxies and their physicals properties. Among them, the most relevant ones are the surface brightness-metallicity, mass-metallicity, and luminosity-metallicity relations \citep{leque79, skill89, vila92, zaritsky94, 2004Tremonti_ApJ613, 2018Barrera_ApJ852}, characteristics vs. integrated abundances \citep{moustakas06}, or characteristic shapes of abundance gradients  \citep[e.g.][]{vila92}. These relations were derived either for integrated properties of galaxies or using limited samples of \HII\ regions (and for a small number of galaxies), limiting our understanding of the chemical evolution derived from them.

In the last years, the new generation of emission-line surveys based on Integral Field Spectroscopy (IFS) instruments and multi-object spectrometers, covering large Fields-of-view (FoVs), offer catalogs of \HII\ regions with hundreds of objects covering the full optical extension of nearby galaxies \citep[e.g., ][]{rosales-ortega10, 2012Sanchez_A&A538A, sanchez14}. The \HII\ region catalogs produced by these surveys are based on large samples of galaxies covering a wide range of morphological types \citep{sanchez12b, sanchez13}. Significant results have been determined using these catalogs: (i) the existence of a typical abundance gradient \citep{sanchez14, 2015Sanchez_AA573, 2018Sanchez-Menguiano_A&A609}, (ii) the deviations from this radial abundance gradient induced by spiral arms  \citep{2020SanchezMenguiano_MNRAS492} and their azimuthal variations \citep{2017Vogt_AA601}, (iii) the correlation between the oxygen abundance of \HII\ regions and the stellar mass densities of underlying stellar populations \citep{rosales-ortega:2012, 2016Barrera_MNRAS463}, (iv) the existence of average patterns in the radial gradient of different physical properties \citep{sanchez12b}, (v) the existence of a large number of \HII\ regions contaminated by supernovae remnants \citep{cid-fernandes21}, (vi) chemical evolution models at kpc scales \citep{2018Barrera_ApJ852}, and (vii) the comparison between the pre-SNe feedback and environmental pressure across the optical extension of spiral galaxies \citep{barnes21}, among the most important ones.

In \cite{2020EspinosaPonce_MNRAS494}, we presented a new catalog of \HII\ regions based on the Calar Alto Legacy Integral Field Area Survey \citep[CALIFA][]{2012Sanchez_A&A538A} data. We used the extended sample (hereafter, eCALIFA) that comprises the CALIFA (692 galaxies) and PISCO samples (232 galaxies) \citep{DR3,2018Galbany_ApJ855}. We use the tool \textsc{pyhiiexplorer}\footnote{https://github.com/cespinosa/pyHIIexplorerV2} to detect, segregate, and extract the main spectroscopic properties of these \HII\ regions. Then, we built the most extensive and homogeneous catalog of spectroscopic properties of \HII\ regions in the nearby Universe. This catalog has allowed us to confirm the correlations between the ionization conditions in \HII\ regions and the properties of the underlying stellar populations. In particular, the relation between [\ion{O}{iii}]$\lambda 5007/$H$\beta$ and [\ion{N}{ii}]$\lambda 6583/$H$\alpha$ ratios with the age and metallicity of the underlying stellar populations \citep[previously found in][]{2015Sanchez_AA574}. More recently we have used this catalog to explore the [$\alpha$/Fe] patterns in galaxies \citep{2020Sanchez_ARA&A58}.

In the current study we derive the main physical properties of the \HII\ regions in this catalog, including different derivations for the oxygen and nitrogen abundances using a wide set of state-of-the art calibrators, together with estimations of the ionization strength, electron density, and dust extinction. We explore the distribution of those properties in the classical diagnostic diagrams, connecting them with stellar population properties. Finally, we explore the gradients of all the derived physical properties and characterize them by the mass and morphology of the galaxies.

The structure of the article is as follows: the data explored in this study are described in Sec. \ref{sec:data}, including a brief description of the sample of galaxies, and the procedure to detect and extract the properties of the \HII\ regions; the analysis and results are presented in Sec. \ref{sec:ana_res}, with the derivations of the properties for the \HII\ regions described in \ref{sec:physical_prop}. The main trends across the diagnostic diagrams are presented in Sec. \ref{sec:trends}, while the radial gradients are shown in Sec. \ref{sec:radialGrads}. Finally, the relations of the oxygen abundance with other physical properties of the \HII\ regions and the properties of the underlying stellar populations are shown in Sec. \ref{sec:rel_OH} and Sec. \ref{sec:OH_st}, respectively; the main conclusions of this study are presented in Sec. \ref{sec:con}.

\section{Data}
\label{sec:data} 

We use the catalog of spectroscopic properties of HII regions described in the previous section. This catalog was extracted from the sample of $\sim1000$ galaxies comprising the eCALIFA survey. This sample was primarily diameter-selected to have a characteristic projected size of $\sim$60$\arcmin$ \citep[][]{walcher14}. It spans through a narrow redshift range centred at $z\sim0.015$, covering a wide range of stellar masses and morphologies. In summary, it is a somehow representative sample of the galaxies in the nearby universe.

The whole dataset was observed at the 3.5m telescope of the Calar Alto Observatory, using the Potsdam Multi Aperture Spectrograph \citep[PMAS; ][]{roth05} in the Pmas fiber PAcK configuration \citep[PPAK; ][]{kelz06}. This configuration provides a sufficient field-of-view ($74''\times 64''$) to map the full optical extent of these galaxies. It is possible to map up to 2.5 effective radii due the diameter sample selection, described before. The observing strategy guaranteed a final spatial resolution of $FWHM \sim 2.4''$ ($\sim 0.8$ kpc) at the sample's average redshift \citep{DR3}. The current sample was observed using the V500 setup, which provides a spectral resolution of $\lambda/\Delta\lambda\sim 850$ and a wavelength range ($3745-7200$\AA). This is sufficient to study the most important ionized gas emission lines in the optical range (from [\ion{O}{ii}]$\lambda 3727$ to [\ion{S}{ii}]$\lambda 6731$, at the average redshift of the explored galaxies). The wavelength range and resolution are good enough to deblend these emission lines from the underlying stellar population, and to derive the main properties of both components \citep[e.g.][]{2013CidFernandez_AA557,cid-fernandes14, kehrig12, 2012Sanchez_A&A538A,2016Sanchez_RMxAA52a}.

The data was reduced with the CALIFA v2.2 pipeline \citep[which is described in detail in ][]{DR3}. The reduction comprises the standard procedures for this kind of data outlined in \citep{sanchez06a}. This procedure provides for each observed galaxy a regular-gridded data-cube, with two axis recording the spatial dimensions (with a sampling size of 1$\arcsec$/spaxel), and the remaining one recording the spectral dimension. These data cubes were analyzed using the \textsc{pipe3d} pipeline in order to derive the spatial resolved properties of the ionized gas emission lines and the stellar population. This tool was developed (and it has been used) to analyze different IFS data, e.g. CALIFA \citep{2016Sanchez-Menguiano_AA587, carlos19}, MaNGA \citep{2018Barrera_ApJ852, ibarra16, 2018Sanchez_RMxAA54}, MUSE \citep{2020LopezCoba_AJ159}, and SAMI \citep{sanchez19}. The underlying stellar population is fitted using the GSD156 single stellar population library \citep{2013CidFernandez_AA557}, which comprises 156 templates: 39 stellar ages (from 1 Myr to 14.1 Gyr), and four metallicities (Z/Z$_\odot$=0.2, 0.4, 1, and 1.5). The derived stellar spectrum model then is subtracted from the original one for each spaxel to obtain gas-pure data-cube, that it is then analyzed to extract the properties of the emission lines. 
The final dataproducts of this analysis is a set of maps that retain the world coordinate system of the original cubes comprising the spatial distribution of each particular quantity. 

The segregation of \HII\ regions and extraction of the corresponding emission lines information is performed by
applying the \textsc{pyhiiexplorer} tool to the dataproducts provided by  \textsc{pipe3d}. This semiautomatic procedure detect and segregate candidates to \HII\ regions based on three assumptions: (i) \HII\ regions have a strong and peaky \Ha\ emission that is clearly recognizable above both the continuum emission and the diffuse ionized gas (DIG) emission across the galaxies; (ii) the typical size of an extragalactic \HII\ regions at this cosmological distance is of the order of the size of the PSF; (iii) the underlying stellar population at the location of the selected regions has enough young stars to produce the observed ionization, as described in detail in \cite{2020EspinosaPonce_MNRAS494}. The input parameters used to identify the ionized regions are: (i) a flux intensity threshold for each ionized region peak emission of $3\times 10^{-17}\,\mathrm{erg}\,\mathrm{s}^{-1}\,\mathrm{cm}^{-2}\,\mathrm{arcsec}^{-1}$, (ii) a minimum relative flux to the peak emission for associated spaxels of the same \HII\ region of 5\%, (iii) a maximum distance to the peak spaxel of $5.5$ arcsec, and (iv) an absolute flux intensity threshold of $0.5\,\mathrm{erg}\,\mathrm{s}^{-1}\,\mathrm{cm}^{-2}\,\mathrm{arcsec}^{-1}$ in the adjacent pixels to associate them to the peak spaxel. The initial sample of ionized regions identified by \textsc{pyHIIexplorer} includes the information of $38\,807$ objects from $924$ galaxies. The $\sim 70\%$ of the total sample corresponds to \HII\ regions, the $6.4\%$ corresponds to DIG powered by hot old low mass evolved stars (HOLMES), and $\sim 3\%$ corresponds to AGN-like \citep[for further details see][]{2020EspinosaPonce_MNRAS494}.

The result of this analysis for the full eCALIFA sample is a catalog that comprises the information of 51 emission lines in the considered wavelength range and the properties of the underlying stellar population for $\sim$26,000 \HII\ regions. However, not all  those regions are suitable for exploring the distributions along the diagnostic diagrams and to derive good quality physical quantities, as we will show in the forthcoming sections. Therefore we select a sub-sample of regions following the procedures that we will describe in Sec \ref{sec:physical_prop}. We use this catalog as the starting point of the current analysis.

\section{Analysis and Results}
\label{sec:ana_res}

We focus in this study on the exploration of the physical properties of \HII\ regions, trying to determine connections between the ionization conditions, the nebular emission and the properties of the underlying stellar populations. Such connection may unveil which properties of \HII\ regions are related to the evolution of different locations in galaxies (and which ones are not).

\subsection{Empirical estimation of physical properties of HII regions}
\label{sec:physical_prop}

The physical properties of \HII\ regions have been frequently explored using the classical diagnostic diagrams \citep{baldwin81, veilleux87, 2001Kewley_ApJ556}. It is known that the location in these diagrams depends on the properties of the ionizing source (a stellar population in the case of an \HII\ region), and the internal physical conditions, structure, and chemical composition of the ionized gas \citep[e.g.][]{evans85, dopita86, 2001Kewley_ApJ556, 2003Kauffmann_mnras346, 2015Sanchez_AA573}. In summary, six main parameters define the location in these diagrams: (i) the shape/hardness of the ionizing spectral energy distribution (SED), that in an \HII\ region is defined by the sum of the spectra of the ionizing stars (a cluster of them at the typical resolution of our data); 
(ii) the mean ionization parameter across the nebula, defined as the Lyman continuum photons with respect to the total amount of hydrogen (defined as $U$, see below); (iii) the chemical abundance of different elements in the ionized gas, relative to hydrogen (e.g., O/H, N/H...); (iv) the electron density of the gas \citep[e.g.,][]{2015Sanchez_AA574, 2016Morisset_aa594A, 2019Kewley_ARA&A57}, denoted as $n_e$; (v) the dust extinction, characterized by the extinction at the V-band (A$_{\rm V}$) and a certain extinction curve and (vi) the optical depth at 13.6~eV (matter-bounded regions show higher ionization, due to the reduction of the low ionization zone). Following \cite{sanchez14}, we investigate the main trends across the diagnostic diagrams of different physical properties of the \HII\ regions: fraction of young stars, ionization parameter, oxygen abundance, nitrogen abundance, electron density, and dust extinction.

The fraction of young stars that contributes to the observed spectrum within the aperture at which the \HII\ region (or cluster) is selected as a gauge of how strong is the contribution of the ionizing OB stars. It is also a proxy of how evolved is the area in the galaxy (or the host galaxy itself), since it weights the relative amount of old-stars with respect to the ionizing ones. Furthermore, it is a proxy of the shape/hardness of the ionizing spectrum, since this will be the product of the mix of the young and old ionizing stars \citep[e.g., hot evolved stars, HOLMES,][]{2011Flores-Fajardo_MNRAS415}. In \citep{2020EspinosaPonce_MNRAS494} we defined this parameter as the fraction of light in the V-band corresponding to stars younger that 350 Myr.

The ionization parameter is defined as $U(r)=Q(H^0)/4\pi r^2 n_H c$, where $Q(H^0)$ is the number of $H^0$-ionizing photons emitted by the source per unit of time, $r$ is the distance between the source and a particular location within the nebula, $n_H$ is the hydrogen density, and $c$ is the speed of light. 
By measuring the ratio between the photons and the hydrogen atom densities, $U$ gives an insight to the efficiency of the radiation that ionize the gas. The higher $U$ is, the easier the radiation can ionize metals. The value of U is straightforward to determine in theoretical models, but difficult to determine from observations. Its mean value within an \HII\ region is usually estimated using different line ratios, like [\ion{O}{ii}]$/$[\ion{O}{III}], as photoionization models unveil a connection with this parameter and O+/O++ \citep[e.g.][]{diaz00}.
However, the reported connection depends strongly on many other properties of the nebulae, in particular on its shape (shell or filled), on the shape of the ionizing SED as well as on the optical depth of the nebula at 13.6~eV. Despite of this issue, different calibrators have been proposed that links this proxy with $U$ \citep[e.g.]{diaz00, dors11, 2016Morisset_aa594A, 2019Kewley_ARA&A57}. We adopted as fiducial calibrator the one presented by \cite{2016Morisset_aa594A}, which is an update of the relation first proposed by \cite{diaz00} and \cite{dors11}. The [\ion{S}{ii}]/[\ion{S}{iii}] line ratio is frequently claimed to get a better estimation of the ionization parameter \citep[][]{2019Kewley_ARA&A57}, but the [\ion{S}{iii}]$\lambda 9069+\lambda 9532$ emission line are not available in the spectral range of our data. Additional estimators of $U$ have been considered along this article, and discussed in Appendix \ref{appx6:OHvsUs}.

The abundance of a particular element (e.g., oxygen),
may be derived in theory by adding up all its ionic abundances. On the one hand, this can be done by measuring the flux of the recombination lines corresponding to the considered element. Unfortunately, most of those lines are extremely weak in the optical range. If those lines are not accessible, it is possible to derive O/H using the direct method, this is by measuring the flux of collisionally excited lines \citep[read][for a review on this topic]{2017Peimbert_PASP129, 2017PerezMontero_PASP12P}. For doing so, it is required to derive the electronic temperature and density of the nebulae. The estimation of the temperature requires to derive the flux ratio between a set of auroral to nebular lines, e.g., [\ion{O}{iii}]$\lambda 4363/\lambda 5007$. Then the O/H is derived computing first certain ionic abundances. This method assumes that the temperature distribution is homogeneous for each considered ion through the ionized extension of the \HII\ region. If this is not the case, it is still possible to apply the so-called $t^2$ correction \citep{peim67}. The direct method relies on auroral lines that are brighter than the recombination ones, but are very weak too (e.g.,  [OIII]$\lambda 4363$). In many cases they are not accessible with the usual depth and resolution of the IFS data of the most recent galaxy surveys \citep[see][as an example]{2013Marino_aa559A}.

When neither the recombination nor the auroral lines are detected or accessible, the estimation of the gas-phase abundance relies on the use of known relations between particular strong emission line ratios and the required physical properties to derive the abundance \citep[e.g., T$_e$,][]{1979Alloin_AA78,2011Pilyugin_MNRAS412}, and/or the abundances themselves \citep{2017Peimbert_PASP129}. These estimations are called strong-lines calibrators, and they can be derived using two approaches: (1) by comparing the ratios of strong emission lines with direct estimation of the oxygen abundances from observational datasets (based either on the direct method or by the use of collisional lines). Then, a polynomial fitting is performed \citep[e.g.]{2013Marino_aa559A, 2011Pilyugin_MNRAS412, 2016Pilyugin_MNRAS457} or another more complex regression techniques \citep[e.g. machine learning][]{2019Ho_MNRAS485}; and, 
(2) using grids of photoionization models to derive the relation between oxygen abundances and emission line ratios \citep[e.g.][]{kewley02, 2018Thomas_ApJ856}. Unfortunately, the derived oxygen abundance using the two methods shows a discrepancy up to $\sim 0.1$-$0.4$ dex \citep{kewley08,2015Blanc_ApJ798}. This discrepancy may be due to (i) the presence of electron temperature fluctuations in HII regions \citep{peim67}, (ii) uncertain ionization corrections factors, (iii) over-simplicity of the photoionization models used to obtain calibrations \citep{2003Kennicutt_ApJ591} or (iv) high energetic electrons following a non-Maxwellian distribution \citep[e.g. the $\kappa$ distribution][]{2012Nicholls_ApJ752}, among other reasons. The most used strong-line indicators are based on the following emission lines: [OII]$\lambda 3727$, H$\beta$, [OIII]$\lambda 5007$, H$\alpha$, [NII]$\lambda 6583$, [SII]$\lambda 6717+31$. They comprise N2 ($\equiv$[NII]/H$\alpha$), O3 ($\equiv$[OIII]/H$\beta$), O2 ($\equiv$[OII]/H$\beta$), S2 ($\equiv$[SII]/H$\alpha$), the combination among them, like R23 ($\equiv$O2+O3), O3N2 ($\equiv$O3/N2), N2O2 ($\equiv$N2/O2) or N2S2 ($\equiv$N2/S2), and more complex combinations, like the $P$ parameter \citep[e.g.][]{2010Pilyugin_ApJ720}. Each one of these line ratios present different strengths and caveats for the oxygen abundance calculation \citep[e.g.][]{2020Curti_MNRAS491}. An additional complication is that many of the adopted emission line ratios present a bi-valuated relation with the oxygen abundance defining upper and lower oxygen branches \citep[e.g. R23][]{2017Peimbert_PASP129}. 

In this work, we calculate the oxygen abundance with different strong-lines calibrators proposed by \cite{kewley02}, \cite{2004Kobulnicky_ApJ617},  \cite{2004Pettini_MNRAS348}, \cite{2004Tremonti_ApJ613}, \cite{2010Pilyugin_ApJ720, 2011Pilyugin_MNRAS412, 2016Pilyugin_MNRAS457}, \cite{2013Marino_aa559A}, \cite{2019Ho_MNRAS485}, and \cite{2020Curti_MNRAS491}. They comprise an heterogeneous compilation of calibrators, both empirical and theoretical, and using different indicators and mathematical procedures to anchor the abundance. 
We adopt the \cite{2019Ho_MNRAS485} calibrator as the fiducial one along this article, although we provide the measurements of the remaining ones in our published catalog (see Appendix \ref{app:desc_catalog}).

As we discussed for the oxygen abundance, the nitrogen abundance determination has similar issues. Moreover, the recombination and auroral nitrogen emission lines are even more difficult to detect than those of the oxygen. 
Nevertheless, there are relations between several strong emission line ratios and the nitrogen abundance that could be used for the abundance determination (i.e. strong-line calibrations). In this work, we estimate the nitrogen abundance (N/H) and nitrogen-to-oxygen relative abundance (N/O) with the calibrations described by \cite{2016Pilyugin_MNRAS457} and \cite{2019Ho_MNRAS485}. We adopt this later one as our fiducial one.

\begin{figure*}
\includegraphics[width=\columnwidth]{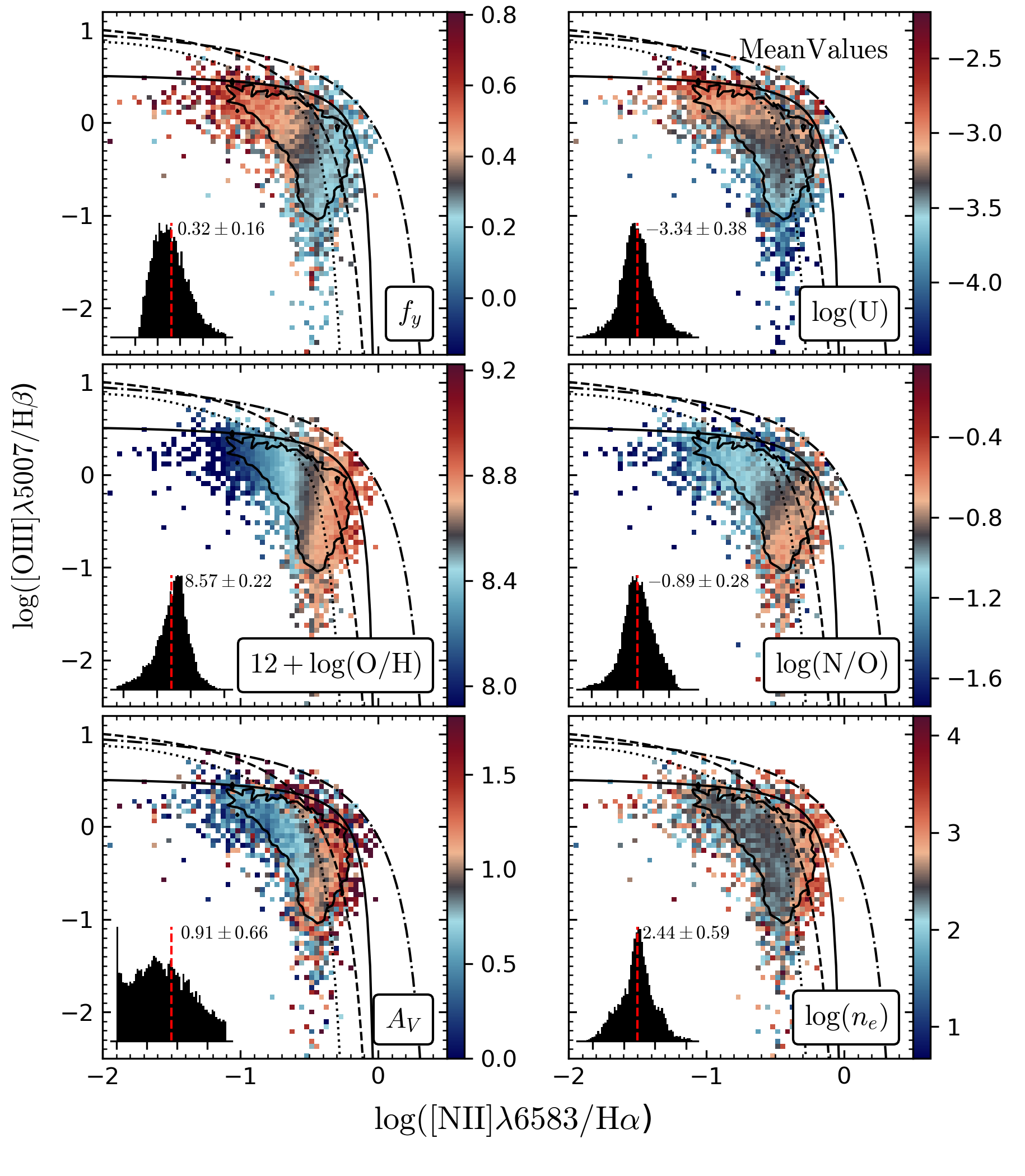}
\includegraphics[width=\columnwidth]{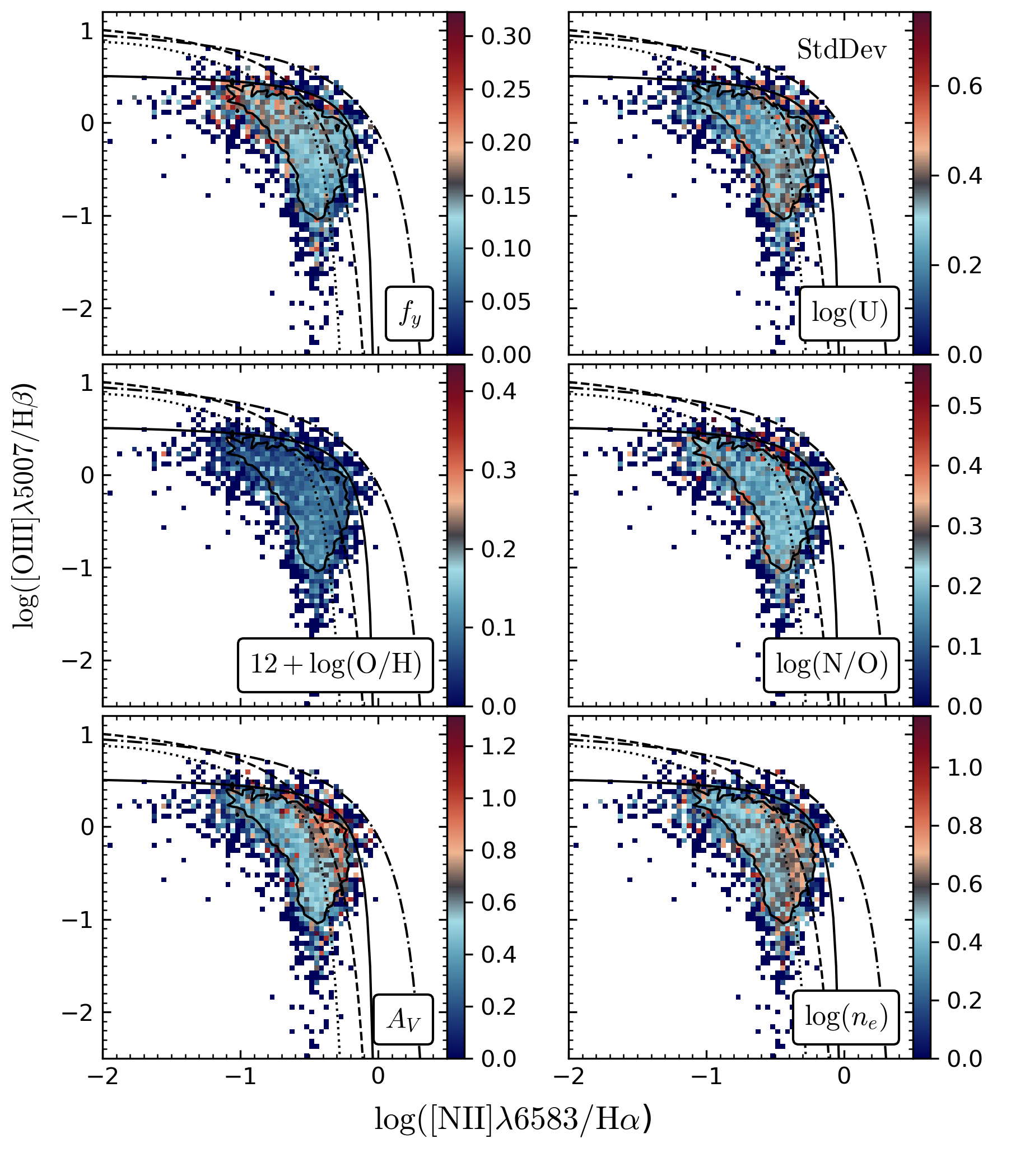}
\caption{Average physical properties (left panel), together with their standard deviation (right panel) for the \HII\ regions at a certain location within the [\ion{O}{iii}]$\lambda5007/$H$\beta$ vs. [\ion{N}{ii}]$\lambda6583/$H$\alpha$ diagnostic diagram. In each panel the color code represents a physical property described in the text: \textit{top left panel}: fraction of young stars to the total luminosity, $f_{y}$; \textit{top right panel}: ionization parameter; \textit{middle left panel}: oxygen abundance; \textit{middle right panel}: nitrogen-to-oxygen abundance; \textit{bottom left panel}: dust extinction ; and \textit{bottom right panel}: electron density ($\log(n_e)$). In each panel,
the scale of the colormap was selected to cover the dynamical range of the physical parameter: $\pm$3$\sigma$ range around the mean value in the left panels, and between zero and 2$\sigma$ in the right panels, for all parameters but $A_{\mathrm{V}}$. For this parameter the dynamical range in the left panel is restricted to 0 and two times the mean value ($\sim$2 mag). The inset within each panel shows the distribution of the corresponding property. 
The contour represent the density distribution enclosing 95$\%$ of the regions.
For all the panels, the dot-dashed line, dashed-line, dotted-line, and solid line represents the demarcation lines proposed by \citet{2001Kewley_ApJ556, 2003Kauffmann_mnras346, 2008Stasinska_MNRAS391, 2020EspinosaPonce_MNRAS494}, respectively.}
\label{fig:BPT_O3N2_physical_properties}
\end{figure*}

The electron density, $n_{e}$, is derived from the ratio of emission lines of the same ion originated from levels with similar energy \citep[e.g.][]{osterbrock89}. In those cases the emission line ratio does not depend strongly on the temperature \citep{2017Peimbert_PASP129}. In particular, we use the [\ion{S}{ii}] doublet, by solving the equation:

\begin{equation}
  \frac{[\ion{S}{II}]\lambda 6717}{[\ion{S}{II}]\lambda 6731} = 1.49 \frac{1+3.77x}{1+12.8x}
\end{equation}

\noindent where $x=10^{-4}n_{e}t^{-1/2}$ and $t$ is the electron temperature in units of $10^{4}$ K \citep{McCa85}. We assume a typical electron temperature of $T=10^{4}$ , that corresponds to the usual conditions in \HII\ regions \citep{osterbrock89}. We should note that, as we indicate before, the dependence with the $t_e$ is very weak. Unfortunately, the [\ion{S}{ii}] doublet ratio is sensitive to electron density only for a limited range of values (from $\sim$50 cm$^-3$ to $\sim$7000 cm$^-3$). Nevertheless, this calibrator is reasonably robust to study the main trends explored in this study.

The dust extinction ($A_{v}$) was estimated using the Balmer line ratio (H$\alpha$/H$\beta$). We assume the extinction law by \cite{cardelli89} using a theoretical value for the unobscured line ratio for case B recombination H$\alpha$/H$\beta=2.86$ (for $T_{e} = 10\,000$ K and $n_{e} = 100$ cm$^{3}$) and a specific dust extinction of $R_{v}=3.1$. Despite the dependence between the H$\alpha$/H$\beta$ and the electron density and the temperature, we consider this variations negligible \citep{2007Sanchez_AA465}.

A catalog with all the physical properties derived for the \HII\ regions is distributed as described in Appendix \ref{app:desc_catalog}. From this catalog we apply different filters to ensure the quality of the derived properties: (i) the [\ion{O}{iii}]/\Hb, [\ion{N}{ii}]/\Ha, [\ion{S}{ii}]/\Ha, [\ion{O}{i}]/\Hb and [\ion{O}{ii}]/\Ha line ratios should be positive and finite. Each of these consecutive cuts implies a reduction
in the original number of $\sim 26,000$ \Hii\ regions by 712, 189, 288, 3326 and 8886 regions, respectively. Therefore, the strongest reduction is due to the presence or absence of [\ion{O}{ii}] ($\sim$30\%) in the wavelength range sampled by the data (due to the redshift), followed by the detectability of [\ion{O}{i}] ($\sim$12\%). The final number of selected \Hii\ regions after applying these filter is $\sim$15,000; (ii) we select those regions for which we can derived all the quantities enlisted along this sub-section (\EWHa, $f_y$, $U$, O/H, N/O, $n_e$ and A$_{\rm V}$), for the fiducial calibrators. This further filter removes just $\sim$500 additional regions; (iii) we exclude those regions above the \citet{2001Kewley_ApJ556} demarcation line in the diagnostic diagrams (that we will show in the forthcoming sections). This filter excludes $\sim$450 regions; finally, (iv) we exclude those regions that are hosted by a galaxies with a  inclination greater than 70$^{\circ}$. This final filter excludes $\sim 3600$ regions corresponding to 177 galaxies. The final catalog once applied all the filters comprises $\sim$11,000 regions. The \HII\ regions sample have a threshold in S/N$>3$ for H$\alpha$ emission line from the initial selection of \HII\ regions \citep{2020EspinosaPonce_MNRAS494}.

\subsection{Trends across the line-ratios diagnostic diagrams}
\label{sec:trends}

In this section, we describe the observed trends of these properties across the emission-line diagnostic diagrams. These diagrams show the distribution of pairs of strong emission line ratios among the ones described in the previous section to classify their primary ionization source. The most frequently used one, known as the BPT diagram, represents the O3 versus N2 line ratios \citep{baldwin81}. Other diagnostic diagrams involving different line ratios were proposed by \citet{veil01}, and a combination of them with the equivalent width of H$\alpha$, EW(H$\alpha$), was first proposed by \citet{cid-fernandes10}. In all those diagrams there have been defined a set of demarcation lines that segregate certain regions associated with particular ionization sources \citep[e.g., AGNs/SF][]{2001Kewley_ApJ556, 2003Kauffmann_mnras346, 2008Stasinska_MNRAS391}, although its ability to distinguish between them is now under discussion \citep[e.g.][]{2021Sanchez_RMxAA57}.

Beside the ionizing source, the location of an \HII\ region on this diagram depends on their physical properties  \citep[see][and reference therein, for a discussion]{2019Kewley_ARA&A57, 2021Sanchez_RMxAA57}. 
In the particular case of \HII\ regions, different theoretical works using photoionization models predict the location on the BPT diagrams according to the metallicity and the ionization parameter \citep[e.g.,][]{kewley02, 2016Morisset_aa594A}. They found a primary trend in which metal-poor \HII\ regions are located in the upper-left zone of the diagram, while metal-rich ones are located in the right-bottom zone.
Unfortunately, the photoionization models present different degeneracies in this diagnostic diagram between the oxygen abundance, the ionization strength, and the relative nitrogen-to-oxygen abundance (among others). As a result, one location in the diagram can be reproduced by a set of different models corresponding to different physical conditions. From the observational point of view other trends have been reported. For instance, \citet{2015Sanchez_AA573} described a clear trend in which \HII\ regions located in regions with older stellar populations are more frequently found in the bottom-right part of the diagram, while those with younger stellar populations are found towards the upper-left zone. This trend, clearly connected with the one described for the metallicity, cannot be explained by photoionization models, being a consequence of the linked co-evolution of the stellar populations and their metal content at a certain location in galaxies. Furthermore, other physical properties show additional trends across the diagnostic diagrams. For instance, there is a clear trend with the electron density easily explained by photoionization models \citep[e. g.][]{2013Kewley_ApJ774}. Finally, it was described a trend with dust extinction, despite the fact that the involved line ratios are almost insensitive to dust \citet{2015Sanchez_AA573}, and even with the gas velocity dispersion \citep[][]{law21}.

\cref{fig:BPT_O3N2_physical_properties} shows the distribution of the physical properties described in Sec. \ref{sec:ana_res} across the BPT diagram for our sample of \HII\ regions. To generate this figure each physical property is averaged (left panels) for those regions at the same location within the diagram. Once derived the mean value we estimate the standard deviation around it (right panels). 
We first explore the trends with the fraction of young stars. The \HII\ regions located on the upper-left zone of the diagram show the largest value of $f_y$; meanwhile, the HII regions located on the bottom-right zone had the lowest one. This result agrees with the trends already reported by \citet{2015Sanchez_AA573}, highlighting the connection between the properties of the underlying stellar population and the observed line ratios. Thus, \HII\ regions generated in (less) evolved regions of a galaxy, those with (younger) older and (less) more metal rich stellar populations, are preferentially found in the areas with (higher) lower $f_y$. Similar trend is observed in the top-right panel of Fig. \ref{fig:BPT_O3N2_physical_properties}, that shows the same distribution color-coded by the ionization parameter. The \HII\ regions with a high value of the ionization parameters are located in the upper zone of the diagram, and regions with low values of $U$ are located in the bottom zone. A connection between $U$, $f_y$, and the metal content in \HII\ regions maybe be the reason for these trends \citep[e.g.][]{2021Ji_arXiv211000612}.

The middle panels of Fig. \ref{fig:BPT_O3N2_physical_properties} show the same distribution color-coded by the oxygen abundance (left) and the nitrogen-to-oxygen relative abundance (right). As expected, the metal-poor  \HII\  regions are located in the upper-left zone of the diagram, and the metal-rich ones are located in the bottom-right zone. A similar trend is found for the nitrogen-oxygen ratio, as a direct consequence of relation between this quantity and the oxygen abundance \citep[e.g.][]{1986Matteucci_MNRAS221, 1993VilaCostas_MNRAS265, 2012Pilyugin_MNRAS421}. Thus, the \HII\ regions with the lowest value of N/O ratio are located on the upper-left zone of the diagram, and those with the largest ratio value are located on the bottom-right zone.

The distribution traced by the electron density ($n_e$) is shown in the bottom-right panel of Fig. \ref{fig:BPT_O3N2_physical_properties}.
The regions with the highest electron density ($\sim$10$^3$cm$^{-3}$) are found on the right-hand area of the diagram, near to the edge of the classical location of \HII\ regions. Then, it is appreciated a sharp transition to values $\sim$10$^2$cm$^{-3}$, distributed along that classical loci of these regions. An increase of $n_e$ towards the upper-right is predicted by photoionization models \citep[e. g.][]{2013Kewley_ApJ774}. However, the described distribution is not that easily replicated based only on these models. We must remark that electron density derived with the [\ion{S}{II}] lines ratio is reliable between $\sim 10^2$ cm$^{-3}$ and $\sim 10^{4}$ cm$^{-3}$. For our current sample we are in any case in the low-density regime for all \HII\ regions.

Finally, the bottom-left panel of Fig. \ref{fig:BPT_O3N2_physical_properties} shows the distribution of  \HII\  regions across the BPT diagram with a color code representing the dust extinction. There is a clear trend across the diagnostic diagram: dust extinction is lower in the \HII\ regions located in the upper left-hand region, and it is higher in those regions at the bottom right-hand end and in the upper right-hand end of the diagram. The regions with the largest dust extinction are located in the intermediate area delimited by \cite{2003Kauffmann_mnras346} and \cite{2001Kewley_ApJ556} curves. Like in the case of some of the distributions discussed before, this result is not predicted by photoionization models which most of the time do not consider this parameter. We should note that the two line ratios involved in this diagram ([\ion{N}{ii}] and \ion{O}{iii}) are expected to be insensitive to dust extinction.


The distribution of the same physical properties for a set of additional diagnostic diagrams ([OIII]$\lambda5007/$H$\beta$ vs. [SII]$\lambda6716+30/$H$\alpha$ and [OIII]$\lambda5007/$H$\beta$ vs. [OI]$\lambda6300/$H$\alpha$) are shown in Appendix \ref{app:diagnostic_diagrams}. In all cases the distributions cannot be reproduced based on predictions from simple photoionization models. Their origin should be related to the properties of the ambient in which the \HII\ regions are formed. In summary, young massive stars born in regions with a large number of old stars (i.e., low $f_y$), would be formed from a gas that has been recycled and enriched by previous stellar generations (i.e., high oxygen abundance and N/O ratio). The gas from which it is formed should present also a larger amount of dust, and therefore a stronger dust extinction. More difficult is to interpret the trends described for the ionization parameter and the electron density. A possible explanation for the trend described by $U$ could be that \HII\ regions of a high metal content gas, ionized by a cluster of stars with a narrow range of ages, may present a lower ionization parameter due to metal blanketing. However, in this case O/H and $U$ should present just opposite trends across the diagram, and this is clearly not the case. On the other hand, for the electron density, the results suggest that regions with higher dust extinction \citep[i.e., corresponding to large gas densities][]{brinchmann04,jkbb20} do have larger n$_e$ too. This would imply that their internal pressure should be also higher, in agreement with the trends recently reported by \citet{barnes21}. In this case, since $U$ depends inversely on the electron density the trend observed for this parameter could be a combination of the trends reported for both O/H and  $n_e$. Another possibility would be that for the same ionizing stellar cluster the ionization $U$ decreases as  $A_{\rm V}$ increases, as a pure consequence of the dust extinction decreases the available UV ionizing photons \citep{2021Ji_arXiv211000612}. Finally, the observed trend for $U$ could be a consequence of all those effects combined together. Besides, this result suggests that the diagnostic diagrams do not provide a proper description of the properties of \HII\ regions, therefore the models need to consider a parameter space with higher dimensions.

\begin{figure*}
  \includegraphics[scale=0.8,trim=0 27 0 0, clip]{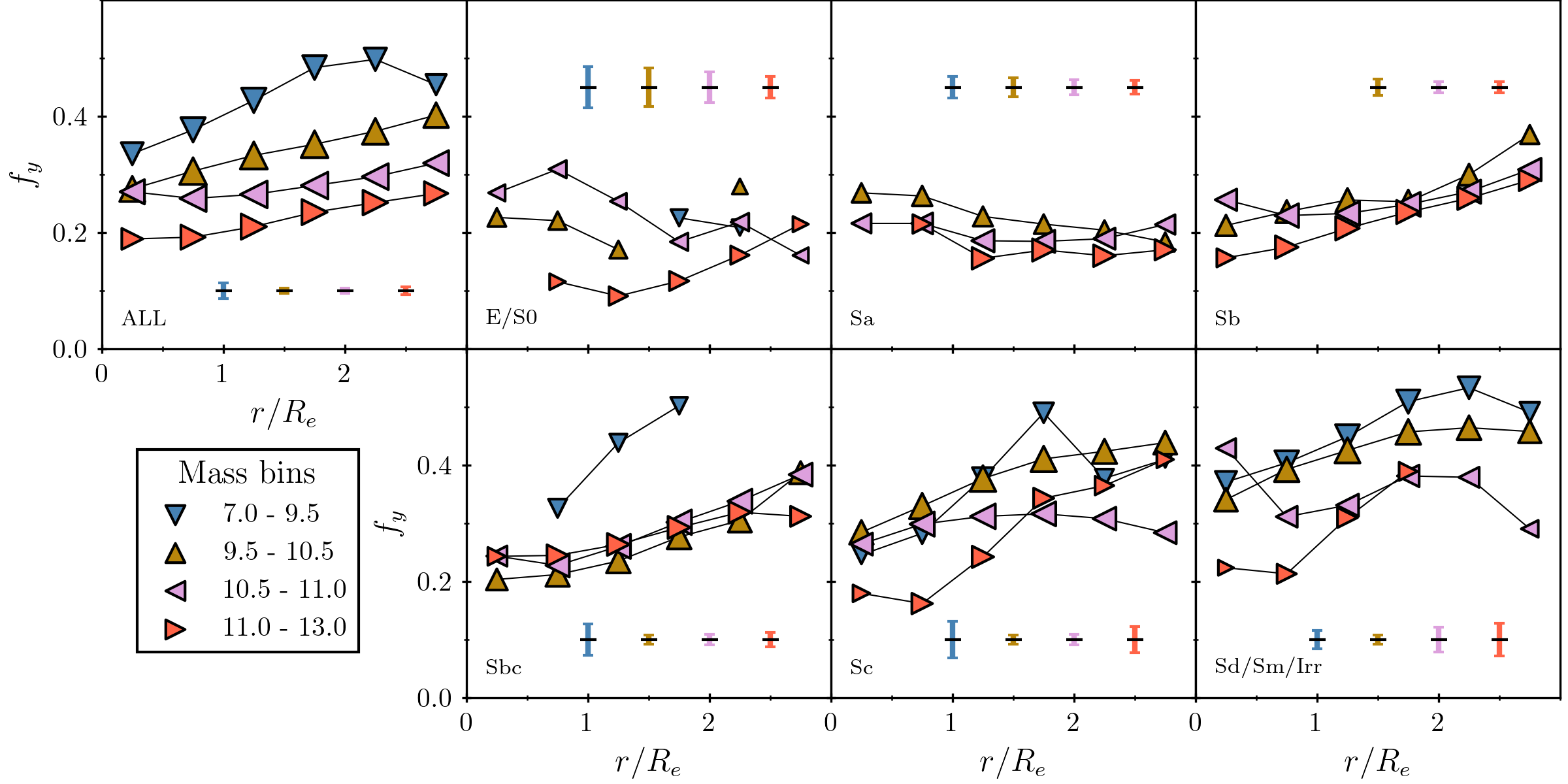}
  \includegraphics[scale=0.8,trim=5 0 0 0, clip]{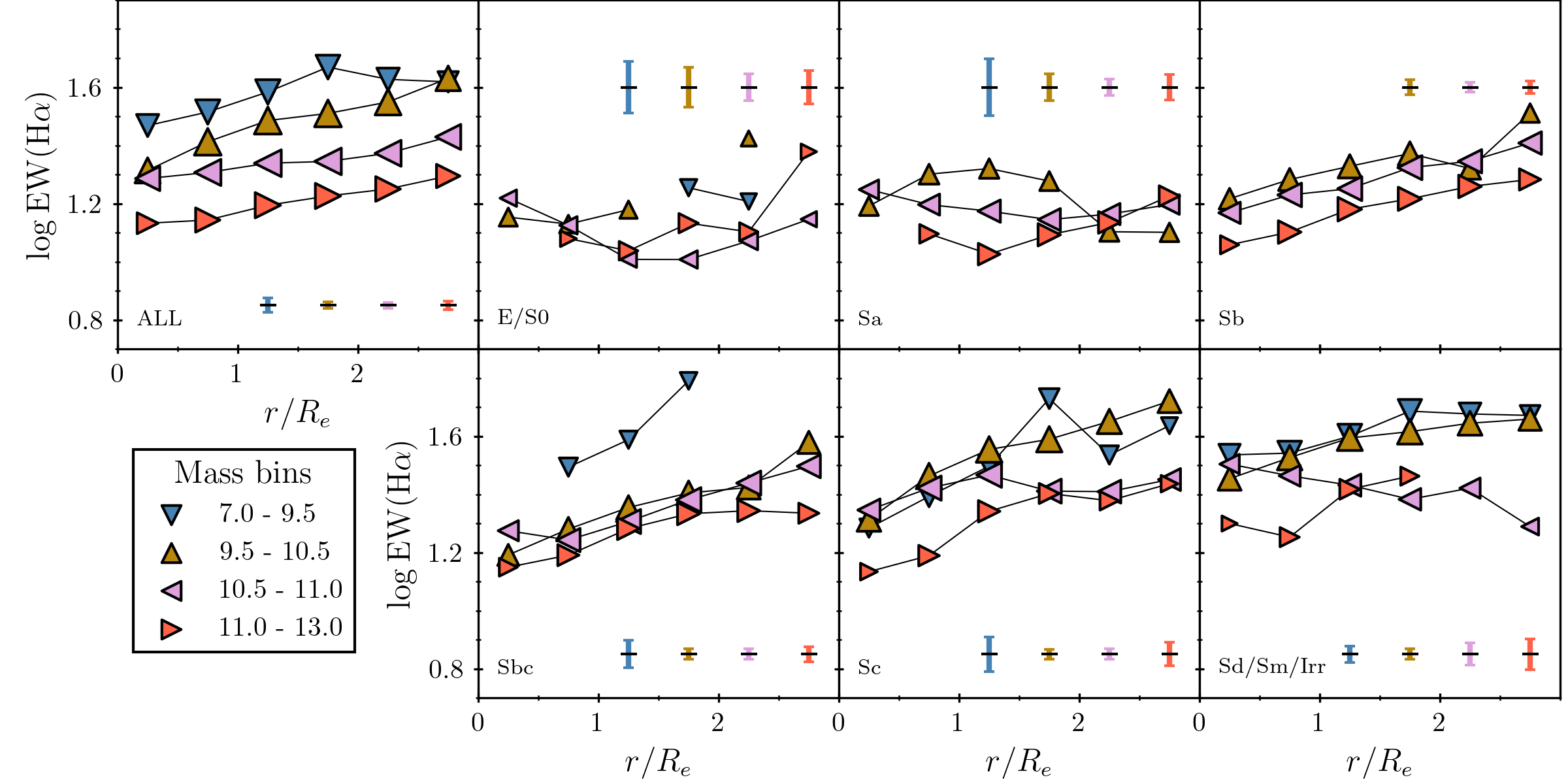}
  \caption{Radial distribution of the $f_{y}$ (top panels) and EW(H$\alpha$) (bottom panels) along the galactocentric distance for the \HII\ regions in our catalog. Each symbol corresponds to the average value of the corresponding parameter in bins of 0.5 effective radius for each stellar mass bin (color coded) and morphology (different panels) of their host galaxy, with its size being proportional to the number of \HII\ regions in each bin. The solid lines link the different values corresponding to the same stellar mass bin. The average standard error of each parameter in each panel is represented with an error bar. }
  \label{fig:DistanceDependenceA}
\end{figure*}

\begin{figure*}
   \includegraphics[scale=0.8,trim=0 27 0 0, clip]{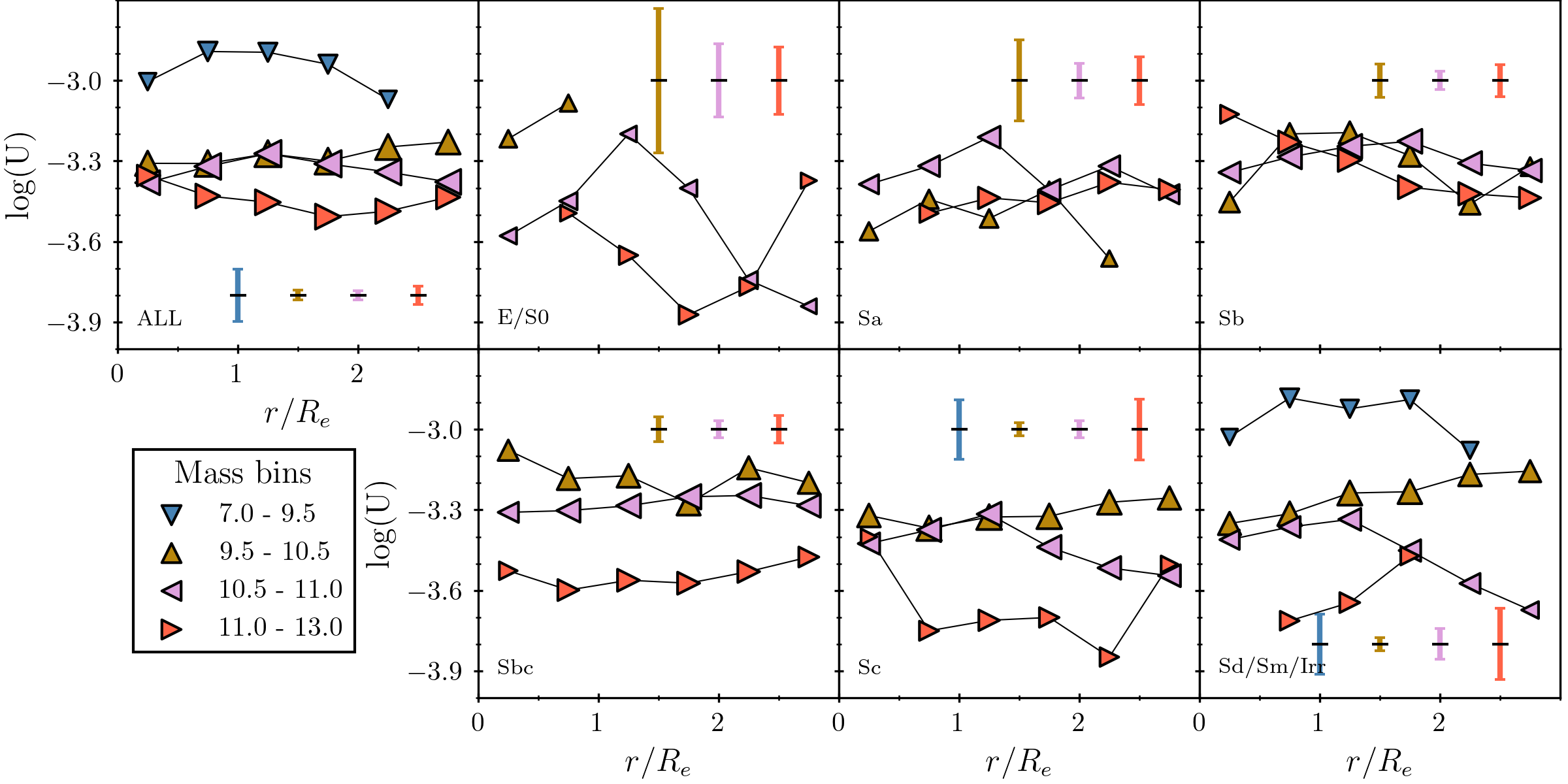}
   \includegraphics[scale=0.8,trim=0 27 0 0, clip]{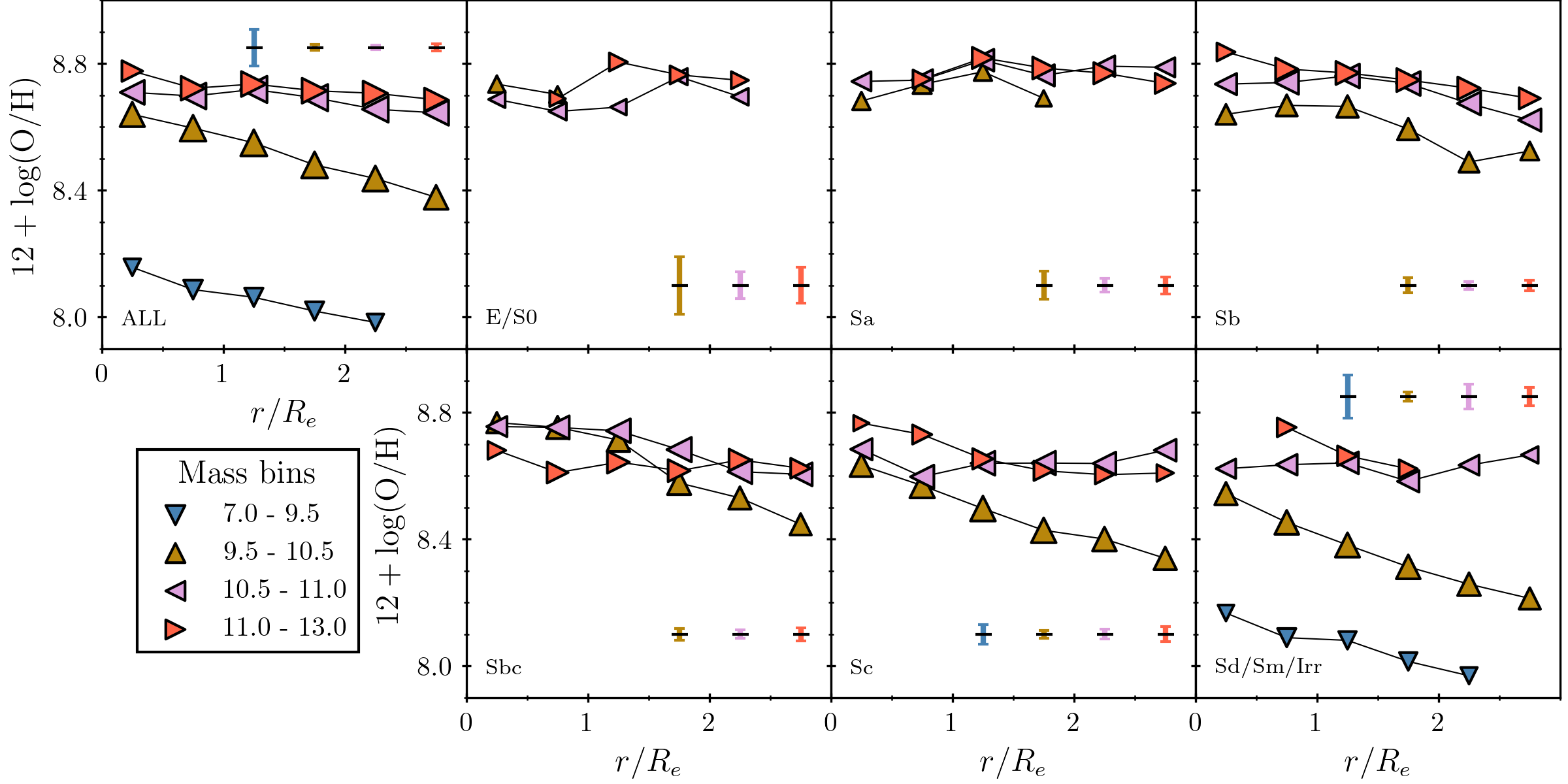}
   \includegraphics[scale=0.8,trim=0 0 0 0, clip]{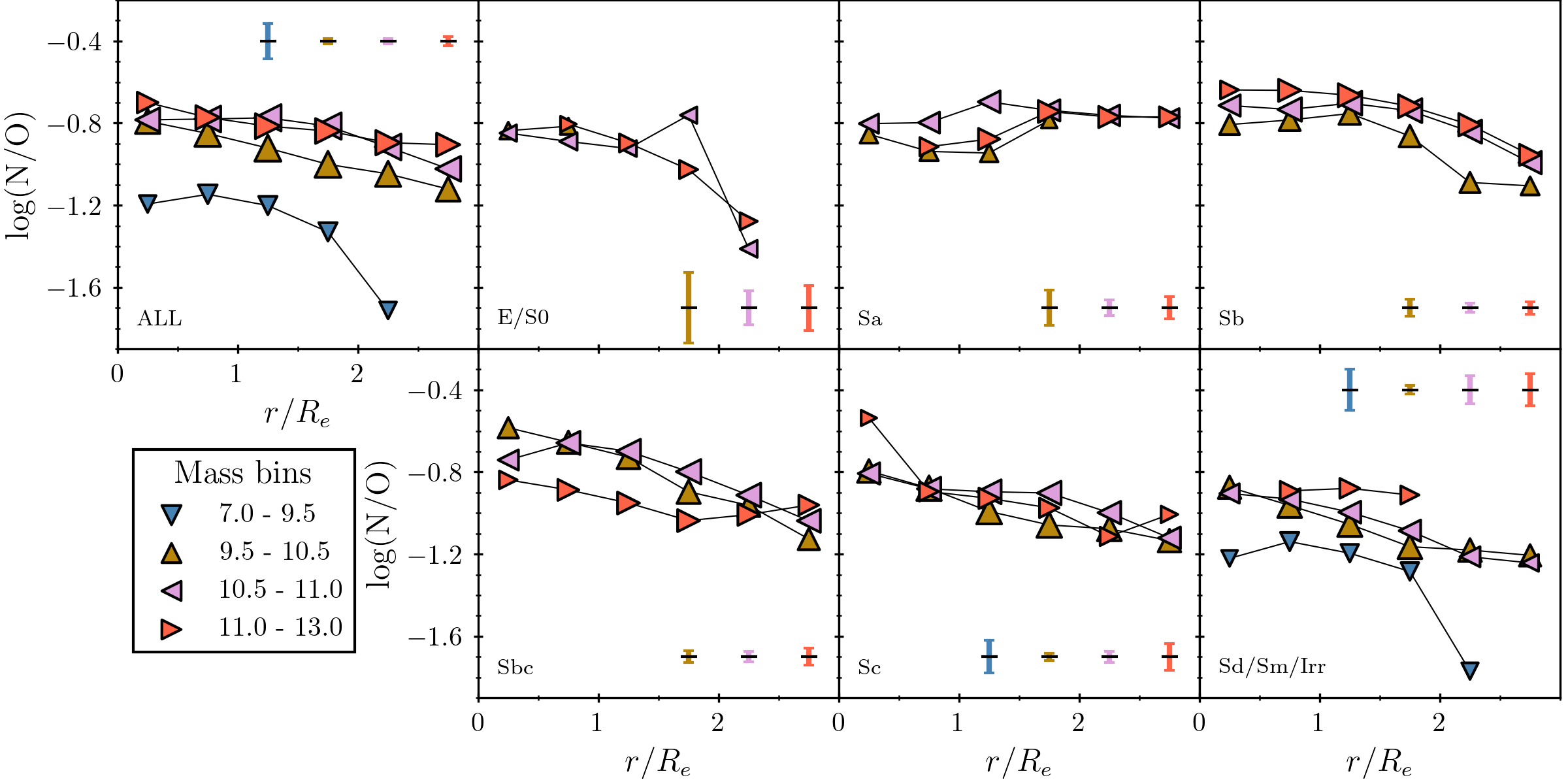}
  \caption{Similar figure as Fig. \ref{fig:DistanceDependenceA}, for the $\log(U)$ (top panels), 12+log(O/H) (middle panels) and log(N/O) (bottom panels) parameters.}
  \label{fig:DistanceDependenceB}
\end{figure*}

The outlined interpretation holds if the described distributions, traced by the average values of the considered parameters, are indeed representative of the bulk sample of \HII\ regions. In other words, if the dispersion around these mean values are small compared to the typical value themselves. As indicated before, the distributions of the standard deviations are shown in the right panels of Fig. \ref{fig:BPT_O3N2_physical_properties} for the same physical parameters shown in the left panels. It is appreciated that in all cases the standard deviation is small with respect to the dynamical range of the data. However, there are differences parameter to parameter. For instance, $f_y$ and O/H present very low standard deviations, of the order of $\sim$0.05-0.1 dex, at almost any location within the diagram, what indicates that indeed the mean values describe well the behavior of the \HII\ regions. Slightly larger dispersions are observed for log(U) and N/O ($\sim$0.2-0.3 dex), both of them
distributed in the same way along the explored diagrams. Thus, despite of the larger spread, for both parameters the mean value seem to describe well the distributions. Much larger scatters are found for $n_e$ and $A_{\rm V}$, with values around $\sim$0.6 dex, for the first parameter, and $\sim$0.3-0.8 mag, for the second one, respectively. Despite the fact that their dynamical ranges are larger, it is clear that the trends reported based on the mean values are just valid at a first order. Furthermore, in these two parameters a trend is observed in the standard deviation, that appears to be smaller at the upper-left area of the diagram than at the bottom-right one. Those regions corresponds to areas with relatively larger electron densities and dust extinction (left panel). Therefore, at this locations the reported trends for $n_e$ and $A_{\rm V}$ should not be interpreted just as an increase of both parameters, but as an enlargement of the range of values for both of them. Accordingly, the outlined trends between both parameters and other physical quantities (O/H and in particular $U$) discussed in the previous paragraph should be revised, considering that they most probably would present larger scatters as $n_e$ and $A_{\rm V}$ increases.

\subsection{Radial distributions of the physical properties}
\label{sec:radialGrads}

\begin{figure*}
   \includegraphics[scale=0.8,trim=0 27 0 0, clip]{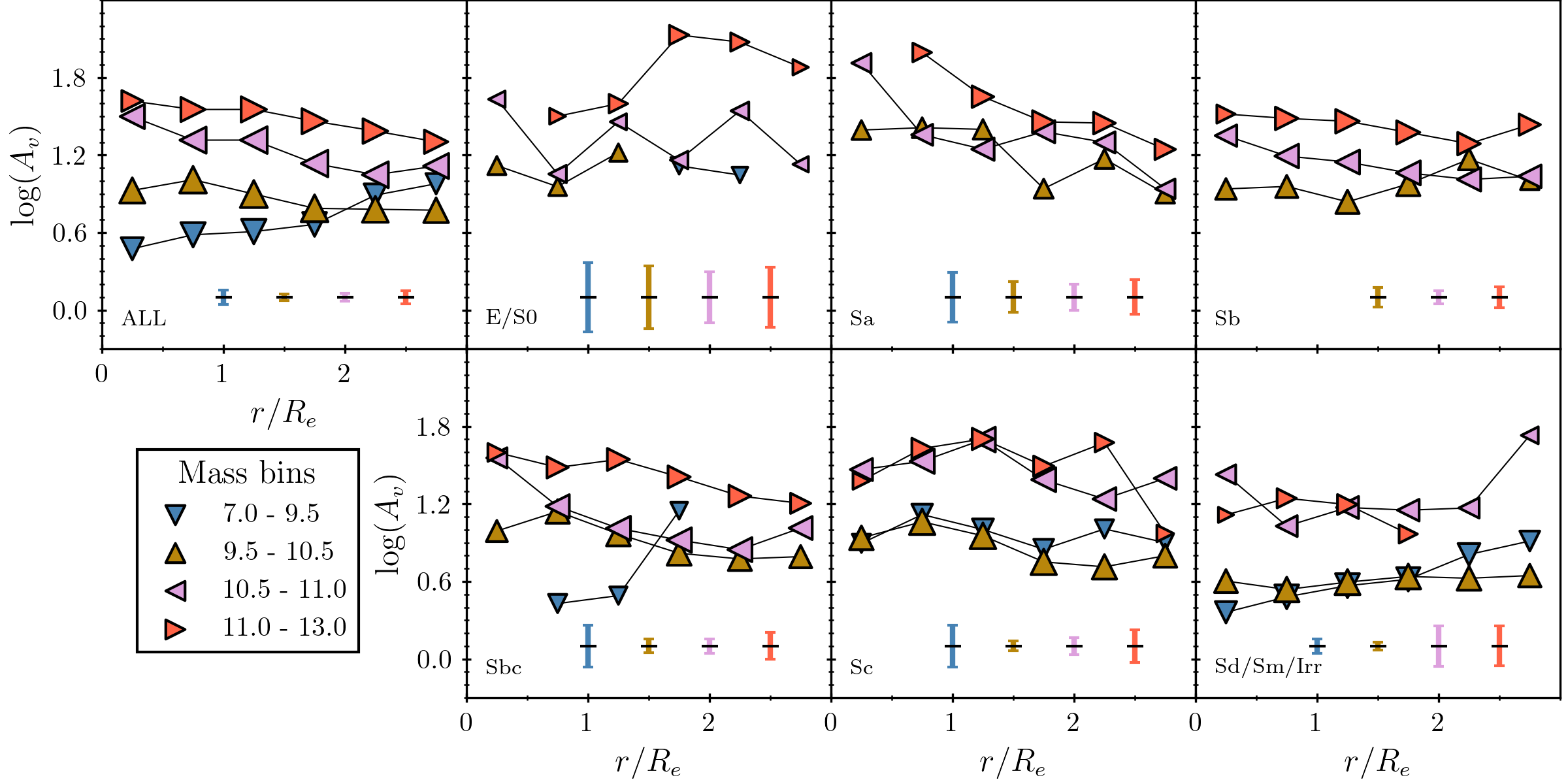}
   \includegraphics[scale=0.8,trim=0 0 0 0, clip]{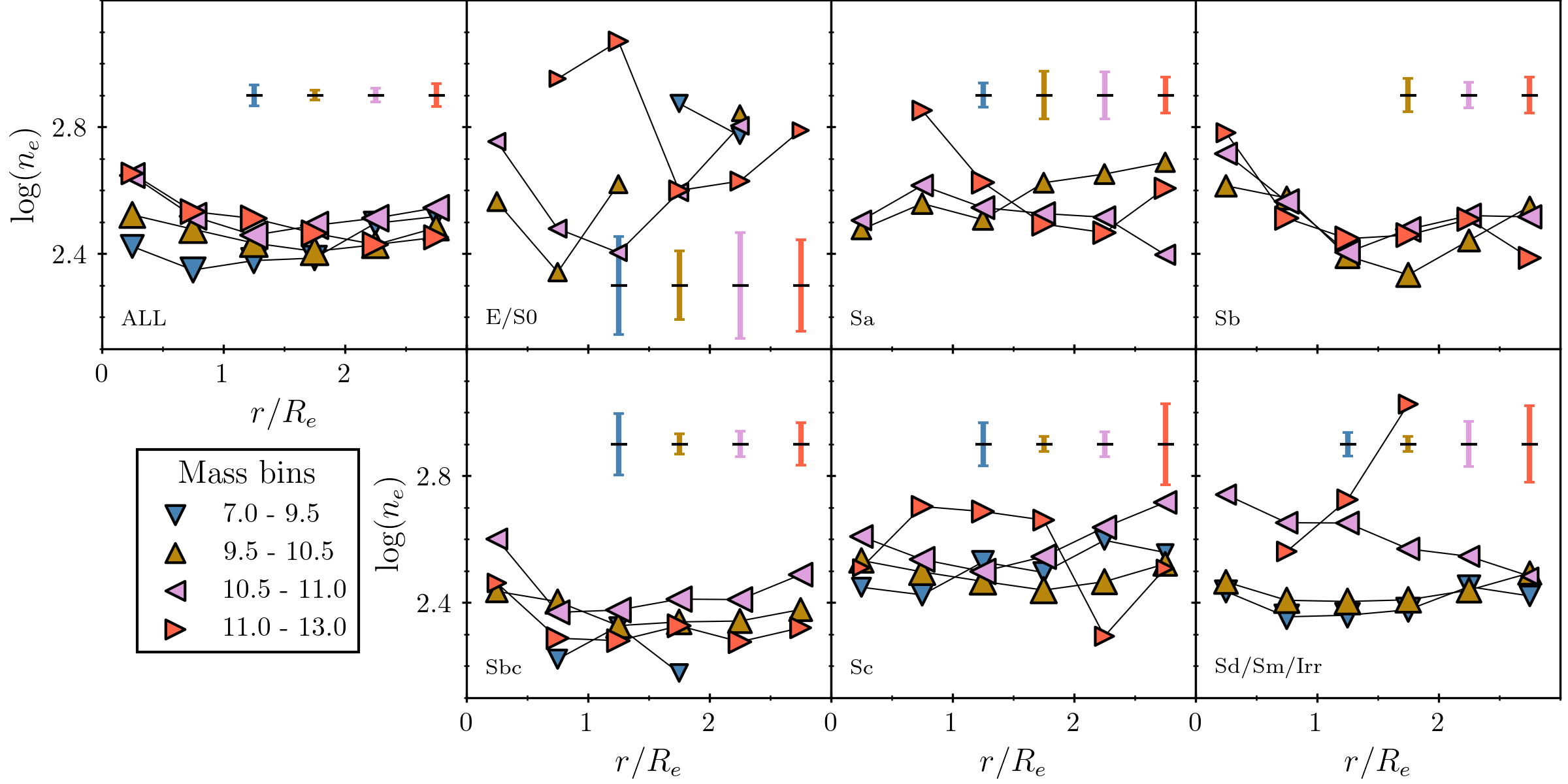}
  \caption{Similar figure as Fig. \ref{fig:DistanceDependenceA}, for the $n_{e}$ (top panels), and A$_V$ (bottom panels) parameters.}
  \label{fig:DistanceDependenceC}
\end{figure*}

The exploration of the distribution of the physical properties of the \HII\ regions across the BPT diagram, shown in the previous section, indicates that the location on this diagram and therefore the observed line ratios present clear dependencies with the physical properties. Since many of these physical quantities are tightly connected with the evolution of the stellar populations (in particular f$_y$, O/H and N/O, but also $A_{\rm V}$), they should present patterns and variations with the same parameters that affects that evolution: (i) galactocentric distances; (ii) stellar mass and (iii) morphology of the host galaxies \citep[e.g.][]{rgb17, 2020Sanchez_ARA&A58}.

In this section we present the radial distribution of the properties described in Sec. \ref{sec:physical_prop} for galaxies of different stellar mass and morphology. For each \HII\ region we derive its galactocentric distance, taking into account the effects of inclination.
Then, we normalize the distance by effective radius of the corresponding host galaxy \citep[$R_e$, derived following ][]{walcher14}. Then, for each galaxy, we obtain the azimuthal average of each property in radial bins of 0.5 $r/R_e$ (a size that guarantees a sufficient number of \HII\ regions within each bin). Finally, we segregate the galaxies in four bins of stellar mass (log(M$_*$/M$_{sun}$=7-9.5, 9.5-10.5, 10.5-11.0 and 11.0-13.0), and six morphological types (E/S0, Sa, Sb, Sbc, Sc and Sd/Sm/Irr). For each sub-sample we obtain the average radial distribution of each physical property. Finally, in order to quantify the variation in the scale and gradient of each property, we perform a linear fitting to each average radial distribution.

\cref{fig:DistanceDependenceA}, \ref{fig:DistanceDependenceB} and \ref{fig:DistanceDependenceC} summarize the result of this analysis. For each physical property we show the average radial profile segregated by the stellar mass of the galaxies for all the sample (first column panels), and for each morphological type (right panels). The values of the zero-point ($c_0$), and slopes ($c_1$), derived from the linear regression to each of these radial distributions are presented in Appendix \ref{app:grads}, listed in Tables  \ref{tab:fy_coeffs}, \ref{tab:EWHa_coeffs}-\ref{tab:Av_coeffs}. The zero-point corresponds to the value of the physical quantity in the central regions of the galaxies while the slope is a gauge of the strength of the radial gradients.

The upper panels of Fig. \ref{fig:DistanceDependenceA} shows the radial profiles for the $f_y$. In general, the HII regions in the inner areas of the galaxies show a lower value of $f_y$ than those regions in the outer ones. In addition, there is a clear trend with the stellar mass. The $f_y$ is higher for low values of the stellar mass. On the contrary, this fraction is lower for those \HII\ regions located in more massive galaxies. We find a similar trend for EW(H$\alpha$), which radial distributions are shown in the lower panels of Fig~.\ref{fig:DistanceDependenceA}.
The \HII\ regions with the largest EW(H$\alpha$) are located in the outer regions of low-mass galaxies, and those with the lowest values of this parameter are located in the inner regions of high-mass galaxies. These trends has been already explored in previous studies \citep[e.g.][]{2016GonzalezDelgado_A&A590,2017Belfiore_MNRAS469,2018Sanchez_RMxAA54}. They are the consequences of the higher star-formation rates with respect to the already accumulated stellar mass (i.e., higher specific star-formation rates, sSFRs) found in the outer regions of low-mass galaxies, in combination with the larger amount of stars formed in earlier star-formation episodes in the center of high-mass galaxies and their lower actual sSFR. In summary, both relations are a consequence of the so-called local-downsizing \citep[][]{eperez13} and the inside-out quenching/ageing \citep[][]{2017Belfiore_MNRAS469}. Those trends with the stellar mass are modulated by the morphology. The few earlier type galaxies with \HII\ regions show lower values of both $f_y$ and EW(H$\alpha$), and a weak negative gradient (i.e., decrease with the galactocentric distance). On the contrary, spiral galaxies present higher values and a clear positive gradient. This suggests that the general trend described before is dominated by late-type galaxies, with early-types presenting a slightly different pattern.

The upper panels of Fig. \ref{fig:DistanceDependenceB} show the radial profiles of the ionization parameter. There is no clear trend of this parameter with the galactocentric distance for any of the explored mass bins. However, the ionization parameter shows a higher value for the lowest stellar mass bin than for the other ones. Although we should note that the mean error for this mass bin is higher than other ones. When segregating by morphology we find a variety of behaviors without a clear pattern. If any, the trend with the stellar mass is much more clear for the very late spiral galaxies (Sd/Sm).

The central panels of Fig. \ref{fig:DistanceDependenceB} show the radial profiles of the oxygen abundance. There is a clear and well-known trend of this parameter with the galactocentric distance and stellar mass. The \HII\ regions with higher oxygen abundances are located in the inner regions of the galaxy, and those with lower oxygen abundances are towards the outer zones. Also, there is a clear dependence between oxygen abundance and stellar mass \citep[a consequence of the well known Mass-Metallicity relation, e.g., ][]{garnett:2002p339,2004Tremonti_ApJ613}. Low-mass galaxies generally host the more metal-poor HII regions, and high-mass galaxies host the metal-rich ones. The negative gradient was first described for the Milky-Way and galaxies of a similar mass by \citet{sear71} and \citet{peim78}, and it has been deeply explored by many previous studies \citep[e.g.,][]{sanchez13,esteban18}. The slope of the gradient presents a dependence with the mass as well, with more massive galaxies presenting flatter gradients \cite[as already reported by][]{2017Belfiore_MNRAS469}. Both trends are modulated by the morphology, with early-type galaxies showing an even flatter gradient, with even positive gradients observed \citep[as noticed by][]{2021Sanchez_RMxAA57}. Finally, the range of values for the oxygen abundances is larger in late-type galaxies than in early-types. 
These results speak against a single characteristic oxygen abundance gradient for all galaxies as proposed by \citet{sanchez13}, where most of the explored galaxies have similar stellar mass as the MW (for which the range of slopes is rather narrow). Indeed, they are more in agreement with the recent results by \citet{2017Belfiore_MNRAS469}, \citet{2020Boardman_MNRAS491} and \citet{2021Zinchenko_AA655}, where it was reported that the slope indeed depends on different properties of the galaxies, including the stellar mass.

A similar trend is found for the radial gradient of the nitrogen-to-oxygen ratio, which radial profiles are found in the bottom panel of \cref{fig:DistanceDependenceB}. In average the \HII\ regions with a high N/O ratio are located in the inner areas of galaxies, and those with lower N/O are located in the outer zones. This result is
in agreement with the most recent explorations of those distributions \citep[][]{2016PerezMontero_AA595,2017Belfiore_MNRAS469,2021Zinchenko_AA655}. Also, there is a clear dependence of the N/O ratio with the stellar mass. The lower N/O ratios are found on \HII\ regions hosted on low-mass galaxies, and the higher values are on regions in high-mass galaxies. This, again, is a consequence of a global relation between this parameter and the stellar mass \citep[e.g.][for a recent study]{2020Schaefer_ApJL890}, that can be a consequence of the N/O-O/H relation via the MZR. A less clear pattern is observed in the slopes of the gradients. For early-type galaxies they seem to be flatter in average. However, for the more massive ones it was reported the highest negative gradient of all the sample (although the low number statistic of this particular subsample does not allow us to make a strong statement in this regards).

\begin{figure*}
\includegraphics[scale=0.80]{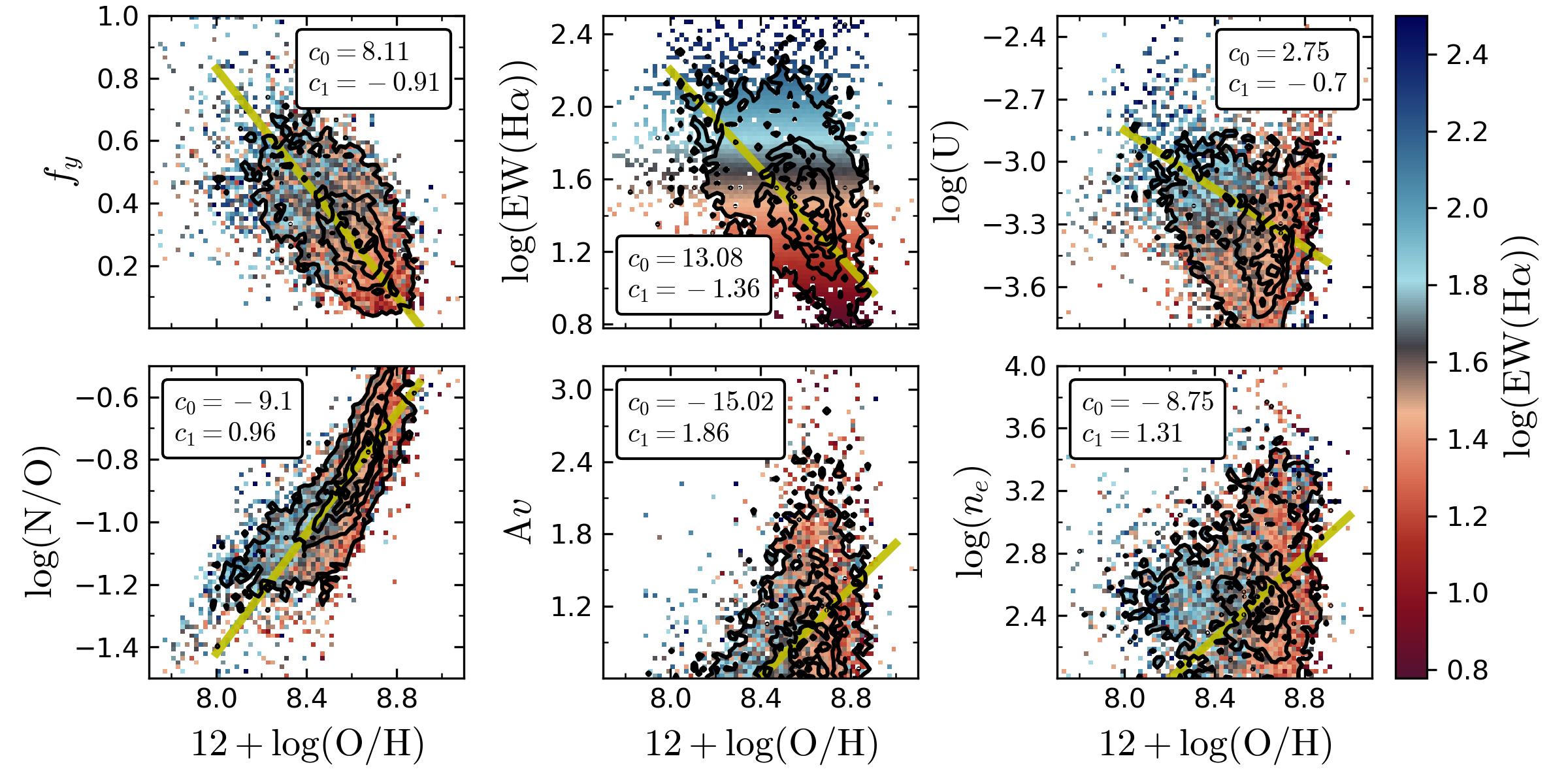}

\caption{Relation between the oxygen abundances and the others physical properties of HII regions. The oxygen abundance is compared with i) $f_{y}$ (\textit{top left}); ii) \EWHa\ (\textit{top middle}); iii) ionization parameter (\textit{top right}); iv) nitrogen-oxygen ratio (\textit{bottom left}); v) dust extinction (\textit{bottom middle}); and, v) electron density (\textit{bottom right}). The contours represents the density distribution, the outermost contour enclose the 90$\%$ of the regions, the middle contour enclose the 75$\%$ of the regions, and  the innermost contour encloses the 60$\%$ of the regions.}
\label{fig:OHvsPP}
\end{figure*}

The top panels of \cref{fig:DistanceDependenceC} show the same distributions for the dust extinction. For this parameter both the stellar mass and the morphology play a clear role. In general the two more massive bins correspond to the higher values of the dust extinction. In both cases they present a negative radial gradient (i.e., more dust towards the center). The highest values are found for the highest mass bin. For the third mass bin the distribution shows a slight decline with the radius too, but less evident than in the other two cases. Finally, for the lowest mass bin there is a positive gradient, with more dust in outer regions. These trends trace the presence of molecular gas in these galaxies via the metallicity \citep[][]{brinchmann04,jkbb20, jkbb21a}, suggesting that in more massive galaxies it is clearly concentrated in the central regions. When exploring the dust content by morphology, it is clear that the most massive Sa galaxies, at any galactocentric distances, and E/S0, in the outer regions, are those with the highest dust content. On the other hand low-mass Sd/Sm/Irr are those with the lowest dust content.  Regarding the gradient, clear negative slopes are found for Sa of any mass and for the more massive late-type galaxies, then the slope seems to become shallower and finally positive for the lowest-mass latest-type galaxies. The gradient does not present a clear pattern for E/S0 galaxies, maybe due to the limited number of \HII\ regions in these galaxies. If any, they present a flat distribution of A$_{\rm V}$.

Finally, the bottom panels of \cref{fig:DistanceDependenceC} show the radial profiles of the electron density. In general, the denser \HII\ regions are located towards the center of the galaxies. This result is expected since at the central regions of the galaxies, with larger values of the stellar and molecular gas surface densities, the pressure is expected to be larger \citep[e.g.][]{jkbb21b,barnes21}. Then the density declines and rise again in the outer regions. There is a slight trend between the electron density and the stellar mass, with denser regions hosted by high-mass galaxies. Again, this could be a consequence of the effect of the pressure. Indeed a similar trend is described with the morphology, with earlier galaxies showing the higher values of this quantity. The pattern with the radial trend is less evident, with slopes covering a wide range of values for any morphology and stellar mass.

\subsection{Relation between oxygen abundance and others physical properties of HII regions}
\label{sec:rel_OH}

So far we have explored how the physical properties of \HII\ regions distribute along the diagnostic diagrams (Sec. \ref{sec:trends} and App. \ref{app:diagnostic_diagrams}), and along the galactocentric distance (Sec. \ref{sec:radialGrads}). In this section we explore the dependence of these parameters among themselves. For doing so we adopted the oxygen abundance as the primary property, studying how the rest of the parameters correlate (or not) with O/H. We quantify the correlations based on the Pearson correlation coefficient between each pair of explored parameters. To characterize the relation between them we explore three different linear regressions: (i) fitting Y vs. X, (ii) fitting X vs. Y, and (iii) deriving the bisector line between both previous relations. We adopt as the best relation the one that minimize the dispersion of the residuals ($\sigma$). Regardless of the procedure that provides the best fit of the three outlined before we report the parameters of the relation between the Y and X axis, i.e., y$=c_1 x+ c_0$.

Figure \ref{fig:OHvsPP} shows the derived distributions for all the parameters described in Sec. \ref{sec:physical_prop}, where clear trends are appreciated. Regarding f$_y$ and \EWHa, they both show a clear decline with the oxygen abundance, with more metal rich regions corresponding to those having a low fraction of young stars and low values of the equivalent width of \Ha. This is expected based on the distributions described for the BPT diagrams shown in Sec. \ref{sec:trends} \citep[and][]{2020EspinosaPonce_MNRAS494} and the radial distributions shown in Sec. \ref{sec:radialGrads}, as well. Regions of high O/H correspond to those located in areas with higher stellar mass density, based on the resolved MZR relation \citep[e.g.,][]{rosales-ortega:2012,2016Barrera_MNRAS463}, i.e., the regions with the brightest continuum emission. Thus, this is the location where f$_y$ and \EWHa\ is expected to be lower, for \HII\ regions ionized by the same kind of stellar clusters with similar \Ha\ luminosities.
In consequence there is a relative significant correlation between both parameters ($\rho=$-0.58 and -0.46, for f$_y$ and \EWHa\ respectively), parametrized by a clear negative slope ($c_1$=$-$0.91 and $-$1.36, respectively), and with a small scatter ($\sigma=$0.12 and 0.19 dex, respectively). Thus, for these two parameters the reported relations are a pure consequence of the differential stellar/chemical evolution of galaxies in different locations.

The relations of O/H with log(U) and the N/O ratio have a different nature. In principle, those relations are not expected to be connected with the properties and evolution of the underlying stellar population in the areas in which the \HII\ regions are formed, like the former two. Early explorations on the properties of \HII\ regions suggest that log(U) declines with O/H \citep[e.g.][]{dopita86}, a trend that has been ample adopted in the literature, in particular when photoionization models are used to derive both parameters \citep[e.g.][]{2018Thomas_ApJ856}. 
Empirically, it has been clearly found with empirical calibrators of log(U) are used \citep[e.g.,][]{2015Sanchez_AA574}. However, using photoionization models a loose relation is found between both parameters, tighter at high oxygen abundances than at low ones \citep[e.g.,][]{2016Morisset_aa594A}. This trend can be characterized using a linear relation only at first order. Consequently we found a weaker trend between both parameters as the one described for the other two ones ($\rho=-0.23$), parametrized by a linear relation with a slope of $c_1=-0.7$, and a large standard deviation compared with the dynamical range of log(U), $\sigma=0.2$ dex. Indeed, it is clearly observed that the negative trend is obtained, for the fiducial calibrators adopted for both parameters, only in the low abundance regime. At high abundance there is an apparent turn-up of log(U), that reach high values. Contrary to other distributions explored in this study we find a strong dependence on the actual adopted calibrators on the observed trends between both parameters. Appendix \ref{appx6:OHvsUs} includes similar explorations for the different ionization parameter calibrators included in our catalog of physical properties of \HII\ regions. The expected decline of $U$ with the abundance is not observed in most of the cases. Only for the $U$ calibrators reported by \citet{dors11} and \citet{2016Morisset_aa594A} there is a weak trend like the one reported here, with the corresponding up-turn. In other cases, like the values provided by the NB code \citep[][respectively]{2018Thomas_ApJ856}, the distribution is discrete and does not show a clear trend.
In other cases, like HCm \citep{2014PerezMontero_MNRAS441} the distribution covers the same range of parameters as the one shown by our fiducial calibrator, but with two clear discrete clouds.
Finally, for the case of the values reported by the IZI code \citep{2015Blanc_ApJ798, 2020Mingozzi_AA636} the ionization parameter seems to be constant.

The nature of this predicted relation is still broadly unclear \citep[see][, for an extensive discussion]{2021Ji_arXiv211000612} . Early explanations involved (i) a correlation between the initial mass function (IMF) and the metallicity, (ii) an effect of the dust absorption, and (iii) an environmental effect in the sense that \HII\ regions that are formed in the inner and outer regions of galaxies present physical differences \citep[e.g.][]{dopita86}. The first of this hypothesis requires that a low metal ISM enhances the upper end of the IMF because of the increased Jeans mass, producing an apparent larger log(U), while metal rich gas, that cool faster, may result in a lower IMF cutoff and therefore a lower apparent log(U). Recent results have suggested that this IMF-Z relation is indeed present, at least for early-type galaxies \citep[][]{navarro16}. On the other hand, the dust absorption may produce a soften of the radiation field, producing low-excitation \HII\ regions in dusty environments. If this is the main driver between the log(U)-O/H relation a decrease (increase) is expected of the ionization parameter (oxygen abundance) with the dust extinction. We will get back to that later on in Sec. \ref{sec:discussion}. Finally, the environmental scenario suggests that \HII\ regions in the central locations of galaxies are more numerous and they have a smaller average surface brightness \citep{dopita86}. The reason for that could be the enhancement of the cooling and fragmentation induced by the metallicity: molecular clouds in regions of larger metal content would fragment in smaller clouds and collapse to form smaller and lower surface-brightness H ii regions. Since both the larger metal content and the highest molecular gas surface density are found in the central regions of galaxies, the log(U)-O/H relation would naturally emanate from the two observed trends.
However, this was never confirmed. More recently it has been suggested that blanketing due to metal content may be responsible of this relation \citep[e.g.][]{2015Sanchez_AA574, 2021Ji_arXiv211000612}, although it is not clear if this effect may produce
a soften or a harder radiation field. All these explanations suggest that log(U) is considered as a tracer of the hardness of the ionizing source. However, this is not
completely true. The issue is due to the fact that the empirical indicators adopted to trace log(U) actually trace the excitation level of the gas, that depends on both log(U) and the shape of the ionizing source (and also changes if the cloud is matter- or radiation-bounded). In summary what theory defines as this parameter and what we actually trace may well be two different things. Most probably there is no simple explanation to this relation, if it exists, and a combination of all the former scenarios is required to describe it. We should note that this correlation is not observed for all calibrators, and using a different combination of them could produce very different results.


\begin{figure*}
\includegraphics[width=\textwidth]{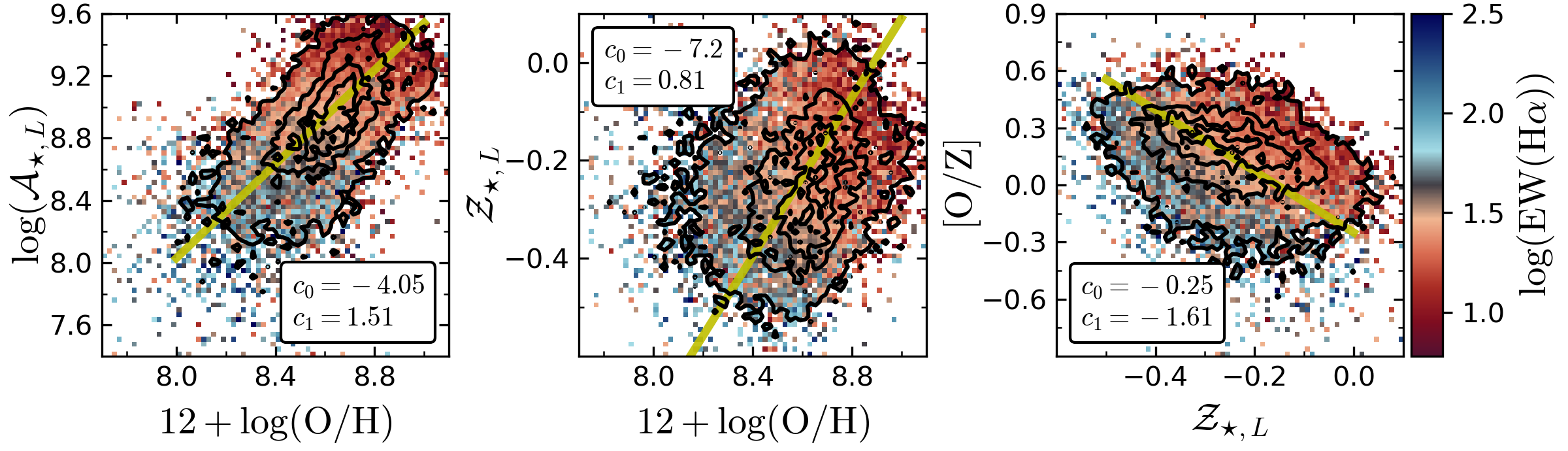}
\caption{Relation between the oxygen abundances and physical properties of underlying stellar populations of \HII\ regions. The 
oxygen is compared with i) luminosity-weighted age, \age\ (\textit{left panel}, ii) luminosity-weighted metallicity, \met\ (\textit{middle panel}) of underlying stellar population. The \textit{right panel} shows the relative oxygen abundance of \HII\ regions with respect to the Z-elements of underlying stellar populations ([O/Z]) vs the metallicity of stellar population, [Z/H]. The density contours are the same that used in \cref{fig:OHvsPP}.}
\label{fig:OHvsStellarProps}
\end{figure*}

The relative nitrogen-to-oxygen abundance presents, as expected, a strong positive correlation with the oxygen abundance ($\rho=$0.83), as seen in Fig. \ref{fig:OHvsPP}.
This distribution is well characterized by a tight ($\sigma=$0.09 dex) linear relation of slope c1$=$0.96 (for the range of abundances covered by the current dataset). This well known relation is a consequence of the differential nature of the oxygen and nitrogen production. Oxygen is a primary element, which production depends only on physical processes \citep[][]{1986Matteucci_MNRAS221,2019Maiolino_A&ARv27}. However, nitrogen has both a primary and secondary nature. At low metallicity, i.e., 12+log(O/H)$\lesssim$8, is may be primary, and therefore the N/O ratio would be independent of O/H. On the contrary, at high metallicity, i.e., 12+log(O/H)$\geq$8.2, its production depends on the initial abundance of other elements, in particular oxygen. Thus, in this regime we observe the reported linear relation between both quantities, corresponding to the regime of oxygen abundances covered by our catalog of \HII\ regions. With the current data we just appreciate that the linear relation is blended at low metallicity, hinting the reported plateau.
This flattening (i.e., constant N/O) is observed at low-O/H \HII\ regions (or galaxies)  \citep[e.g.,][]{izotov12,2016Vincenzo_MNRAS458}. Recent result have shown that there are secondary trends in this relation, being modulated by the global stellar mass and the local star-formation efficiency \citep[SFE,][]{2020Schaefer_ApJL890}, although they could be due to systematics in the adopted calibrators. In our data it is observed a clear segregation of the relation between N/O and O/H as a function of the \EWHa. This segregation may be connected with the one reported by \citet{2020Schaefer_ApJL890}, once considered the relation between \EWHa\ and SFE \citep[e.g.][]{colombo18,2021Sanchez_RMxAA57}. Despite of this difference both results suggest a possible lack of universality of the N/O-O/H relation, or they highlight the effects of the systematics in the adopted calibrators (that were different in both studies).

The distribution of the dust extinction of the ionized gas as a function of the oxygen abundance presented in Fig \ref{fig:OHvsPP} shows that both parameters do not present a clear correlation ($c_0=$0.2). Both parameters present a fuzzy trend that can be characterized by the linear relation with a positive slope of $c_1=$1.86 mag/dex. Consistently with the weakness of the relation the scatter is considerably larger than the one reported for other relations described between the explored parameters ($\sigma$=0.36 mag), apart from the log(U)-O/H one. This weak trend is better appreciated when considering the radial gradients observed for both the oxygen abundance and the dust extinction. Those azimuthal averaged values for galaxies of the same mass and morphology (Fig. \ref{fig:DistanceDependenceB} and \ref{fig:DistanceDependenceC}), present very similar radial trends that would directly produce a relation between both quantities. Despite of this average trends, the dispersion around this relation for individual \HII\ regions is large. Taking into account the direct relation between the dust extinction and the gas column density \citep[][]{schmidt59,jkbb20}, this relation may suggest that regions with higher metallicity are those with higher molecular gas surface density, what would be against one of the lines of reasoning by \citet{dopita86} to explain the relation between log(U) and O/H, discussed before. Again, this relation highlights the strong connection between different properties of the \HII\ regions and the local/resolved properties of the location in which they are formed in galaxies. The fact that it is weaker/loose than other ones described in here, like the N/O-O/H or the \EWHa- or $f_y$-O/H one, indicates that there is a clear hierarchy between the explored relations. The N/O-O/H relation seems to be more universal (i.e., independent of properties external to the nebulae itself), despite the possible secondary trend reported. On the other hand, the relations with \EWHa, $f_y$ and $A_V$ seem to be deeply related with the location in the galaxy in which the \HII\ region is formed, and the imprints that the evolution and chemical enrichment have left in the properties of the ISM. The fact that the relation with $A_V$ is the weakest indicates that this relation is strongly modulated by the dependence of both parameters with other physical properties. This is not a total surprise. Based on Eq. 13 of \citet{brinchmann04}, the relation between $A_V$ and O/H should depends at least on the molecular gas surface density and on the dust-to-metal ratio of the ionized gas, two parameters that are clearly non constant across the extension of a galaxy.

Finally, Fig. \ref{fig:OHvsPP} shows that the distribution of the electron density is essentially insensitive to the oxygen abundance (bottom-middle panel). Both parameters present the weakest correlation coefficient among the explored ones ($\rho=$0.16), with a large dispersion around the estimated linear regression ($\sigma$=0.2 dex). Contrary to all the remaining parameters, the electron density is clearly unconnected with the oxygen abundance. Furthermore, it seems to be totally insensitive to the described connection with the evolution of the stellar population or the location at which the \HII\ regions are formed, observed in all the properties explored so far.

\subsection{Relation between the oxygen abundance and physical properties of underlying stellar populations}
\label{sec:OH_st}

Most of the relations between the different explored parameters described in Sec. \ref{sec:rel_OH} suggest a direct connection between the oxygen abundance of \HII\ and the properties of the surrounding underlying stellar population. Such connection is behind the recently proposed relations between O/H and different local/resolved properties in galaxies, such as the stellar surface mass density \citep[rMZR][]{rosales-ortega:2012,2016Barrera_MNRAS463}, the gas fraction and/or the escape velocity \citep[][]{2018Barrera_ApJ852}.
Based on these results, we explore in this section the relation between the properties of the underlying stellar population and the oxygen abundance for our sample of \HII\ regions.

Fig. \ref{fig:OHvsStellarProps}, left-panel, shows the distribution of \age\ as a function of O/H. There is a strong correlation between both parameters, with a correlation coefficient of $\rho=$0.66. We estimate the best linear relation between the two parameters by exploring the linear regressions assuming that either the \age\ or the O/H is the independent parameter (i.e., fitting Y vs. X, or X vs. Y), or neither of them (i.e., deriving the bisector line between themselves. The best linear relation for the \age-O/H distribution results from the fitting of the later (X-axis) versus the former (Y-axis), corresponding to a slope of $c_1=$1.51 (where y$=c_1 x+ c0$, see Fig.  \ref{fig:OHvsStellarProps}), and a standard deviation in the Y-axis of $\sigma=$0.15 dex. Since the age of the ionizing stellar populations in an \HII\ region is very similar in all of them, being limited to less than 10 Myr, the reported \age\ is essentially tracing the distribution of ages of the underlying stellar population. Therefore, this correlation indicates that the more metal-rich \HII\ regions are formed in areas of the galaxies which bulk stellar population was formed long ago in cosmological times. On the contrary, metal poor \HII\ regions are formed in areas where the stellar population has been formed along a wider range of times, and therefore they present a broader range of ages. Since the oxygen abundance is tightly connected with the amount of stars formed, as indicated before, the reason behind that correlation is that older stellar populations were formed in star-formation processes that were stronger and sharper, peaked in earlier cosmological times. This is the basis of local downsizing \citep[e.g.][]{eperez13}, a proved scenario in which regions of larger stellar mass density formed they stars earlier and following more abrupt SFHs than regions of smaller stellar mass density \citep[e.g.][]{rgb17,2020Sanchez_ARA&A58}. As a consequence these locations have both older stellar populations and a higher metal enrichment. Indeed the distribution of \EWHa\ along this figure follows almost the same distribution of both O/H and \age, highlighting the fact that regions located in areas with relative low sSFR nowadays (low \EWHa) are those that have accumulated their stellar mass (and metals) long before, and therefore they have the older stellar populations. On the contrary, areas with relative large sSFR (high \EWHa) are those that are still actively forming stars, having accumulated their stellar mass (and metals) more recently.

This scenario implies a connection between the oxygen abundance of the \HII\ regions and the metallicity of the underlying stellar population as well, since they are the result of the local metal enrichment. Fig. \ref{fig:OHvsStellarProps}, central panel, shows the distribution of these two parameters. As expected, the stellar metallicity, \met, presents a very loose trend with O/H ($\rho=0.27$). Once again, the best linear relation is derived by fitting the oxygen abundance (X-axis) to the \met (Y-axis), corresponding to a slope of $c_1=$0.81, and a standard deviation of $\sigma=$0.15 dex. However, the correlation is much weaker than the one found between O/H and \age. It is worth noticing that \met is the average of the metallicities of all the stellar populations in which the observed spectra is decomposed. By construction it essentially traces iron trapped on the surviving stars. On the other hand, O/H traces the oxygen abundance, i.e., the abundance of an $\alpha$ element, at the ISM of the \HII\ region.
This abundance should be the same as the one of the young ionizing stars, recently formed ($\sim$4Myr) from this gas. Therefore both parameters trace a different type of element at different time scales and for different families of stars. It is important to recall that iron is mostly produced and expelled to the media by SNIa, being the product of the collapse of binary systems of average massive stars, while oxygen is essentially produced by core-collapsed supernovae (CCSN) produced at the death of very massive stars. Thus, \met and O/H may present significant differences in their corresponding chemical enrichment histories \citep[see][for a review on the topic]{2019Maiolino_A&ARv27}. The current LW stellar metallicity depends on both the final stellar mass and the shape of the star-formation histories \citep[SFH, e.g.,][]{camps20}. On the other hand, the oxygen abundance follows more tightly the strength rather than the shape of the SFH, being more strongly tied to the final stellar mass \citep[e.g.][]{lacerda20} and the stellar mass surface density \citep[e.g.][]{gallazzi05,2016GonzalezDelgado_A&A590,2020Sanchez_ARA&A58}. The relative weakness of the observed correlation and the large standard deviation suggest that both parameters are connected, but without one fully following the other. 

Finally, right panel of Fig. \ref{fig:OHvsStellarProps} shows the distribution of the [O/Z] ratio along \met. This ratio was constructed by subtracting \met\ to the O/H normalized to the solar value \citep[assuming a 12+log(O/H)$_\odot$=8.69][]{aspl09}, as described in detail in \citet{2021Sanchez_A&A652}. In this recent letter we demonstrated that this parameters is a good tracer of the $\alpha$-enhancement at the location of \HII\ regions in galaxies. As reported in \citet{2021Sanchez_A&A652}, [O/Z] declines with \met, following a weak correlation ($\rho=$-0.35), that, it is better described by the linear regression of the former parameter (Y-axis) versus the later one (X-axis). This relation shows a slope of $c_1=$-1.61. The dispersion around this trend is similar to the one reported for the relation between O/H and \met\ ($\sigma=$0.13 dex), despite the fact that the current correlation is slightly stronger for this one. Furthermore, as explicitly discussed in \citet{2021Sanchez_A&A652}, the dispersion is modulated by a secondary trend with the \age\ of the underlying stellar population, that in the current figure is coded by the \EWHa\ (based on the correspondence between both parameters discussed before). It is appreciated that a sequence of relations between [O/Z] and \met\ with a different zero-point (and maybe different slopes) for each value of \EWHa\ would produce a significant lower scatter than a single relation for the bulk of \HII\ regions. This is in agreement with the strength of the relation between O/H and \age\ reported before, that, based on the results presented in \citet{2021Sanchez_A&A652} and the distribution observed in here is most probably equally modulated by [Z/H], highlighting the connection between both parameters. It is worth noticing that these trends are fully compatible with those found for individual stars in the MW \citep[e.g.][]{2015Hayden_ApJ808} and galaxy wide integrated populations \citep[e.g.][]{walcher14}. The straightforward interpretation of this relation is a consequence of the differential enrichment histories between the $\alpha$ and iron peak elements \citep[e.g.][]{2015Hayden_ApJ808,walcher16, 2021Yu_ApJ912}.
As discussed before, this relation is heavily modulated by either the shape IMF, the shape of the star-formation histories, or a combination of both, as predicted by the chemical evolution models \citep[e.g.][and references therein]{carigi19,wein17}.

\section{Discussion}
\label{sec:discussion}

Along this study we have explored the physical properties of the catalog of $\sim$26,000 \HII\ regions and aggregations extracted from the CALIFA IFS data presented by \citet{2020EspinosaPonce_MNRAS494}. First, we studied how these properties distribute across the diagnostic diagrams (Sec. \ref{sec:trends}, and App.\ref{app:diagnostic_diagrams}). As already shown in \citet{2020EspinosaPonce_MNRAS494}, our selection of \HII\ regions, purely based on basic properties of these regions\footnote{(1) to present a clumpy/peaky shape in the emission line maps; (2) to present a minimum $f_y$; and (3) to present and \EWHa\ compatible with that $f_y$}, recovers the well-known distributions along the diagnostic diagrams for these objects. Just a small fraction of the finally selected regions ($<$4\%) are located above the classical demarcation lines proposed by \citet{2001Kewley_ApJ556}, based on theoretical considerations and photoinization models. Thus, the proposed selection reproduce the expectations by these models, without requiring a prior selection based on line ratios.

Any set of photoinization models aimed to reproduce the emission line ratios produced by the ionization due to young massive OB stars \citep[e.g.][]{2001Kewley_ApJ556,2016Morisset_aa594A} cover the observed distribution \citep[e.g.][]{2015Sanchez_AA574}. However, it is well known that (i) these models are heavily degenerated, i.e., a set of line ratios (e.g., O3 vs N2) can be reproduced by a different set of physical properties associated with different kind of modeled \HII\ regions (for instance, two values of log(U) and O/H), and (ii) physically possible photoionization models cover a much wider region of the diagrams than the actually covered by the observed regions. As a consequence photoinization models, {\it per se}, cannot predict univocally the physical properties of these regions, like the oxygen abundance or the ionization parameter, purely based on the line ratios involved in the diagnostic diagrams. However, there are clear trends between the physical parameters and the observed line ratios. In Sec. \ref{sec:trends} we show that the actual values of several physical parameters of \HII\ regions are well defined by the location on these diagrams. Although we observed a certain degree of data scattering, in general, it is much lower than the dynamical range of the explored parameter: e.g., Fig. \ref{fig:BPT_O3N2_physical_properties} (and also in Fig. \ref{fig:BPT_O3S2_physical_properties}, \ref{fig:BPT_O3O2_physical_properties}, \ref{fig:BPT_O3O1_physical_properties}). This result was already reported in \citet{2015Sanchez_AA574}, using a smaller sample of \HII\ regions. Furthermore, the connection between different physical properties and the location in the diagnostic diagram is not putative of \HII\ regions. It is observed in other sources of ionization, as a connection between line ratios and the dynamical stage of the underlying stellar population \citep[e.g.,][]{law21}, the velocity dispersion of the ionized gas \citep[e.g.][]{dagos21}, or the morphology, the stellar mass and the galactocentric distance \citep[e.g.][]{2020Sanchez_ARA&A58}. Again, none of those trends can be predicted by photoinization models. The reason behind this discrepancy is that photoinization models are physical models of the ionized regions, but they do not include the astrophysical context in which the ionization is produced. Thus, given a certain ionizing source (e.g., an OB star), a certain geometry and density distribution of the ionized nebulae, and a certain chemical composition of this nebulae, photoionization models are able to predict a set of line ratios. However, they are unable to determine which type of ionizing source, which type of geometry or which chemical composition is found at a certain location within a certain type of galaxy. Essentially, they ignore the constrains in which star-formation and chemical enrichment took place in galaxies (and within galaxies).

The reported trends along the diagnostic diagrams are not a consequence of the predictions of how the ionization happens. They are the consequence of the spatially resolved evolution of the stellar populations, the consequent chemical enrichment at the location in which \HII\ regions are formed, and maybe the kind of nebulae that can be formed at different locations. Because of this, there are well defined gradients in all physical properties (e.g., O/H, N/O, A$_V$, as seen in Fig. \ref{fig:DistanceDependenceB} and Fig. \ref{fig:DistanceDependenceC}), well described in the literature \citep[e.g.][]{sear71,peim78,vila92,sanchez14,2017Belfiore_MNRAS469,esteban18}.
These gradients have been frequently used to explore the star-formation histories and chemical enrichment histories in other galaxies \citep[e.g.,][]{tissera:2013aa,belfiore19} and in the Milky-Way \citep[e.g.][]{carigi19}. However, it is not so frequently acknowledged that they left imprints in the physical properties of \HII\ regions that are fossil records of those processes. This perspective was proposed by \citet{2015Sanchez_AA574}, once it was observed the direct connection between the line ratios observed in \HII\ regions and the physical properties of the underlying stellar populations \citep[like the B-V color][]{2012Sanchez_A&A538A}. These results were confirmed by \citet{2020EspinosaPonce_MNRAS494}, where it was explored the direct relation between those line ratios and physical properties of the stellar populations, like the \age\ and \met. 

As indicated before, photoinization models require to be combined with empirical relations between different physical quantities, like the N/O-O/H and the log(U)-O/H relations explored in Sec \ref{sec:rel_OH} \citep[][]{dopita13, 2016Asari_MNRAS460}, to make predictions on those properties based on the observed line ratios. On the contrary, they may lead to significant errors and miss-interpretations. 
For instance, the average physical properties of those models that reproduce the observed line ratios are not observed previously in the literature or are not in general representative of the real physical properties, if they are heavily degenerated. Indeed, in many cases they are substantially wrong. The most simple example of this situation is the well known bi-valuate relation between the R23 line ratio and the oxygen abundance \citep[e.g.][]{pagel79}.

The nature of these relations between N/O, log(O/H) and log(U) is astrophysical as well. As indicated before, the N/O-O/H relation is a consequence of the primary production of oxygen and the primary and secondary production of nitrogen \citep{1978Edmuns_MNRAS185, 1979Alloin_AA78, 2005Molla_MNRAS358}. In particular, the secondary production, that depends on the previous production of oxygen (and carbon), is the mechanism behind the relation above 12+log(O/H)$>$8.2 observed in Fig. \ref{fig:OHvsPP}. The flatenning at lower oxygen abundances is due to the primary phases of the production of nitrogen, that is observed when the location in which the \HII\ region is form has not reached the required oxygen abundance needed to ignite the secondary production in the newly formed stars. Thus, it depends on the location within galaxies, the stellar mass of those galaxies (a gauge of the oxygen abundance), and their morphological type \citep[e.g.][Schaefer et al. in prep]{belfiore17a}. On the other hand, the nature of the log(U)-O/H relation is less clear. Different mechanisms were already proposed in early explorations by \citet{evans85} and \citet{dopita86}, as discussed in Sec. \ref{sec:rel_OH}. All of them require a connection between this relation and the location in which the \HII\ is formed: (i) a change in the IMF and its connection with the oxygen abundance; (ii) an effect of the dust or the metal blanketing on the hardness of the ionization and (iii) an environmental bias on which kind of nebulae is formed. 

As indicated before, a non universal or variable IMF and its connection with galaxy (or local) properties is a topic of discussion. \citet{chab03} reviewed the possible variability of the IMF depending on the environment within a galaxy, although some recent results indicate that this many not be the case \citep[e.g.][]{wegg21}. A possible time evolution from early cosmological times has been proposed too \citep[e.g.][]{vazdekis96}, what would explain the differential $\alpha$-enhancement in different regions within our Galaxy \citep[e.g.][]{carigi19} or among different galaxies \citep[e.g.][]{walcher14}.
Since in this period the star-formation was dominated by present day early-type galaxies \citep[e.g.][]{sanchez19}, or in the bulge/center of early spirals \citep[e.g.][]{cheng15,rgb18}, this time change would imply an environmental  dependency of the IMF too. Recent explorations have shown further relations between the shape of the IMF and other galaxy (or regions within galaxies) properties, such as the stellar metallicity \citep[e.g.][]{navarro16} or the velocity dispersion \citep[][]{mcdermid15}. Those relations may be well interconnected, and connected with the time evolution discussed before. In any case, if this is the cause of the log(U)-O/H relation, the connection would be deeply related on how the star-formation happens due to how the metallicity enhances the cooling, fragmentation and collapse of the molecular clouds \citep{dopita86}. Thus, this relation would be a consequence of the enrichment process, but only through this physical connection. The possible effect of the dust and/or metal blanketing on the hardness of the radiation has a similar origin, since those regions with stronger metal enrichment would be those suffering this effect more clearly. 

The final scenario proposed for the observed log(U)-O/H by \citet{dopita86} has a totally different origin. They proposed that \HII\ regions are formed in locations within a galaxy with larger molecular gas densities (i.e., the center) may have a smaller surface brightness. The reason for that could be, one more time, the enhancement of the cooling and fragmentation induced by the metallicity: molecular clouds in regions of larger metal content would fragment in smaller clouds and collapse to form smaller and lower surface-brightness \HII\ regions. Since both the larger metal content and the highest molecular gas surface density are found in the central regions of galaxies, the log(U)-O/H relation would naturally emanate from the two observed trends. The local pressure may also play a role in this regards, in addition to the direct effect of the metal content. Radial declines in the pressure are suggested based on the relations proposed by \citet{jkbb21b} and the direct estimations reported by \citet{barnes21}. A high pressure may facilitate the fragmentation and collapse of smaller molecular clouds and the formation of lower surface brightness \HII\ regions. We should not exclude that the log(U)-O/H anti-correlation is the consequence of a combined effect of all the reported scenarios that operate at a different level but simultaneously. This could be the reason why the relation is weaker and less well defined as other trends found along this study, due to its multi-parametric nature.

Finally, we should acknowledge that not all the properties of the \HII\ regions depends on the star-formation and chemical evolution of galaxies. It is obvious that the age of the ionizing population is independent of those processes by definition of these regions (i.e., ionized by young massive OB-stars), although their metallicity depends on the evolution. The electron density seems to be decoupled of the evolution processes in galaxies as well, at least to the precision in which we can measure this parameter with our current dataset. We should highlight that the average values reported for this parameter ($n_e\sim$300), although they are similar to the ones reported using IFS data with similar spatial and spectral resolutions \citep[e.g.][]{sanchez12b}, they are larger than the ones recently reported for similar data with better spatial and spectral resolutions \citep{barnes21}. Thus, we need to explore the reported trends with better quality data and other ionic line rations to understand these discrepancies and confirm (or not) our results.

\section{Conclusions}
\label{sec:con}

Along this study we explored the connection between the physical properties of a large catalog of \HII\ regions extracted from a representative sample of the galaxies in the nearby universe, finding that:

\begin{itemize}
    \item The location of the \HII\ regions across the classical diagnostic diagrams, and therefore, the values of the line ratios involved on those diagrams are tightly connected with their physical properties, and the properties of the underlying stellar populations. 
    \item This correspondence between observed and physical properties is univocal in most of the cases (within a certain dispersion), contrary to the predictions of photoionization models when they cover a Cartesian grid in O/H, N/O, age and log(U). But if additional relations between these parameters are used to restrict the grid, models may recover the observed behavior.
    \item Most of the explored physical properties present clear radial gradients with patterns that depend on the mass and morphology of the host galaxies, relations that in many cases have been reported and discussed before in the literature. Once again those trends are tightly related to those of the underlying stellar populations (e.g., f$_y$).
    \item The physical properties of the \HII\ regions present clear relations between themselves and with the properties of the underlying stellar populations, that in some cases are more evident when exploring the azimutal averaged radial patterns discussed before. 
\end{itemize}

In conclusion we confirm the results from previous explorations suggesting that the astrophysical context in which \HII\ regions are generated are of a fundamental importance in shaping their observed properties. Indeed, most of those properties are the result from the local chemical evolution of the stellar populations at the locations in which those regions are found. In summary, \HII\ regions are a fundamental proxy of the chemical evolution of galaxies that have left clear imprints in their observed properties.

\section*{Acknowledgements}

We are very grateful with the referee for the comments and suggestions that have improved the quality of this paper.

CM acknowledges support from grant UNAM / PAPIIT - IN101220.
We are grateful for the support of the PAPIIT-DGAPA-IG100622 and PAPIIT-DGAPA-IN112620 (UNAM) projects.

RGB acknowledges additional financial support from the State Agency for Research of the Spanish MCIU through the Center of Excellence Severo Ochoa award to the Instituto de Astrof\'isica de Andaluc\'ia (SEV-2017-0709), grants PID2019- 109067GB-I00 (MCIU) and P18-FRJ-2595 (Junta de Andaluc\'ia).

L.G. acknowledges financial support from the Spanish Ministerio de Ciencia e Innovaci\'on (MCIN), the Agencia Estatal de Investigaci\'on (AEI) 10.13039/501100011033, and the European Social Fund (ESF) "Investing in your future" under the 2019 Ram\'on y Cajal program RYC2019-027683-I and the PID2020-115253GA-I00 HOSTFLOWS project, and from Centro Superior de Investigaciones Cient\'ificas (CSIC) under the PIE project 20215AT016.

\section*{Data Availability}

The \HII\ regions catalog used in this paper are presented in \cite{2020EspinosaPonce_MNRAS494}. The estimated physical properties of \HII\ regions are available online at \url{http://ifs.astroscu.unam.mx/CALIFA/HII_regions/}. An example code to read and use the data is available at \url{https://github.com/cespinosa/HII_regions_catalog}.


\bibliographystyle{mnras}
\bibliography{references,CALIFAI,library,my_bib}

\newcommand{\noop}[1]{}
\begin{thebibliography}{}
\makeatletter
\relax
\def\mn@urlcharsother{\let\do\@makeother \do\$\do\&\do\#\do\^\do\_\do\%\do\~}
\def\mn@doi{\begingroup\mn@urlcharsother \@ifnextchar [ {\mn@doi@}
  {\mn@doi@[]}}
\def\mn@doi@[#1]#2{\def\@tempa{#1}\ifx\@tempa\@empty \href
  {http://dx.doi.org/#2} {doi:#2}\else \href {http://dx.doi.org/#2} {#1}\fi
  \endgroup}
\def\mn@eprint#1#2{\mn@eprint@#1:#2::\@nil}
\def\mn@eprint@arXiv#1{\href {http://arxiv.org/abs/#1} {{\tt arXiv:#1}}}
\def\mn@eprint@dblp#1{\href {http://dblp.uni-trier.de/rec/bibtex/#1.xml}
  {dblp:#1}}
\def\mn@eprint@#1:#2:#3:#4\@nil{\def\@tempa {#1}\def\@tempb {#2}\def\@tempc
  {#3}\ifx \@tempc \@empty \let \@tempc \@tempb \let \@tempb \@tempa \fi \ifx
  \@tempb \@empty \def\@tempb {arXiv}\fi \@ifundefined
  {mn@eprint@\@tempb}{\@tempb:\@tempc}{\expandafter \expandafter \csname
  mn@eprint@\@tempb\endcsname \expandafter{\@tempc}}}

\bibitem[\protect\citeauthoryear{{Alloin}, {Collin-Souffrin}, {Joly}  \&
  {Vigroux}}{{Alloin} et~al.}{1979}]{1979Alloin_AA78}
{Alloin} D.,  {Collin-Souffrin} S.,  {Joly} M.,   {Vigroux} L.,  1979, \aap,
  \href {https://ui.adsabs.harvard.edu/abs/1979A&A....78..200A} {78, 200}

\bibitem[\protect\citeauthoryear{{Asplund}, {Grevesse}, {Sauval}  \&
  {Scott}}{{Asplund} et~al.}{2009}]{aspl09}
{Asplund} M.,  {Grevesse} N.,  {Sauval} A.~J.,   {Scott} P.,  2009, \mn@doi
  [\araa] {10.1146/annurev.astro.46.060407.145222}, \href
  {https://ui.adsabs.harvard.edu/abs/2009ARA&A..47..481A} {47, 481}

\bibitem[\protect\citeauthoryear{{Baldwin}, {Phillips}  \&
  {Terlevich}}{{Baldwin} et~al.}{1981}]{baldwin81}
{Baldwin} J.~A.,  {Phillips} M.~M.,   {Terlevich} R.,  1981, \mn@doi [\pasp]
  {10.1086/130766}, \href {http://adsabs.harvard.edu/abs/1981PASP...93....5B}
  {93, 5}

\bibitem[\protect\citeauthoryear{{Barnes} et~al.,}{{Barnes}
  et~al.}{2021}]{barnes21}
{Barnes} A.~T.,  et~al., 2021, \mn@doi [\mnras] {10.1093/mnras/stab2958}, \href
  {https://ui.adsabs.harvard.edu/abs/2021MNRAS.508.5362B} {508, 5362}

\bibitem[\protect\citeauthoryear{{Barrera-Ballesteros}
  et~al.,}{{Barrera-Ballesteros} et~al.}{2016}]{2016Barrera_MNRAS463}
{Barrera-Ballesteros} J.~K.,  et~al., 2016, \mn@doi [\mnras]
  {10.1093/mnras/stw1984}, \href
  {https://ui.adsabs.harvard.edu/abs/2016MNRAS.463.2513B} {463, 2513}

\bibitem[\protect\citeauthoryear{{Barrera-Ballesteros}
  et~al.,}{{Barrera-Ballesteros} et~al.}{2018}]{2018Barrera_ApJ852}
{Barrera-Ballesteros} J.~K.,  et~al., 2018, \mn@doi [\apj]
  {10.3847/1538-4357/aa9b31}, \href
  {https://ui.adsabs.harvard.edu/abs/2018ApJ...852...74B} {852, 74}

\bibitem[\protect\citeauthoryear{{Barrera-Ballesteros}
  et~al.,}{{Barrera-Ballesteros} et~al.}{2020}]{jkbb20}
{Barrera-Ballesteros} J.~K.,  et~al., 2020, \mn@doi [\mnras]
  {10.1093/mnras/stz3553}, \href
  {https://ui.adsabs.harvard.edu/abs/2020MNRAS.492.2651B} {492, 2651}

\bibitem[\protect\citeauthoryear{{Barrera-Ballesteros}
  et~al.,}{{Barrera-Ballesteros} et~al.}{2021a}]{jkbb21b}
{Barrera-Ballesteros} J.~K.,  et~al., 2021a, \mn@doi [\mnras]
  {10.1093/mnras/stab755}, \href
  {https://ui.adsabs.harvard.edu/abs/2021MNRAS.503.3643B} {503, 3643}

\bibitem[\protect\citeauthoryear{{Barrera-Ballesteros}
  et~al.,}{{Barrera-Ballesteros} et~al.}{2021b}]{jkbb21a}
{Barrera-Ballesteros} J.~K.,  et~al., 2021b, \mn@doi [\apj]
  {10.3847/1538-4357/abd855}, \href
  {https://ui.adsabs.harvard.edu/abs/2021ApJ...909..131B} {909, 131}

\bibitem[\protect\citeauthoryear{{Belfiore} et~al.,}{{Belfiore}
  et~al.}{2017a}]{belfiore17a}
{Belfiore} F.,  et~al., 2017a, \mn@doi [\mnras] {10.1093/mnras/stw3211}, \href
  {http://adsabs.harvard.edu/abs/2017MNRAS.466.2570B} {466, 2570}

\bibitem[\protect\citeauthoryear{{Belfiore} et~al.,}{{Belfiore}
  et~al.}{2017b}]{2017Belfiore_MNRAS469}
{Belfiore} F.,  et~al., 2017b, \mn@doi [\mnras] {10.1093/mnras/stx789}, \href
  {https://ui.adsabs.harvard.edu/abs/2017MNRAS.469..151B} {469, 151}

\bibitem[\protect\citeauthoryear{{Belfiore}, {Vincenzo}, {Maiolino}  \&
  {Matteucci}}{{Belfiore} et~al.}{2019}]{belfiore19}
{Belfiore} F.,  {Vincenzo} F.,  {Maiolino} R.,   {Matteucci} F.,  2019, \mn@doi
  [\mnras] {10.1093/mnras/stz1165}, \href
  {https://ui.adsabs.harvard.edu/abs/2019MNRAS.487..456B} {487, 456}

\bibitem[\protect\citeauthoryear{{Blanc}, {Kewley}, {Vogt}  \&
  {Dopita}}{{Blanc} et~al.}{2015}]{2015Blanc_ApJ798}
{Blanc} G.~A.,  {Kewley} L.,  {Vogt} F. P.~A.,   {Dopita} M.~A.,  2015, \mn@doi
  [\apj] {10.1088/0004-637X/798/2/99}, \href
  {https://ui.adsabs.harvard.edu/abs/2015ApJ...798...99B} {798, 99}

\bibitem[\protect\citeauthoryear{{Boardman} et~al.,}{{Boardman}
  et~al.}{2020}]{2020Boardman_MNRAS491}
{Boardman} N.,  et~al., 2020, \mn@doi [\mnras] {10.1093/mnras/stz3126}, \href
  {https://ui.adsabs.harvard.edu/abs/2020MNRAS.491.3672B} {491, 3672}

\bibitem[\protect\citeauthoryear{{Brinchmann}, {Charlot}, {White}, {Tremonti},
  {Kauffmann}, {Heckman}  \& {Brinkmann}}{{Brinchmann}
  et~al.}{2004}]{brinchmann04}
{Brinchmann} J.,  {Charlot} S.,  {White} S.~D.~M.,  {Tremonti} C.,  {Kauffmann}
  G.,  {Heckman} T.,   {Brinkmann} J.,  2004, \mn@doi [\mnras]
  {10.1111/j.1365-2966.2004.07881.x}, \href
  {http://adsabs.harvard.edu/abs/2004MNRAS.351.1151B} {351, 1151}

\bibitem[\protect\citeauthoryear{{Byler}, {Dalcanton}, {Conroy}  \&
  {Johnson}}{{Byler} et~al.}{2017}]{2017Byler_ApJ840}
{Byler} N.,  {Dalcanton} J.~J.,  {Conroy} C.,   {Johnson} B.~D.,  2017, \mn@doi
  [\apj] {10.3847/1538-4357/aa6c66}, \href
  {https://ui.adsabs.harvard.edu/abs/2017ApJ...840...44B} {840, 44}

\bibitem[\protect\citeauthoryear{{Camps-Fari{\~n}a}, {Sanchez}, {Lacerda},
  {Carigi}, {Garc{\'\i}a-Benito}, {Mast}  \& {Galbany}}{{Camps-Fari{\~n}a}
  et~al.}{2021}]{camps20}
{Camps-Fari{\~n}a} A.,  {Sanchez} S.~F.,  {Lacerda} E.~A.~D.,  {Carigi} L.,
  {Garc{\'\i}a-Benito} R.,  {Mast} D.,   {Galbany} L.,  2021, \mn@doi [\mnras]
  {10.1093/mnras/stab1018}, \href
  {https://ui.adsabs.harvard.edu/abs/2021MNRAS.504.3478C} {504, 3478}

\bibitem[\protect\citeauthoryear{{Cardelli}, {Clayton}  \& {Mathis}}{{Cardelli}
  et~al.}{1989}]{cardelli89}
{Cardelli} J.~A.,  {Clayton} G.~C.,   {Mathis} J.~S.,  1989, \mn@doi [\apj]
  {10.1086/167900}, \href {http://adsabs.harvard.edu/abs/1989ApJ...345..245C}
  {345, 245}

\bibitem[\protect\citeauthoryear{{Carigi}, {Peimbert}  \& {Peimbert}}{{Carigi}
  et~al.}{2019}]{carigi19}
{Carigi} L.,  {Peimbert} M.,   {Peimbert} A.,  2019, \mn@doi [\apj]
  {10.3847/1538-4357/aaf28e}, \href
  {https://ui.adsabs.harvard.edu/abs/2019ApJ...873..107C} {873, 107}

\bibitem[\protect\citeauthoryear{{Chabrier}}{{Chabrier}}{2003}]{chab03}
{Chabrier} G.,  2003, \mn@doi [\pasp] {10.1086/376392}, \href
  {http://adsabs.harvard.edu/abs/2003PASP..115..763C} {115, 763}

\bibitem[\protect\citeauthoryear{{Cid Fernandes}, {Stasi{\'n}ska},
  {Schlickmann}, {Mateus}, {Vale Asari}, {Schoenell}  \& {Sodr{\'e}}}{{Cid
  Fernandes} et~al.}{2010}]{cid-fernandes10}
{Cid Fernandes} R.,  {Stasi{\'n}ska} G.,  {Schlickmann} M.~S.,  {Mateus} A.,
  {Vale Asari} N.,  {Schoenell} W.,   {Sodr{\'e}} L.,  2010, \mn@doi [\mnras]
  {10.1111/j.1365-2966.2009.16185.x}, \href
  {http://cdsads.u-strasbg.fr/abs/2010MNRAS.403.1036C} {403, 1036}

\bibitem[\protect\citeauthoryear{{Cid Fernandes} et~al.,}{{Cid Fernandes}
  et~al.}{2013}]{2013CidFernandez_AA557}
{Cid Fernandes} R.,  et~al., 2013, \mn@doi [\aap]
  {10.1051/0004-6361/201220616}, \href
  {http://adsabs.harvard.edu/abs/2013A%26A...557A..86C} {557, A86}

\bibitem[\protect\citeauthoryear{{Cid Fernandes} et~al.,}{{Cid Fernandes}
  et~al.}{2014}]{cid-fernandes14}
{Cid Fernandes} R.,  et~al., 2014, \mn@doi [\aap]
  {10.1051/0004-6361/201321692}, \href
  {http://adsabs.harvard.edu/abs/2014A%26A...561A.130C} {561, A130}

\bibitem[\protect\citeauthoryear{{Cid Fernandes}, {Carvalho}, {S{\'a}nchez},
  {de Amorim}  \& {Ruschel-Dutra}}{{Cid Fernandes}
  et~al.}{2021}]{cid-fernandes21}
{Cid Fernandes} R.,  {Carvalho} M.~S.,  {S{\'a}nchez} S.~F.,  {de Amorim} A.,
  {Ruschel-Dutra} D.,  2021, \mn@doi [\mnras] {10.1093/mnras/stab059}, \href
  {https://ui.adsabs.harvard.edu/abs/2021MNRAS.502.1386C} {502, 1386}

\bibitem[\protect\citeauthoryear{{Colombo} et~al.,}{{Colombo}
  et~al.}{2018}]{colombo18}
{Colombo} D.,  et~al., 2018, \mn@doi [\mnras] {10.1093/mnras/stx3233}, \href
  {http://adsabs.harvard.edu/abs/2018MNRAS.475.1791C} {475, 1791}

\bibitem[\protect\citeauthoryear{{Curti}, {Mannucci}, {Cresci}  \&
  {Maiolino}}{{Curti} et~al.}{2020}]{2020Curti_MNRAS491}
{Curti} M.,  {Mannucci} F.,  {Cresci} G.,   {Maiolino} R.,  2020, \mn@doi
  [\mnras] {10.1093/mnras/stz2910}, \href
  {https://ui.adsabs.harvard.edu/abs/2020MNRAS.491..944C} {491, 944}

\bibitem[\protect\citeauthoryear{{D'Agostino}, {Kewley}, {Groves}, {Medling},
  {Dopita}  \& {Thomas}}{{D'Agostino} et~al.}{2019}]{dagos21}
{D'Agostino} J.~J.,  {Kewley} L.~J.,  {Groves} B.~A.,  {Medling} A.,  {Dopita}
  M.~A.,   {Thomas} A.~D.,  2019, \mn@doi [\mnras] {10.1093/mnrasl/slz028},
  \href {https://ui.adsabs.harvard.edu/abs/2019MNRAS.485L..38D} {485, L38}

\bibitem[\protect\citeauthoryear{{D{\'\i}az}, {Castellanos}, {Terlevich}  \&
  {Luisa Garc{\'{\i}}a-Vargas}}{{D{\'\i}az} et~al.}{2000}]{diaz00}
{D{\'\i}az} A.~I.,  {Castellanos} M.,  {Terlevich} E.,   {Luisa
  Garc{\'{\i}}a-Vargas} M.,  2000, \mn@doi [\mnras]
  {10.1046/j.1365-8711.2000.03737.x}, \href
  {http://adsabs.harvard.edu/abs/2000MNRAS.318..462D} {318, 462}

\bibitem[\protect\citeauthoryear{{Dopita} \& {Evans}}{{Dopita} \&
  {Evans}}{1986}]{dopita86}
{Dopita} M.~A.,  {Evans} I.~N.,  1986, \mn@doi [\apj] {10.1086/164432}, \href
  {http://adsabs.harvard.edu/abs/1986ApJ...307..431D} {307, 431}

\bibitem[\protect\citeauthoryear{{Dopita}, {Sutherland}, {Nicholls}, {Kewley}
  \& {Vogt}}{{Dopita} et~al.}{2013}]{dopita13}
{Dopita} M.~A.,  {Sutherland} R.~S.,  {Nicholls} D.~C.,  {Kewley} L.~J.,
  {Vogt} F.~P.~A.,  2013, \mn@doi [\apjs] {10.1088/0067-0049/208/1/10}, \href
  {http://adsabs.harvard.edu/abs/2013ApJS..208...10D} {208, 10}

\bibitem[\protect\citeauthoryear{{Dors}, {Krabbe}, {H{\"a}gele}  \&
  {P{\'e}rez-Montero}}{{Dors} et~al.}{2011}]{dors11}
{Dors} Jr. O.~L.,  {Krabbe} A.,  {H{\"a}gele} G.~F.,   {P{\'e}rez-Montero} E.,
  2011, \mn@doi [\mnras] {10.1111/j.1365-2966.2011.18978.x}, \href
  {http://adsabs.harvard.edu/abs/2011MNRAS.415.3616D} {415, 3616}

\bibitem[\protect\citeauthoryear{{Edmunds} \& {Pagel}}{{Edmunds} \&
  {Pagel}}{1978}]{1978Edmuns_MNRAS185}
{Edmunds} M.~G.,  {Pagel} B.~E.~J.,  1978, \mn@doi [\mnras]
  {10.1093/mnras/185.1.77P}, \href
  {https://ui.adsabs.harvard.edu/abs/1978MNRAS.185P..77E} {185, 77P}

\bibitem[\protect\citeauthoryear{{Espinosa-Ponce}, {S{\'a}nchez}, {Morisset},
  {Barrera-Ballesteros}, {Galbany}, {Garc{\'\i}a-Benito}, {Lacerda}  \&
  {Mast}}{{Espinosa-Ponce} et~al.}{2020}]{2020EspinosaPonce_MNRAS494}
{Espinosa-Ponce} C.,  {S{\'a}nchez} S.~F.,  {Morisset} C.,
  {Barrera-Ballesteros} J.~K.,  {Galbany} L.,  {Garc{\'\i}a-Benito} R.,
  {Lacerda} E.~A.~D.,   {Mast} D.,  2020, \mn@doi [\mnras]
  {10.1093/mnras/staa782}, \href
  {https://ui.adsabs.harvard.edu/abs/2020MNRAS.494.1622E} {494, 1622}

\bibitem[\protect\citeauthoryear{{Esteban} \& {Garc{\'\i}a-Rojas}}{{Esteban} \&
  {Garc{\'\i}a-Rojas}}{2018}]{esteban18}
{Esteban} C.,  {Garc{\'\i}a-Rojas} J.,  2018, \mn@doi [\mnras]
  {10.1093/mnras/sty1168}, \href
  {https://ui.adsabs.harvard.edu/abs/2018MNRAS.478.2315E} {478, 2315}

\bibitem[\protect\citeauthoryear{{Evans} \& {Dopita}}{{Evans} \&
  {Dopita}}{1985}]{evans85}
{Evans} I.~N.,  {Dopita} M.~A.,  1985, \mn@doi [\apjs] {10.1086/191032}, \href
  {http://adsabs.harvard.edu/abs/1985ApJS...58..125E} {58, 125}

\bibitem[\protect\citeauthoryear{{Flores-Fajardo}, {Morisset}, {Stasi{\'n}ska}
  \& {Binette}}{{Flores-Fajardo} et~al.}{2011}]{2011Flores-Fajardo_MNRAS415}
{Flores-Fajardo} N.,  {Morisset} C.,  {Stasi{\'n}ska} G.,   {Binette} L.,
  2011, \mn@doi [\mnras] {10.1111/j.1365-2966.2011.18848.x}, \href
  {http://adsabs.harvard.edu/abs/2011MNRAS.415.2182F} {415, 2182}

\bibitem[\protect\citeauthoryear{{Galbany} et~al.,}{{Galbany}
  et~al.}{2018}]{2018Galbany_ApJ855}
{Galbany} L.,  et~al., 2018, \mn@doi [\apj] {10.3847/1538-4357/aaaf20}, \href
  {http://adsabs.harvard.edu/abs/2018ApJ...855..107G} {855, 107}

\bibitem[\protect\citeauthoryear{{Gallazzi}, {Charlot}, {Brinchmann}, {White}
  \& {Tremonti}}{{Gallazzi} et~al.}{2005}]{gallazzi05}
{Gallazzi} A.,  {Charlot} S.,  {Brinchmann} J.,  {White} S.~D.~M.,   {Tremonti}
  C.~A.,  2005, \mn@doi [\mnras] {10.1111/j.1365-2966.2005.09321.x}, \href
  {http://adsabs.harvard.edu/abs/2005MNRAS.362...41G} {362, 41}

\bibitem[\protect\citeauthoryear{{Garc{\'{\i}}a-Benito}
  et~al.,}{{Garc{\'{\i}}a-Benito} et~al.}{2017}]{rgb17}
{Garc{\'{\i}}a-Benito} R.,  et~al., 2017, \mn@doi [\aap]
  {10.1051/0004-6361/201731357}, \href
  {http://adsabs.harvard.edu/abs/2017A%26A...608A..27G} {608, A27}

\bibitem[\protect\citeauthoryear{{Garc{\'\i}a-Benito}, {Gonz{\'a}lez Delgado},
  {P{\'e}rez}, {Cid Fernandes}, {S{\'a}nchez}  \& {de
  Amorim}}{{Garc{\'\i}a-Benito} et~al.}{2019}]{rgb18}
{Garc{\'\i}a-Benito} R.,  {Gonz{\'a}lez Delgado} R.~M.,  {P{\'e}rez} E.,  {Cid
  Fernandes} R.,  {S{\'a}nchez} S.~F.,   {de Amorim} A.~L.,  2019, \mn@doi
  [\aap] {10.1051/0004-6361/201833993}, \href
  {https://ui.adsabs.harvard.edu/abs/2019A&A...621A.120G} {621, A120}

\bibitem[\protect\citeauthoryear{{Garnett}}{{Garnett}}{2002}]{garnett:2002p339}
{Garnett} D.~R.,  2002, \mn@doi [\apj] {10.1086/344301}, \href
  {http://adsabs.harvard.edu/abs/2002ApJ...581.1019G} {581, 1019}

\bibitem[\protect\citeauthoryear{{Gomes} et~al.,}{{Gomes}
  et~al.}{2016}]{gomes16b}
{Gomes} J.~M.,  et~al., 2016, \mn@doi [\aap] {10.1051/0004-6361/201525974},
  \href {http://adsabs.harvard.edu/abs/2016A%26A...585A..92G} {585, A92}

\bibitem[\protect\citeauthoryear{{Gonz{\'a}lez Delgado} et~al.,}{{Gonz{\'a}lez
  Delgado} et~al.}{2016}]{2016GonzalezDelgado_A&A590}
{Gonz{\'a}lez Delgado} R.~M.,  et~al., 2016, \mn@doi [\aap]
  {10.1051/0004-6361/201628174}, \href
  {https://ui.adsabs.harvard.edu/abs/2016A&A...590A..44G} {590, A44}

\bibitem[\protect\citeauthoryear{{Hayden} et~al.,}{{Hayden}
  et~al.}{2015}]{2015Hayden_ApJ808}
{Hayden} M.~R.,  et~al., 2015, \mn@doi [\apj] {10.1088/0004-637X/808/2/132},
  \href {https://ui.adsabs.harvard.edu/abs/2015ApJ...808..132H} {808, 132}

\bibitem[\protect\citeauthoryear{{Ho}}{{Ho}}{2019}]{2019Ho_MNRAS485}
{Ho} I.~T.,  2019, \mn@doi [\mnras] {10.1093/mnras/stz649}, \href
  {https://ui.adsabs.harvard.edu/abs/2019MNRAS.485.3569H} {485, 3569}

\bibitem[\protect\citeauthoryear{{Ibarra-Medel} et~al.,}{{Ibarra-Medel}
  et~al.}{2016}]{ibarra16}
{Ibarra-Medel} H.~J.,  et~al., 2016, \mn@doi [\mnras] {10.1093/mnras/stw2126},
  \href {http://adsabs.harvard.edu/abs/2016MNRAS.463.2799I} {463, 2799}

\bibitem[\protect\citeauthoryear{{Izotov}, {Thuan}  \& {Guseva}}{{Izotov}
  et~al.}{2012}]{izotov12}
{Izotov} Y.~I.,  {Thuan} T.~X.,   {Guseva} N.~G.,  2012, \mn@doi [\aap]
  {10.1051/0004-6361/201219733}, \href
  {https://ui.adsabs.harvard.edu/abs/2012A&A...546A.122I} {546, A122}

\bibitem[\protect\citeauthoryear{{Ji} \& {Yan}}{{Ji} \&
  {Yan}}{2021}]{2021Ji_arXiv211000612}
{Ji} X.,  {Yan} R.,  2021, arXiv e-prints, \href
  {https://ui.adsabs.harvard.edu/abs/2021arXiv211000612J} {p. arXiv:2110.00612}

\bibitem[\protect\citeauthoryear{{Kauffmann} et~al.,}{{Kauffmann}
  et~al.}{2003}]{2003Kauffmann_mnras346}
{Kauffmann} G.,  et~al., 2003, \mn@doi [\mnras]
  {10.1111/j.1365-2966.2003.07154.x}, \href
  {http://adsabs.harvard.edu/abs/2003MNRAS.346.1055K} {346, 1055}

\bibitem[\protect\citeauthoryear{{Kehrig} et~al.,}{{Kehrig}
  et~al.}{2012}]{kehrig12}
{Kehrig} C.,  et~al., 2012, \mn@doi [\aap] {10.1051/0004-6361/201118357}, \href
  {http://adsabs.harvard.edu/abs/2012A%26A...540A..11K} {540, A11}

\bibitem[\protect\citeauthoryear{{Kelz} et~al.,}{{Kelz} et~al.}{2006}]{kelz06}
{Kelz} A.,  et~al., 2006, \mn@doi [\pasp] {10.1086/497455}, \href
  {http://adsabs.harvard.edu/abs/2006PASP..118..129K} {118, 129}

\bibitem[\protect\citeauthoryear{{Kennicutt} \& {Garnett}}{{Kennicutt} \&
  {Garnett}}{1996}]{1996Kennicutt_ApJ456}
{Kennicutt} Robert~C. J.,  {Garnett} D.~R.,  1996, \mn@doi [\apj]
  {10.1086/176675}, \href
  {https://ui.adsabs.harvard.edu/abs/1996ApJ...456..504K} {456, 504}

\bibitem[\protect\citeauthoryear{{Kennicutt} \& {Hodge}}{{Kennicutt} \&
  {Hodge}}{1980}]{kennicutt80}
{Kennicutt} R.~C.,  {Hodge} P.~W.,  1980, \mn@doi [\apj] {10.1086/158372},
  \href {https://ui.adsabs.harvard.edu/abs/1980ApJ...241..573K} {241, 573}

\bibitem[\protect\citeauthoryear{{Kennicutt}, {Bresolin}  \&
  {Garnett}}{{Kennicutt} et~al.}{2003}]{2003Kennicutt_ApJ591}
{Kennicutt} Robert~C. J.,  {Bresolin} F.,   {Garnett} D.~R.,  2003, \mn@doi
  [\apj] {10.1086/375398}, \href
  {https://ui.adsabs.harvard.edu/abs/2003ApJ...591..801K} {591, 801}

\bibitem[\protect\citeauthoryear{{Kewley} \& {Dopita}}{{Kewley} \&
  {Dopita}}{2002}]{kewley02}
{Kewley} L.~J.,  {Dopita} M.~A.,  2002, \mn@doi [\apjs] {10.1086/341326}, \href
  {http://adsabs.harvard.edu/abs/2002ApJS..142...35K} {142, 35}

\bibitem[\protect\citeauthoryear{{Kewley} \& {Ellison}}{{Kewley} \&
  {Ellison}}{2008}]{kewley08}
{Kewley} L.~J.,  {Ellison} S.~L.,  2008, \mn@doi [\apj] {10.1086/587500}, \href
  {http://adsabs.harvard.edu/abs/2008ApJ...681.1183K} {681, 1183}

\bibitem[\protect\citeauthoryear{{Kewley}, {Dopita}, {Sutherland}, {Heisler}
  \& {Trevena}}{{Kewley} et~al.}{2001}]{2001Kewley_ApJ556}
{Kewley} L.~J.,  {Dopita} M.~A.,  {Sutherland} R.~S.,  {Heisler} C.~A.,
  {Trevena} J.,  2001, \mn@doi [\apj] {10.1086/321545}, \href
  {http://adsabs.harvard.edu/abs/2001ApJ...556..121K} {556, 121}

\bibitem[\protect\citeauthoryear{{Kewley}, {Groves}, {Kauffmann}  \&
  {Heckman}}{{Kewley} et~al.}{2006}]{kewley06}
{Kewley} L.~J.,  {Groves} B.,  {Kauffmann} G.,   {Heckman} T.,  2006, \mn@doi
  [\mnras] {10.1111/j.1365-2966.2006.10859.x}, \href
  {http://adsabs.harvard.edu/abs/2006MNRAS.372..961K} {372, 961}

\bibitem[\protect\citeauthoryear{{Kewley}, {Dopita}, {Leitherer}, {Dav{\'e}},
  {Yuan}, {Allen}, {Groves}  \& {Sutherland}}{{Kewley}
  et~al.}{2013}]{2013Kewley_ApJ774}
{Kewley} L.~J.,  {Dopita} M.~A.,  {Leitherer} C.,  {Dav{\'e}} R.,  {Yuan} T.,
  {Allen} M.,  {Groves} B.,   {Sutherland} R.,  2013, \mn@doi [\apj]
  {10.1088/0004-637X/774/2/100}, \href
  {https://ui.adsabs.harvard.edu/abs/2013ApJ...774..100K} {774, 100}

\bibitem[\protect\citeauthoryear{{Kewley}, {Nicholls}  \&
  {Sutherland}}{{Kewley} et~al.}{2019}]{2019Kewley_ARA&A57}
{Kewley} L.~J.,  {Nicholls} D.~C.,   {Sutherland} R.~S.,  2019, \mn@doi [\araa]
  {10.1146/annurev-astro-081817-051832}, \href
  {https://ui.adsabs.harvard.edu/abs/2019ARA&A..57..511K} {57, 511}

\bibitem[\protect\citeauthoryear{{Kobulnicky} \& {Kewley}}{{Kobulnicky} \&
  {Kewley}}{2004}]{2004Kobulnicky_ApJ617}
{Kobulnicky} H.~A.,  {Kewley} L.~J.,  2004, \mn@doi [\apj] {10.1086/425299},
  \href {https://ui.adsabs.harvard.edu/abs/2004ApJ...617..240K} {617, 240}

\bibitem[\protect\citeauthoryear{{Lacerda}, {S{\'a}nchez}, {Cid Fernandes},
  {L{\'o}pez-Cob{\'a}}, {Espinosa-Ponce}  \& {Galbany}}{{Lacerda}
  et~al.}{2020}]{lacerda20}
{Lacerda} E. A.~D.,  {S{\'a}nchez} S.~F.,  {Cid Fernandes} R.,
  {L{\'o}pez-Cob{\'a}} C.,  {Espinosa-Ponce} C.,   {Galbany} L.,  2020, \mn@doi
  [\mnras] {10.1093/mnras/staa008}, \href
  {https://ui.adsabs.harvard.edu/abs/2020MNRAS.492.3073L} {492, 3073}

\bibitem[\protect\citeauthoryear{{Law} et~al.,}{{Law} et~al.}{2021}]{law21}
{Law} D.~R.,  et~al., 2021, \mn@doi [\apj] {10.3847/1538-4357/abfe0a}, \href
  {https://ui.adsabs.harvard.edu/abs/2021ApJ...915...35L} {915, 35}

\bibitem[\protect\citeauthoryear{{Lequeux}, {Peimbert}, {Rayo}, {Serrano}  \&
  {Torres-Peimbert}}{{Lequeux} et~al.}{1979}]{leque79}
{Lequeux} J.,  {Peimbert} M.,  {Rayo} J.~F.,  {Serrano} A.,   {Torres-Peimbert}
  S.,  1979, \aap, \href {http://adsabs.harvard.edu/abs/1979A%26A....80..155L}
  {80, 155}

\bibitem[\protect\citeauthoryear{{Levesque}, {Kewley}  \& {Larson}}{{Levesque}
  et~al.}{2010}]{2010Levesque_AJ139}
{Levesque} E.~M.,  {Kewley} L.~J.,   {Larson} K.~L.,  2010, \mn@doi [\aj]
  {10.1088/0004-6256/139/2/712}, \href
  {https://ui.adsabs.harvard.edu/abs/2010AJ....139..712L} {139, 712}

\bibitem[\protect\citeauthoryear{{Li} et~al.,}{{Li} et~al.}{2015}]{cheng15}
{Li} C.,  et~al., 2015, \mn@doi [\apj] {10.1088/0004-637X/804/2/125}, \href
  {https://ui.adsabs.harvard.edu/abs/2015ApJ...804..125L} {804, 125}

\bibitem[\protect\citeauthoryear{{L{\'o}pez-Cob{\'a}}, {S{\'a}nchez}, {Bland
  -Hawthorn}, {Moiseev}, {Cruz-Gonz{\'a}lez}, {Garc{\'\i}a-Benito},
  {Barrera-Ballesteros}  \& {Galbany}}{{L{\'o}pez-Cob{\'a}}
  et~al.}{2019}]{carlos19}
{L{\'o}pez-Cob{\'a}} C.,  {S{\'a}nchez} S.~F.,  {Bland -Hawthorn} J.,
  {Moiseev} A.~V.,  {Cruz-Gonz{\'a}lez} I.,  {Garc{\'\i}a-Benito} R.,
  {Barrera-Ballesteros} J.~K.,   {Galbany} L.,  2019, \mn@doi [\mnras]
  {10.1093/mnras/sty2960}, \href
  {https://ui.adsabs.harvard.edu/abs/2019MNRAS.482.4032L} {482, 4032}

\bibitem[\protect\citeauthoryear{{L{\'o}pez-Cob{\'a}}
  et~al.,}{{L{\'o}pez-Cob{\'a}} et~al.}{2020}]{2020LopezCoba_AJ159}
{L{\'o}pez-Cob{\'a}} C.,  et~al., 2020, \mn@doi [\aj]
  {10.3847/1538-3881/ab7848}, \href
  {https://ui.adsabs.harvard.edu/abs/2020AJ....159..167L} {159, 167}

\bibitem[\protect\citeauthoryear{{Maiolino} \& {Mannucci}}{{Maiolino} \&
  {Mannucci}}{2019}]{2019Maiolino_A&ARv27}
{Maiolino} R.,  {Mannucci} F.,  2019, \mn@doi [\aapr]
  {10.1007/s00159-018-0112-2}, \href
  {https://ui.adsabs.harvard.edu/abs/2019A&ARv..27....3M} {27, 3}

\bibitem[\protect\citeauthoryear{{Marino} et~al.,}{{Marino}
  et~al.}{2013}]{2013Marino_aa559A}
{Marino} R.~A.,  et~al., 2013, \mn@doi [\aap] {10.1051/0004-6361/201321956},
  \href {http://adsabs.harvard.edu/abs/2013A%26A...559A.114M} {559, A114}

\bibitem[\protect\citeauthoryear{{Mart{\'{\i}}n-Navarro}
  et~al.,}{{Mart{\'{\i}}n-Navarro} et~al.}{2015}]{navarro16}
{Mart{\'{\i}}n-Navarro} I.,  et~al., 2015, \mn@doi [\apjl]
  {10.1088/2041-8205/806/2/L31}, \href
  {http://adsabs.harvard.edu/abs/2015ApJ...806L..31M} {806, L31}

\bibitem[\protect\citeauthoryear{{Matteucci}}{{Matteucci}}{1986}]{1986Matteucci_MNRAS221}
{Matteucci} F.,  1986, \mn@doi [\mnras] {10.1093/mnras/221.4.911}, \href
  {https://ui.adsabs.harvard.edu/abs/1986MNRAS.221..911M} {221, 911}

\bibitem[\protect\citeauthoryear{{McCall}, {Rybski}  \& {Shields}}{{McCall}
  et~al.}{1985}]{McCa85}
{McCall} M.~L.,  {Rybski} P.~M.,   {Shields} G.~A.,  1985, \mn@doi [\apjs]
  {10.1086/190994}, \href {http://adsabs.harvard.edu/abs/1985ApJS...57....1M}
  {57, 1}

\bibitem[\protect\citeauthoryear{{McDermid} et~al.,}{{McDermid}
  et~al.}{2015}]{mcdermid15}
{McDermid} R.~M.,  et~al., 2015, \mn@doi [\mnras] {10.1093/mnras/stv105}, \href
  {http://adsabs.harvard.edu/abs/2015MNRAS.448.3484M} {448, 3484}

\bibitem[\protect\citeauthoryear{{Mingozzi} et~al.,}{{Mingozzi}
  et~al.}{2020}]{2020Mingozzi_AA636}
{Mingozzi} M.,  et~al., 2020, \mn@doi [\aap] {10.1051/0004-6361/201937203},
  \href {https://ui.adsabs.harvard.edu/abs/2020A&A...636A..42M} {636, A42}

\bibitem[\protect\citeauthoryear{{Moll{\'a}} \& {D{\'\i}az}}{{Moll{\'a}} \&
  {D{\'\i}az}}{2005}]{2005Molla_MNRAS358}
{Moll{\'a}} M.,  {D{\'\i}az} A.~I.,  2005, \mn@doi [\mnras]
  {10.1111/j.1365-2966.2005.08782.x}, \href
  {https://ui.adsabs.harvard.edu/abs/2005MNRAS.358..521M} {358, 521}

\bibitem[\protect\citeauthoryear{{Morisset} et~al.,}{{Morisset}
  et~al.}{2016}]{2016Morisset_aa594A}
{Morisset} C.,  et~al., 2016, \mn@doi [\aap] {10.1051/0004-6361/201628559},
  \href {http://adsabs.harvard.edu/abs/2016A%26A...594A..37M} {594, A37}

\bibitem[\protect\citeauthoryear{{Moustakas} \& {Kennicutt}}{{Moustakas} \&
  {Kennicutt}}{2006}]{moustakas06}
{Moustakas} J.,  {Kennicutt} Jr. R.~C.,  2006, \mn@doi [\apjs]
  {10.1086/500971}, \href {http://adsabs.harvard.edu/abs/2006ApJS..164...81M}
  {164, 81}

\bibitem[\protect\citeauthoryear{{Nicholls}, {Dopita}  \&
  {Sutherland}}{{Nicholls} et~al.}{2012}]{2012Nicholls_ApJ752}
{Nicholls} D.~C.,  {Dopita} M.~A.,   {Sutherland} R.~S.,  2012, \mn@doi [\apj]
  {10.1088/0004-637X/752/2/148}, \href
  {https://ui.adsabs.harvard.edu/abs/2012ApJ...752..148N} {752, 148}

\bibitem[\protect\citeauthoryear{{Osterbrock}}{{Osterbrock}}{1989}]{osterbrock89}
{Osterbrock} D.~E.,  1989, {Astrophysics of gaseous nebulae and active galactic
  nuclei}.
University Science Books

\bibitem[\protect\citeauthoryear{{Pagel}, {Edmunds}, {Blackwell}, {Chun}  \&
  {Smith}}{{Pagel} et~al.}{1979}]{pagel79}
{Pagel} B.~E.~J.,  {Edmunds} M.~G.,  {Blackwell} D.~E.,  {Chun} M.~S.,
  {Smith} G.,  1979, \mnras, \href
  {http://adsabs.harvard.edu/abs/1979MNRAS.189...95P} {189, 95}

\bibitem[\protect\citeauthoryear{{Peimbert}}{{Peimbert}}{1967}]{peim67}
{Peimbert} M.,  1967, \mn@doi [\apj] {10.1086/149385}, \href
  {http://adsabs.harvard.edu/abs/1967ApJ...150..825P} {150, 825}

\bibitem[\protect\citeauthoryear{{Peimbert}, {Torres-Peimbert}  \&
  {Rayo}}{{Peimbert} et~al.}{1978}]{peim78}
{Peimbert} M.,  {Torres-Peimbert} S.,   {Rayo} J.~F.,  1978, \mn@doi [\apj]
  {10.1086/155933}, \href {http://adsabs.harvard.edu/abs/1978ApJ...220..516P}
  {220, 516}

\bibitem[\protect\citeauthoryear{{Peimbert}, {Peimbert}  \&
  {Delgado-Inglada}}{{Peimbert} et~al.}{2017}]{2017Peimbert_PASP129}
{Peimbert} M.,  {Peimbert} A.,   {Delgado-Inglada} G.,  2017, \mn@doi [\pasp]
  {10.1088/1538-3873/aa72c3}, \href
  {https://ui.adsabs.harvard.edu/abs/2017PASP..129h2001P} {129, 082001}

\bibitem[\protect\citeauthoryear{{P{\'e}rez-Montero}}{{P{\'e}rez-Montero}}{2014}]{2014PerezMontero_MNRAS441}
{P{\'e}rez-Montero} E.,  2014, \mn@doi [\mnras] {10.1093/mnras/stu753}, \href
  {https://ui.adsabs.harvard.edu/abs/2014MNRAS.441.2663P} {441, 2663}

\bibitem[\protect\citeauthoryear{{P{\'e}rez-Montero}}{{P{\'e}rez-Montero}}{2017}]{2017PerezMontero_PASP12P}
{P{\'e}rez-Montero} E.,  2017, \mn@doi [\pasp] {10.1088/1538-3873/aa5abb},
  \href {https://ui.adsabs.harvard.edu/abs/2017PASP..129d3001P} {129, 043001}

\bibitem[\protect\citeauthoryear{{P{\'e}rez-Montero}
  et~al.,}{{P{\'e}rez-Montero} et~al.}{2016}]{2016PerezMontero_AA595}
{P{\'e}rez-Montero} E.,  et~al., 2016, \mn@doi [\aap]
  {10.1051/0004-6361/201628601}, \href
  {https://ui.adsabs.harvard.edu/abs/2016A&A...595A..62P} {595, A62}

\bibitem[\protect\citeauthoryear{{P{\'e}rez} et~al.,}{{P{\'e}rez}
  et~al.}{2013}]{eperez13}
{P{\'e}rez} E.,  et~al., 2013, \mn@doi [\apjl] {10.1088/2041-8205/764/1/L1},
  \href {http://adsabs.harvard.edu/abs/2013ApJ...764L...1P} {764, L1}

\bibitem[\protect\citeauthoryear{{Pettini} \& {Pagel}}{{Pettini} \&
  {Pagel}}{2004}]{2004Pettini_MNRAS348}
{Pettini} M.,  {Pagel} B. E.~J.,  2004, \mn@doi [\mnras]
  {10.1111/j.1365-2966.2004.07591.x}, \href
  {https://ui.adsabs.harvard.edu/abs/2004MNRAS.348L..59P} {348, L59}

\bibitem[\protect\citeauthoryear{{Pilyugin} \& {Grebel}}{{Pilyugin} \&
  {Grebel}}{2016}]{2016Pilyugin_MNRAS457}
{Pilyugin} L.~S.,  {Grebel} E.~K.,  2016, \mn@doi [\mnras]
  {10.1093/mnras/stw238}, \href
  {https://ui.adsabs.harvard.edu/abs/2016MNRAS.457.3678P} {457, 3678}

\bibitem[\protect\citeauthoryear{{Pilyugin} \& {Mattsson}}{{Pilyugin} \&
  {Mattsson}}{2011}]{2011Pilyugin_MNRAS412}
{Pilyugin} L.~S.,  {Mattsson} L.,  2011, \mn@doi [\mnras]
  {10.1111/j.1365-2966.2010.17970.x}, \href
  {https://ui.adsabs.harvard.edu/abs/2011MNRAS.412.1145P} {412, 1145}

\bibitem[\protect\citeauthoryear{{Pilyugin}, {V{\'\i}lchez}  \&
  {Thuan}}{{Pilyugin} et~al.}{2010}]{2010Pilyugin_ApJ720}
{Pilyugin} L.~S.,  {V{\'\i}lchez} J.~M.,   {Thuan} T.~X.,  2010, \mn@doi [\apj]
  {10.1088/0004-637X/720/2/1738}, \href
  {https://ui.adsabs.harvard.edu/abs/2010ApJ...720.1738P} {720, 1738}

\bibitem[\protect\citeauthoryear{{Pilyugin}, {V{\'\i}lchez}, {Mattsson}  \&
  {Thuan}}{{Pilyugin} et~al.}{2012}]{2012Pilyugin_MNRAS421}
{Pilyugin} L.~S.,  {V{\'\i}lchez} J.~M.,  {Mattsson} L.,   {Thuan} T.~X.,
  2012, \mn@doi [\mnras] {10.1111/j.1365-2966.2012.20420.x}, \href
  {https://ui.adsabs.harvard.edu/abs/2012MNRAS.421.1624P} {421, 1624}

\bibitem[\protect\citeauthoryear{{Rosales-Ortega}, {Kennicutt}, {S{\'a}nchez},
  {D{\'{\i}}az}, {Pasquali}, {Johnson}  \& {Hao}}{{Rosales-Ortega}
  et~al.}{2010}]{rosales-ortega10}
{Rosales-Ortega} F.~F.,  {Kennicutt} R.~C.,  {S{\'a}nchez} S.~F.,
  {D{\'{\i}}az} A.~I.,  {Pasquali} A.,  {Johnson} B.~D.,   {Hao} C.~N.,  2010,
  \mn@doi [\mnras] {10.1111/j.1365-2966.2010.16498.x}, \href
  {http://adsabs.harvard.edu/abs/2010MNRAS.405..735R} {405, 735}

\bibitem[\protect\citeauthoryear{{Rosales-Ortega}, {S{\'a}nchez},
  {Iglesias-P{\'a}ramo}, {D{\'{\i}}az}, {V{\'{\i}}lchez}, {Bland-Hawthorn},
  {Husemann}  \& {Mast}}{{Rosales-Ortega} et~al.}{2012}]{rosales-ortega:2012}
{Rosales-Ortega} F.~F.,  {S{\'a}nchez} S.~F.,  {Iglesias-P{\'a}ramo} J.,
  {D{\'{\i}}az} A.~I.,  {V{\'{\i}}lchez} J.~M.,  {Bland-Hawthorn} J.,
  {Husemann} B.,   {Mast} D.,  2012, \mn@doi [\apjl]
  {10.1088/2041-8205/756/2/L31}, \href
  {http://adsabs.harvard.edu/abs/2012ApJ...756L..31R} {756, L31}

\bibitem[\protect\citeauthoryear{{Roth} et~al.,}{{Roth} et~al.}{2005}]{roth05}
{Roth} M.~M.,  et~al., 2005, \mn@doi [\pasp] {10.1086/429877}, \href
  {http://adsabs.harvard.edu/abs/2005PASP..117..620R} {117, 620}

\bibitem[\protect\citeauthoryear{{S{\'a}nchez}}{{S{\'a}nchez}}{2006}]{sanchez06a}
{S{\'a}nchez} S.~F.,  2006, \mn@doi [Astronomische Nachrichten]
  {10.1002/asna.200610643}, \href
  {https://ui.adsabs.harvard.edu/abs/2006AN....327..850S} {327, 850}

\bibitem[\protect\citeauthoryear{{S{\'a}nchez}}{{S{\'a}nchez}}{2020}]{2020Sanchez_ARA&A58}
{S{\'a}nchez} S.~F.,  2020, \mn@doi [\araa]
  {10.1146/annurev-astro-012120-013326}, \href
  {https://ui.adsabs.harvard.edu/abs/2020ARA&A..58...99S} {58, 99}

\bibitem[\protect\citeauthoryear{{S{\'a}nchez-Menguiano}
  et~al.,}{{S{\'a}nchez-Menguiano} et~al.}{2016}]{2016Sanchez-Menguiano_AA587}
{S{\'a}nchez-Menguiano} L.,  et~al., 2016, \mn@doi [\aap]
  {10.1051/0004-6361/201527450}, \href
  {http://adsabs.harvard.edu/abs/2016A%26A...587A..70S} {587, A70}

\bibitem[\protect\citeauthoryear{{S{\'a}nchez-Menguiano}
  et~al.,}{{S{\'a}nchez-Menguiano} et~al.}{2018}]{2018Sanchez-Menguiano_A&A609}
{S{\'a}nchez-Menguiano} L.,  et~al., 2018, \mn@doi [\aap]
  {10.1051/0004-6361/201731486}, \href
  {http://adsabs.harvard.edu/abs/2018A%26A...609A.119S} {609, A119}

\bibitem[\protect\citeauthoryear{{S{\'a}nchez-Menguiano}, {S{\'a}nchez},
  {P{\'e}rez}, {Ruiz-Lara}, {Galbany}, {Anderson}  \&
  {Kuncarayakti}}{{S{\'a}nchez-Menguiano}
  et~al.}{2020}]{2020SanchezMenguiano_MNRAS492}
{S{\'a}nchez-Menguiano} L.,  {S{\'a}nchez} S.~F.,  {P{\'e}rez} I.,  {Ruiz-Lara}
  T.,  {Galbany} L.,  {Anderson} J.~P.,   {Kuncarayakti} H.,  2020, \mn@doi
  [\mnras] {10.1093/mnras/staa088}, \href
  {https://ui.adsabs.harvard.edu/abs/2020MNRAS.492.4149S} {492, 4149}

\bibitem[\protect\citeauthoryear{{S{\'a}nchez}, {Cardiel}, {Verheijen},
  {Mart{\'\i}n-Gord{\'o}n}, {Vilchez}  \& {Alves}}{{S{\'a}nchez}
  et~al.}{2007}]{2007Sanchez_AA465}
{S{\'a}nchez} S.~F.,  {Cardiel} N.,  {Verheijen} M.~A.~W.,
  {Mart{\'\i}n-Gord{\'o}n} D.,  {Vilchez} J.~M.,   {Alves} J.,  2007, \mn@doi
  [\aap] {10.1051/0004-6361:20066620}, \href
  {https://ui.adsabs.harvard.edu/abs/2007A&A...465..207S} {465, 207}

\bibitem[\protect\citeauthoryear{{S{\'a}nchez} et~al.,}{{S{\'a}nchez}
  et~al.}{2012a}]{2012Sanchez_A&A538A}
{S{\'a}nchez} S.~F.,  et~al., 2012a, \mn@doi [\aap]
  {10.1051/0004-6361/201117353}, \href
  {http://adsabs.harvard.edu/abs/2012A%26A...538A...8S} {538, A8}

\bibitem[\protect\citeauthoryear{{S{\'a}nchez} et~al.,}{{S{\'a}nchez}
  et~al.}{2012b}]{sanchez12b}
{S{\'a}nchez} S.~F.,  et~al., 2012b, \mn@doi [\aap]
  {10.1051/0004-6361/201219578}, \href
  {http://adsabs.harvard.edu/abs/2012A%26A...546A...2S} {546, A2}

\bibitem[\protect\citeauthoryear{{S{\'a}nchez} et~al.,}{{S{\'a}nchez}
  et~al.}{2013}]{sanchez13}
{S{\'a}nchez} S.~F.,  et~al., 2013, \mn@doi [\aap]
  {10.1051/0004-6361/201220669}, \href
  {http://adsabs.harvard.edu/abs/2013A%26A...554A..58S} {554, A58}

\bibitem[\protect\citeauthoryear{{S{\'a}nchez} et~al.,}{{S{\'a}nchez}
  et~al.}{2014}]{sanchez14}
{S{\'a}nchez} S.~F.,  et~al., 2014, \mn@doi [\aap]
  {10.1051/0004-6361/201322343}, \href
  {http://adsabs.harvard.edu/abs/2014A%26A...563A..49S} {563, A49}

\bibitem[\protect\citeauthoryear{{S{\'a}nchez} et~al.,}{{S{\'a}nchez}
  et~al.}{2015a}]{2015Sanchez_AA573}
{S{\'a}nchez} S.~F.,  et~al., 2015a, \mn@doi [\aap]
  {10.1051/0004-6361/201424950}, \href
  {https://ui.adsabs.harvard.edu/abs/2015A&A...573A.105S} {573, A105}

\bibitem[\protect\citeauthoryear{{S{\'a}nchez} et~al.,}{{S{\'a}nchez}
  et~al.}{2015b}]{2015Sanchez_AA574}
{S{\'a}nchez} S.~F.,  et~al., 2015b, \mn@doi [\aap]
  {10.1051/0004-6361/201424873}, \href
  {https://ui.adsabs.harvard.edu/abs/2015A&A...574A..47S} {574, A47}

\bibitem[\protect\citeauthoryear{{S{\'a}nchez} et~al.,}{{S{\'a}nchez}
  et~al.}{2016a}]{2016Sanchez_RMxAA52a}
{S{\'a}nchez} S.~F.,  et~al., 2016a, \rmxaa, \href
  {http://adsabs.harvard.edu/abs/2016RMxAA..52...21S} {52, 21}

\bibitem[\protect\citeauthoryear{{S{\'a}nchez} et~al.,}{{S{\'a}nchez}
  et~al.}{2016b}]{DR3}
{S{\'a}nchez} S.~F.,  et~al., 2016b, \mn@doi [\aap]
  {10.1051/0004-6361/201628661}, \href
  {http://adsabs.harvard.edu/abs/2016A%26A...594A..36S} {594, A36}

\bibitem[\protect\citeauthoryear{{S{\'a}nchez} et~al.,}{{S{\'a}nchez}
  et~al.}{2018}]{2018Sanchez_RMxAA54}
{S{\'a}nchez} S.~F.,  et~al., 2018, \rmxaa, \href
  {https://ui.adsabs.harvard.edu/abs/2018RMxAA..54..217S} {54, 217}

\bibitem[\protect\citeauthoryear{{S{\'a}nchez} et~al.,}{{S{\'a}nchez}
  et~al.}{2019}]{sanchez19}
{S{\'a}nchez} S.~F.,  et~al., 2019, \mn@doi [\mnras] {10.1093/mnras/stz019},
  \href {http://adsabs.harvard.edu/abs/2019MNRAS.484.3042S} {484, 3042}

\bibitem[\protect\citeauthoryear{{S{\'a}nchez}, {Walcher}, {Lopez-Cob{\'a}},
  {Barrera-Ballesteros}, {Mej{\'\i}a-Narv{\'a}ez}, {Espinosa-Ponce}  \&
  {Camps-Fari{\~n}a}}{{S{\'a}nchez} et~al.}{2021a}]{2021Sanchez_RMxAA57}
{S{\'a}nchez} S.~F.,  {Walcher} C.~J.,  {Lopez-Cob{\'a}} C.,
  {Barrera-Ballesteros} J.~K.,  {Mej{\'\i}a-Narv{\'a}ez} A.,  {Espinosa-Ponce}
  C.,   {Camps-Fari{\~n}a} A.,  2021a, \mn@doi [\rmxaa]
  {10.22201/ia.01851101p.2021.57.01.01}, \href
  {https://ui.adsabs.harvard.edu/abs/2021RMxAA..57....3S} {57, 3}

\bibitem[\protect\citeauthoryear{{S{\'a}nchez} et~al.,}{{S{\'a}nchez}
  et~al.}{2021b}]{2021Sanchez_A&A652}
{S{\'a}nchez} S.~F.,  et~al., 2021b, \mn@doi [\aap]
  {10.1051/0004-6361/202141225}, \href
  {https://ui.adsabs.harvard.edu/abs/2021A&A...652L..10S} {652, L10}

\bibitem[\protect\citeauthoryear{{Schaefer}, {Tremonti}, {Belfiore}, {Pace},
  {Bershady}, {Andrews}  \& {Drory}}{{Schaefer}
  et~al.}{2020}]{2020Schaefer_ApJL890}
{Schaefer} A.~L.,  {Tremonti} C.,  {Belfiore} F.,  {Pace} Z.,  {Bershady}
  M.~A.,  {Andrews} B.~H.,   {Drory} N.,  2020, \mn@doi [\apjl]
  {10.3847/2041-8213/ab6f06}, \href
  {https://ui.adsabs.harvard.edu/abs/2020ApJ...890L...3S} {890, L3}

\bibitem[\protect\citeauthoryear{{Schmidt}}{{Schmidt}}{1959}]{schmidt59}
{Schmidt} M.,  1959, \mn@doi [\apj] {10.1086/146614}, \href
  {http://adsabs.harvard.edu/abs/1959ApJ...129..243S} {129, 243}

\bibitem[\protect\citeauthoryear{{Searle}}{{Searle}}{1971}]{sear71}
{Searle} L.,  1971, \mn@doi [\apj] {10.1086/151090}, \href
  {http://adsabs.harvard.edu/abs/1971ApJ...168..327S} {168, 327}

\bibitem[\protect\citeauthoryear{{Skillman}, {Kennicutt}  \&
  {Hodge}}{{Skillman} et~al.}{1989}]{skill89}
{Skillman} E.~D.,  {Kennicutt} R.~C.,   {Hodge} P.~W.,  1989, \mn@doi [\apj]
  {10.1086/168178}, \href {http://adsabs.harvard.edu/abs/1989ApJ...347..875S}
  {347, 875}

\bibitem[\protect\citeauthoryear{{Stasi{\'n}ska} et~al.,}{{Stasi{\'n}ska}
  et~al.}{2008}]{2008Stasinska_MNRAS391}
{Stasi{\'n}ska} G.,  et~al., 2008, \mn@doi [\mnras]
  {10.1111/j.1745-3933.2008.00550.x}, \href
  {http://adsabs.harvard.edu/abs/2008MNRAS.391L..29S} {391, L29}

\bibitem[\protect\citeauthoryear{{Str{\"o}mgren}}{{Str{\"o}mgren}}{1939}]{strom39}
{Str{\"o}mgren} B.,  1939, \mn@doi [\apj] {10.1086/144074}, \href
  {http://adsabs.harvard.edu/abs/1939ApJ....89..526S} {89, 526}

\bibitem[\protect\citeauthoryear{{Thomas}, {Dopita}, {Kewley}, {Groves},
  {Sutherland}, {Hopkins}  \& {Blanc}}{{Thomas}
  et~al.}{2018}]{2018Thomas_ApJ856}
{Thomas} A.~D.,  {Dopita} M.~A.,  {Kewley} L.~J.,  {Groves} B.~A.,
  {Sutherland} R.~S.,  {Hopkins} A.~M.,   {Blanc} G.~A.,  2018, \mn@doi [\apj]
  {10.3847/1538-4357/aab3db}, \href
  {https://ui.adsabs.harvard.edu/abs/2018ApJ...856...89T} {856, 89}

\bibitem[\protect\citeauthoryear{{Tissera}, {Scannapieco}, {Beers}  \&
  {Carollo}}{{Tissera} et~al.}{2013}]{tissera:2013aa}
{Tissera} P.~B.,  {Scannapieco} C.,  {Beers} T.~C.,   {Carollo} D.,  2013,
  \mn@doi [\mnras] {10.1093/mnras/stt691}, \href
  {http://adsabs.harvard.edu/abs/2013MNRAS.432.3391T} {432, 3391}

\bibitem[\protect\citeauthoryear{{Tremonti} et~al.,}{{Tremonti}
  et~al.}{2004}]{2004Tremonti_ApJ613}
{Tremonti} C.~A.,  et~al., 2004, \mn@doi [\apj] {10.1086/423264}, \href
  {https://ui.adsabs.harvard.edu/abs/2004ApJ...613..898T} {613, 898}

\bibitem[\protect\citeauthoryear{{Vale Asari}, {Stasi{\'n}ska}, {Morisset}  \&
  {Cid Fernandes}}{{Vale Asari} et~al.}{2016}]{2016Asari_MNRAS460}
{Vale Asari} N.,  {Stasi{\'n}ska} G.,  {Morisset} C.,   {Cid Fernandes} R.,
  2016, \mn@doi [\mnras] {10.1093/mnras/stw971}, \href
  {https://ui.adsabs.harvard.edu/abs/2016MNRAS.460.1739V} {460, 1739}

\bibitem[\protect\citeauthoryear{{Vazdekis}, {Casuso}, {Peletier}  \&
  {Beckman}}{{Vazdekis} et~al.}{1996}]{vazdekis96}
{Vazdekis} A.,  {Casuso} E.,  {Peletier} R.~F.,   {Beckman} J.~E.,  1996,
  \mn@doi [\apjs] {10.1086/192340}, \href
  {https://ui.adsabs.harvard.edu/abs/1996ApJS..106..307V} {106, 307}

\bibitem[\protect\citeauthoryear{{Veilleux} \& {Osterbrock}}{{Veilleux} \&
  {Osterbrock}}{1987}]{veilleux87}
{Veilleux} S.,  {Osterbrock} D.~E.,  1987, \mn@doi [\apjs] {10.1086/191166},
  \href {http://adsabs.harvard.edu/abs/1987ApJS...63..295V} {63, 295}

\bibitem[\protect\citeauthoryear{{Veilleux}, {Shopbell}  \&
  {Miller}}{{Veilleux} et~al.}{2001}]{veil01}
{Veilleux} S.,  {Shopbell} P.~L.,   {Miller} S.~T.,  2001, \mn@doi [\aj]
  {10.1086/318046}, \href
  {https://ui.adsabs.harvard.edu/abs/2001AJ....121..198V} {121, 198}

\bibitem[\protect\citeauthoryear{{Vila-Costas} \& {Edmunds}}{{Vila-Costas} \&
  {Edmunds}}{1992}]{vila92}
{Vila-Costas} M.~B.,  {Edmunds} M.~G.,  1992, \mnras, \href
  {http://adsabs.harvard.edu/abs/1992MNRAS.259..121V} {259, 121}

\bibitem[\protect\citeauthoryear{{Vila-Costas} \& {Edmunds}}{{Vila-Costas} \&
  {Edmunds}}{1993}]{1993VilaCostas_MNRAS265}
{Vila-Costas} M.~B.,  {Edmunds} M.~G.,  1993, \mn@doi [\mnras]
  {10.1093/mnras/265.1.199}, \href
  {https://ui.adsabs.harvard.edu/abs/1993MNRAS.265..199V} {265, 199}

\bibitem[\protect\citeauthoryear{{Vincenzo}, {Belfiore}, {Maiolino},
  {Matteucci}  \& {Ventura}}{{Vincenzo} et~al.}{2016}]{2016Vincenzo_MNRAS458}
{Vincenzo} F.,  {Belfiore} F.,  {Maiolino} R.,  {Matteucci} F.,   {Ventura} P.,
   2016, \mn@doi [\mnras] {10.1093/mnras/stw532}, \href
  {https://ui.adsabs.harvard.edu/abs/2016MNRAS.458.3466V} {458, 3466}

\bibitem[\protect\citeauthoryear{{Vogt}, {P{\'e}rez}, {Dopita},
  {Verdes-Montenegro}  \& {Borthakur}}{{Vogt} et~al.}{2017}]{2017Vogt_AA601}
{Vogt} F.~P.~A.,  {P{\'e}rez} E.,  {Dopita} M.~A.,  {Verdes-Montenegro} L.,
  {Borthakur} S.,  2017, \mn@doi [\aap] {10.1051/0004-6361/201629853}, \href
  {https://ui.adsabs.harvard.edu/abs/2017A&A...601A..61V} {601, A61}

\bibitem[\protect\citeauthoryear{{Walcher} et~al.,}{{Walcher}
  et~al.}{2014}]{walcher14}
{Walcher} C.~J.,  et~al., 2014, \mn@doi [\aap] {10.1051/0004-6361/201424198},
  \href {http://adsabs.harvard.edu/abs/2014A%26A...569A...1W} {569, A1}

\bibitem[\protect\citeauthoryear{{Walcher} et~al.,}{{Walcher}
  et~al.}{2016}]{walcher16}
{Walcher} C.~J.,  et~al., 2016, \mn@doi [\aap] {10.1051/0004-6361/201528019},
  \href {https://ui.adsabs.harvard.edu/abs/2016A&A...594A..61W} {594, A61}

\bibitem[\protect\citeauthoryear{{Wegg}, {Gerhard}  \& {Portail}}{{Wegg}
  et~al.}{2017}]{wegg21}
{Wegg} C.,  {Gerhard} O.,   {Portail} M.,  2017, \mn@doi [\apjl]
  {10.3847/2041-8213/aa794e}, \href
  {https://ui.adsabs.harvard.edu/abs/2017ApJ...843L...5W} {843, L5}

\bibitem[\protect\citeauthoryear{{Weinberg}, {Andrews}  \&
  {Freudenburg}}{{Weinberg} et~al.}{2017}]{wein17}
{Weinberg} D.~H.,  {Andrews} B.~H.,   {Freudenburg} J.,  2017, \mn@doi [\apj]
  {10.3847/1538-4357/837/2/183}, \href
  {https://ui.adsabs.harvard.edu/abs/2017ApJ...837..183W} {837, 183}

\bibitem[\protect\citeauthoryear{{Yu} et~al.,}{{Yu}
  et~al.}{2021}]{2021Yu_ApJ912}
{Yu} Z.,  et~al., 2021, \mn@doi [\apj] {10.3847/1538-4357/abf098}, \href
  {https://ui.adsabs.harvard.edu/abs/2021ApJ...912..106Y} {912, 106}

\bibitem[\protect\citeauthoryear{{Zaritsky}, {Kennicutt}  \&
  {Huchra}}{{Zaritsky} et~al.}{1994}]{zaritsky94}
{Zaritsky} D.,  {Kennicutt} Jr. R.~C.,   {Huchra} J.~P.,  1994, \mn@doi [\apj]
  {10.1086/173544}, \href {http://adsabs.harvard.edu/abs/1994ApJ...420...87Z}
  {420, 87}

\bibitem[\protect\citeauthoryear{{Zinchenko}, {V{\'\i}lchez},
  {P{\'e}rez-Montero}, {Sukhorukov}, {Sobolenko}  \& {Duarte
  Puertas}}{{Zinchenko} et~al.}{2021}]{2021Zinchenko_AA655}
{Zinchenko} I.~A.,  {V{\'\i}lchez} J.~M.,  {P{\'e}rez-Montero} E.,
  {Sukhorukov} A.~V.,  {Sobolenko} M.,   {Duarte Puertas} S.,  2021, \mn@doi
  [\aap] {10.1051/0004-6361/202141522}, \href
  {https://ui.adsabs.harvard.edu/abs/2021A&A...655A..58Z} {655, A58}

\makeatother
\end{thebibliography}

\appendix

\begin{table*}
    \centering
    \begin{tabular}{ccc}
    \hline
    \hline
    Column & Keyword & Description \\
    \hline
    \MakeUppercase{column1} & \texttt{HIIREGID} & Ionized region ID \\
    \MakeUppercase{column2} & \texttt{f\textunderscore y} & Fraction of young stars to the total luminosity \\
    \MakeUppercase{column3} & \texttt{dist} & Galactocentric distance normalized by $R_e$\\
    \MakeUppercase{column4} & \texttt{Av} & Dust extinction \\
    \MakeUppercase{column5} & \texttt{OH\textunderscore author\textunderscore elines\textunderscore ftype 1} & Oxygen abundance  \\
    \MakeUppercase{column6} & \texttt{OH\textunderscore author\textunderscore elines\textunderscore ftype 2} & Oxygen abundance  \\
    $\vdots$ & $\vdots$ & $\vdots$ \\
    \MakeUppercase{column29} & \texttt{OH\textunderscore author\textunderscore elines\textunderscore ftype 26} & Oxygen abundance  \\
    \MakeUppercase{column30} & \texttt{OH\textunderscore IZI\textunderscore value\textunderscore model 1} & Oxygen abundance  \\
    $\vdots$ & $\vdots$ & $\vdots$ \\
    \MakeUppercase{column44} & \texttt{OH\textunderscore IZI\textunderscore value\textunderscore model 15} & Oxygen abundance  \\
    \MakeUppercase{column45} & \texttt{U\textunderscore author\textunderscore elines\textunderscore ftype 1} & Ionization Parameter (adimensional) \\
    $\vdots$ & $\vdots$ & $\vdots$ \\
    \MakeUppercase{column52} & \texttt{U\textunderscore author\textunderscore elines\textunderscore ftype 8} & Ionization Parameter (adimensional) \\
    \MakeUppercase{column53} & \texttt{q\textunderscore IZI\textunderscore value\textunderscore model 1} & Ionization Parameter \\
    $\vdots$ & $\vdots$ & $\vdots$ \\
    \MakeUppercase{column67} & \texttt{q\textunderscore IZI\textunderscore value\textunderscore model 15} & Ionization Parameter \\
    \MakeUppercase{column68} & \texttt{NH\textunderscore author\textunderscore ftype} & Nitrogen abundance\\
    \MakeUppercase{column69} & \texttt{NO\textunderscore HCm\textunderscore ftype 1} & Nitrogen-Oxygen ratio\\
    \MakeUppercase{column70} & \texttt{NO\textunderscore HCm\textunderscore ftype 2} & Nitrogen-Oxygen ratio\\
    \MakeUppercase{column71} & \texttt{Ne\textunderscore Oster\textunderscore S} & Electronic density ratio\\
    \MakeUppercase{column72} & \texttt{eAv} & Error of dust extinction \\
    \MakeUppercase{column73} & \texttt{eOH\textunderscore author\textunderscore elines\textunderscore ftype 1} & Error of oxygen abundance  \\
    $\vdots$ & $\vdots$ & $\vdots$ \\
    \MakeUppercase{column172} & \texttt{eNe\textunderscore Oster\textunderscore S} & Error of electronic density ratio\\
    \hline
    \end{tabular}
    \caption{Description of the distributed files comprising the derived physical properties for the \HII\ regions. Each ionized region is identified by their \texttt{HIIREGID}, therefore, these new tables can be used with the previous ones published by \citep{2020EspinosaPonce_MNRAS494}, that include only the observational properties the regions.}
    \label{tab:file_desc}
\end{table*}

\section{Description of the Catalogue}
\label{app:desc_catalog}

The catalog of physical properties of HII regions used along this study is publicly available. It is distributed through the same webpage as the original catalog of \HII\ regions published \cite{2020EspinosaPonce_MNRAS494}\footnote{\url{http://ifs.astroscu.unam.mx/CALIFA/HII_regions/}}, comprising a set of additional files. Each file, labeled by the name of the host galaxy (\texttt{HII.GALNAME.phys\textunderscore props.fits}), contains the full set of parameters derived for each region in a FITs-table format. Each row in the table corresponds to one \HII\ region, labeled with the same ID used in the original HII regions catalog (\texttt{HIREGID}): the galaxy name (\textsc{GALNAME}) plus the corresponding index of its segmentation map, \texttt{seg\textunderscore Ha.GALNAME.fits}. 

The general structure of each file is show in Table \ref{tab:file_desc}. We include the \texttt{HIREGID}, the fraction of young stars to the total luminosity, the galactocentric distance, and the dust extinction for each object. The next set of properties correspond to the oxygen abundances, ionization parameters and nitrogen/nitrogen-to-oxygen abundances derived using several calibrators described in the literature. The keyword of each one includes a description of the considered physical property (e.g., OH for O/H) and a reference to the article from which it was taken. In most keywords is included the lines ratios used for the respective calibrations and, if it were the case, the type of calibration (i.e. linear or polynomial fitting).

As indicated before, we include several values of oxygen abundances calculated with different calibrators, all them listed in Table \ref{tab:OH_calibrators}. We use the v5.1 of HCm code, the v0.9.9 of NebulaBayes code, and the python version of IZI code. 

For the ionization parameter, $U$, we include the results from several studies too: \cite{dors11}, \citet[Mor16]{2016Morisset_aa594A}, \citet[NB]{2018Thomas_ApJ856} and \citet[HCm]{2014PerezMontero_MNRAS441}. The nitrogen abundance was calculated with the calibration described by \citet[Pil16]{2016Pilyugin_MNRAS457}. In the later case N/O relative abundance was not included, however, it could be calculated with the oxygen and nitrogen abundances described above. In addition, we include the N/O ratio estimated by the \textsc{HII-CHI-mistry} code \citep{2014PerezMontero_MNRAS441}. The electron density described in the Section \ref{sec:physical_prop} is included in this table too. Finally, we include the complete output of IZI code \citep{2020Mingozzi_AA636} for both the oxygen abundance and ionization parameter, $q$. In order to run the IZI code, we use the photionization models set presented in \citet[labeled as d13\textunderscore k\textunderscore 20 and d13\textunderscore k\textunderscore inf]{dopita13}, \citet[labeled as levesque]{2010Levesque_AJ139} and \citet[labeled as byler and byler\textunderscore CSFR]{2017Byler_ApJ840}. The description of each of those photoionization grids and their respective results derived by \textsc{IZI} are fully explored in \cite{2015Blanc_ApJ798} and \cite{2020Mingozzi_AA636}.

\section{Physical properties on other diagnostic diagrams}
\label{app:diagnostic_diagrams}

\begin{figure}
\includegraphics[width=\columnwidth]{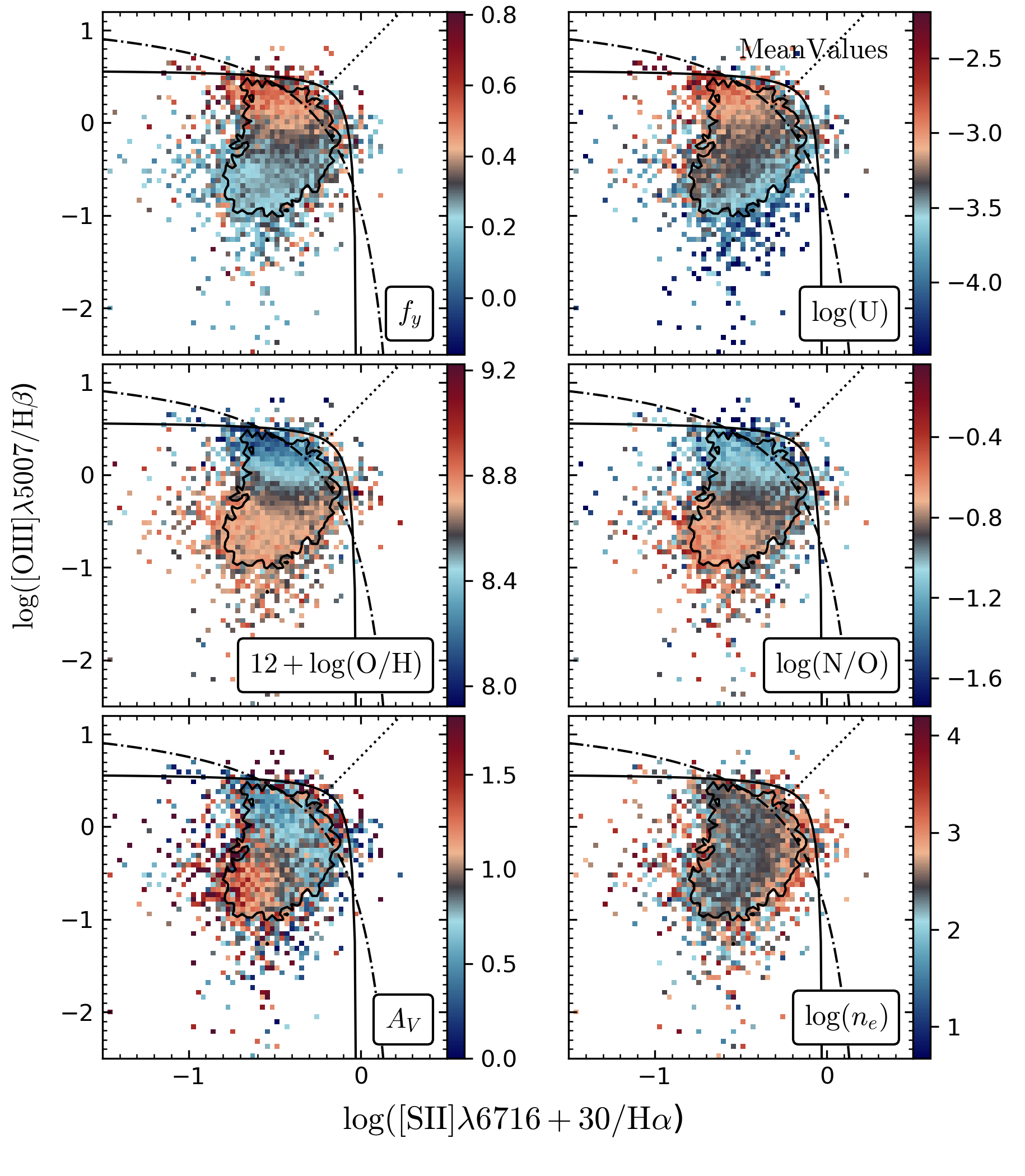}
\caption{Physical properties of the \HII\ regions explored along this study across the [OIII]$\lambda5007/$H$\beta$ vs. [SII]$\lambda6716+30/$H$\alpha$ diagnostic diagram. In each panel the color code represents one of the physical properties described in the text: \textit{top left panel}: fraction of young stars to the total luminosity, $f_{y}$; \textit{top right panel}: ionization parameter; \textit{middle left panel}: oxygen abundance; \textit{middle right panel}: nitrogen-to-oxygen abundance; \textit{bottom left panel}: dust extinction; and \textit{bottom right panel} electron density. The density contours are similar to those shown  \cref{fig:BPT_O3N2_physical_properties}. The most frequently used demarcation curves \citep{kewley06} for this diagram are shown. In each panel the solid-line corresponds to the demarcation line proposed by \citet{2020EspinosaPonce_MNRAS494}, the dashed-dotted line to the AGN demarcation line and the dotted-line to the LINER/Sy2 demarcation line. }
\label{fig:BPT_O3S2_physical_properties}
\end{figure}
\begin{figure}
\includegraphics[width=\columnwidth]{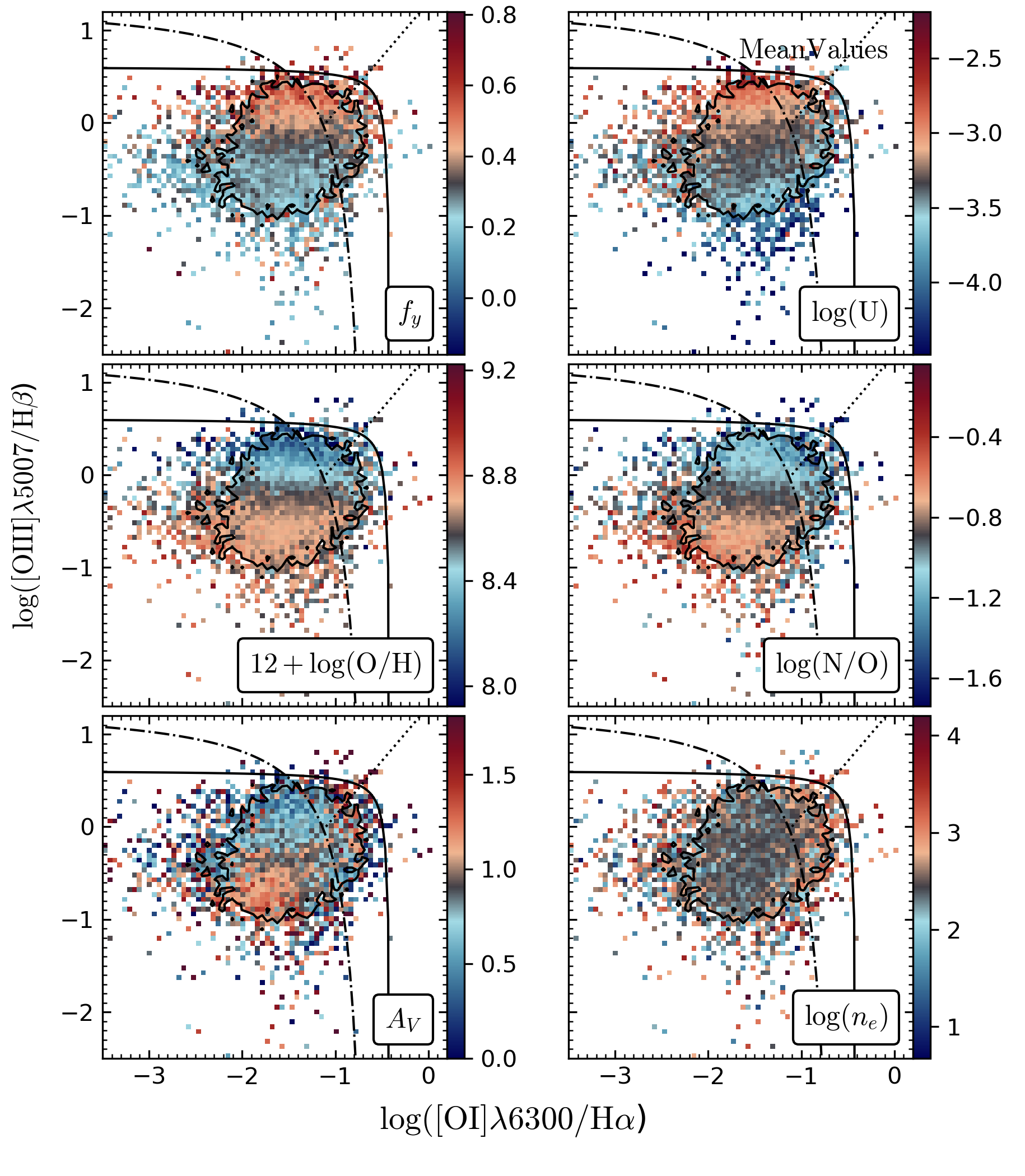}
\caption{Similar plots as those shown in \cref{fig:BPT_O3S2_physical_properties} for the [OIII]$\lambda5007/$H$\beta$ vs. [OI]$\lambda6300/$H$\alpha$ diagnostic diagram, using the same representation for the parameters, similar color-codes, contours and meaning for the plotted lines.}
\label{fig:BPT_O3O1_physical_properties}
\end{figure}

\begin{figure}
\includegraphics[width=\columnwidth]{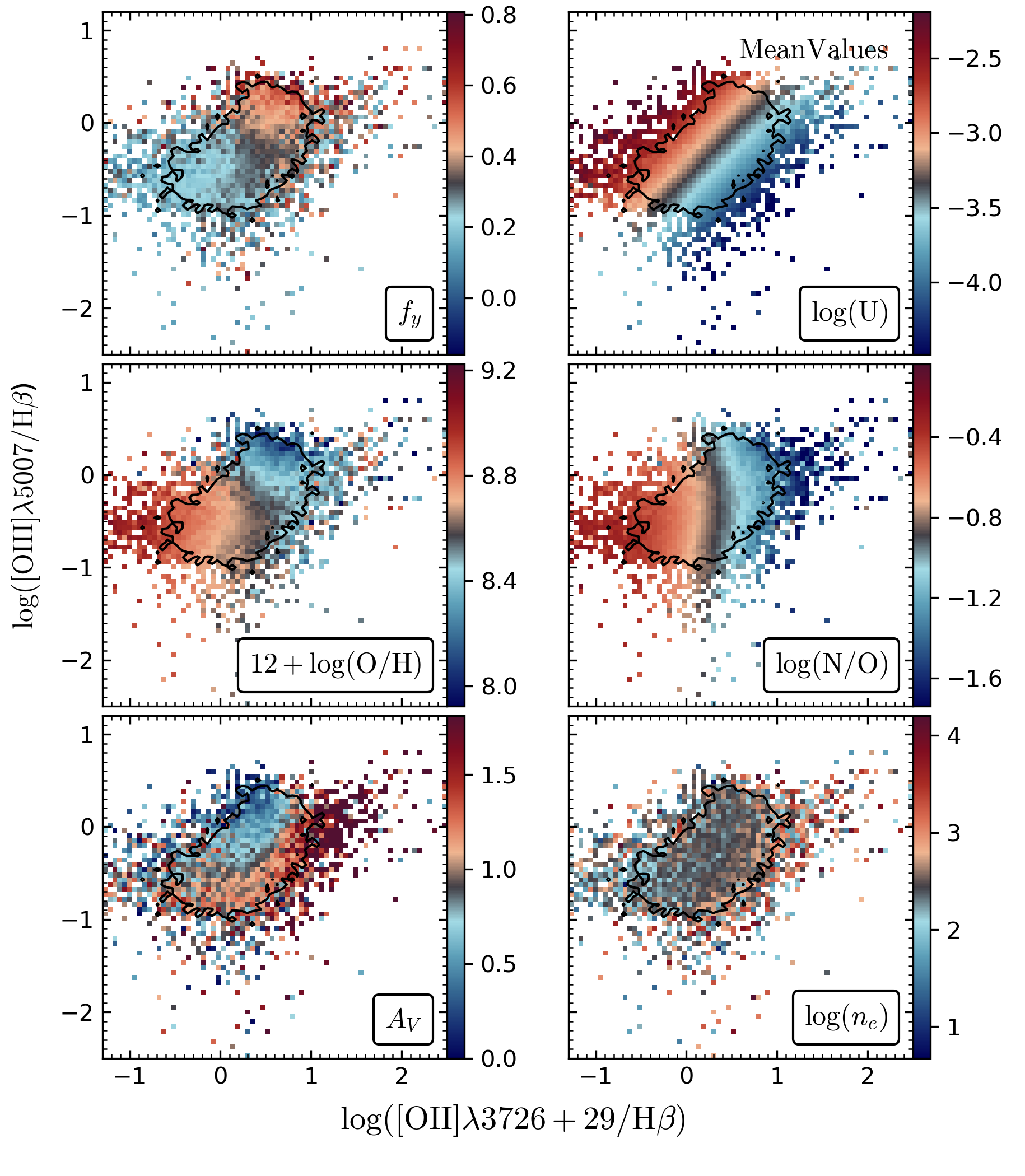}
\caption{Similar plots as those shown in \cref{fig:BPT_O3S2_physical_properties} for the [OIII]$\lambda5007/$H$\beta$ vs.  [OII]$\lambda3727/$H$\beta$ diagnostic diagram, using the same representation for the parameters, similar color-codes, contours and meaning for the plotted lines.}
\label{fig:BPT_O3O2_physical_properties}
\end{figure}


As we show in \cref{sec:trends}, the physical properties described in \cref{sec:physical_prop} present evident trends across the BPT diagram. In this appendix, we include the distributions those properties in other most frequently used diagnostic diagrams: \cref{fig:BPT_O3S2_physical_properties} and \cref{fig:BPT_O3O1_physical_properties} show the distribution across the [OIII]$\lambda5007/$H$\beta$ vs. [SII]$\lambda6716+30/$H$\alpha$ and [OIII]$\lambda5007/$H$\beta$ vs. [OI]$\lambda6300/$H$\alpha$ diagrams, respectively. In addition, we include the [\ion{O}{II}]$\lambda 3726+29$/H$\beta$ vs. [\ion{O}{III}]$\lambda 5007$/H$\beta$ diagnostic diagram in \cref{fig:BPT_O3O2_physical_properties}. 
Some trends are clearly highlighted in the observed distributions: (i) \HII\ regions with high $f_y$ values are always found in the upper range of the three diagrams, with a decreasing gradient along the Y-axis for the first two diagrams, and a diagonal gradient for the third one (i.e., the larger values are at the upper-right end of this diagram); (ii) regions with higher values of log(U) are also located in the upper range of the diagrams, with a diagonal distribution from the upper-left (high values) to the lower-right (low values) in the first two diagrams, and an inverse diagonal distribution, from the upper-right (high values) to the lower-left (low values) for the third diagram; (iii) O/H and N/O present similar trends, as expected due to the strong correlation between both parameters. In the first two diagrams the gradient is vertical, with both abundances increasing from the upper to the lower values, which in the third diagram the gradient is diagonal (or even horizontal), with abundances declining from the lower-left to the upper-right distribution of this diagrams; (iv) the trends for  $A_{\rm V}$ and $n_e$ are the less evident ones. For the dust extinction there is a clearer diagonal trend in the third diagram, with lower values at the upper-left regime and higher values at the lower-right locations. A similar trend, weaker, but still appreciated, is observed for the $n_e$ in this diagram too. For the other two diagrams trends are much weaker or absent. For $A_{\rm V}$ it is observed a loose trend to lower values in the upper range of the distribution. Finally, no trend is observed for $n_e$.

\section{Physical properties of edge-on galaxies}
\label{app:edge_on_diag}
In this section, we study how the galaxy inclination affects the trends described for the different physical properties on the diagnostic diagrams. For doing so we select those \HII\ regions in edge-on galaxies, i.e., with an inclination $> 70^{\circ}$. There are $\sim$100 of these galaxies in which we detect $3,600$ \HII\ regions.

\begin{figure}
\includegraphics[width=\columnwidth]{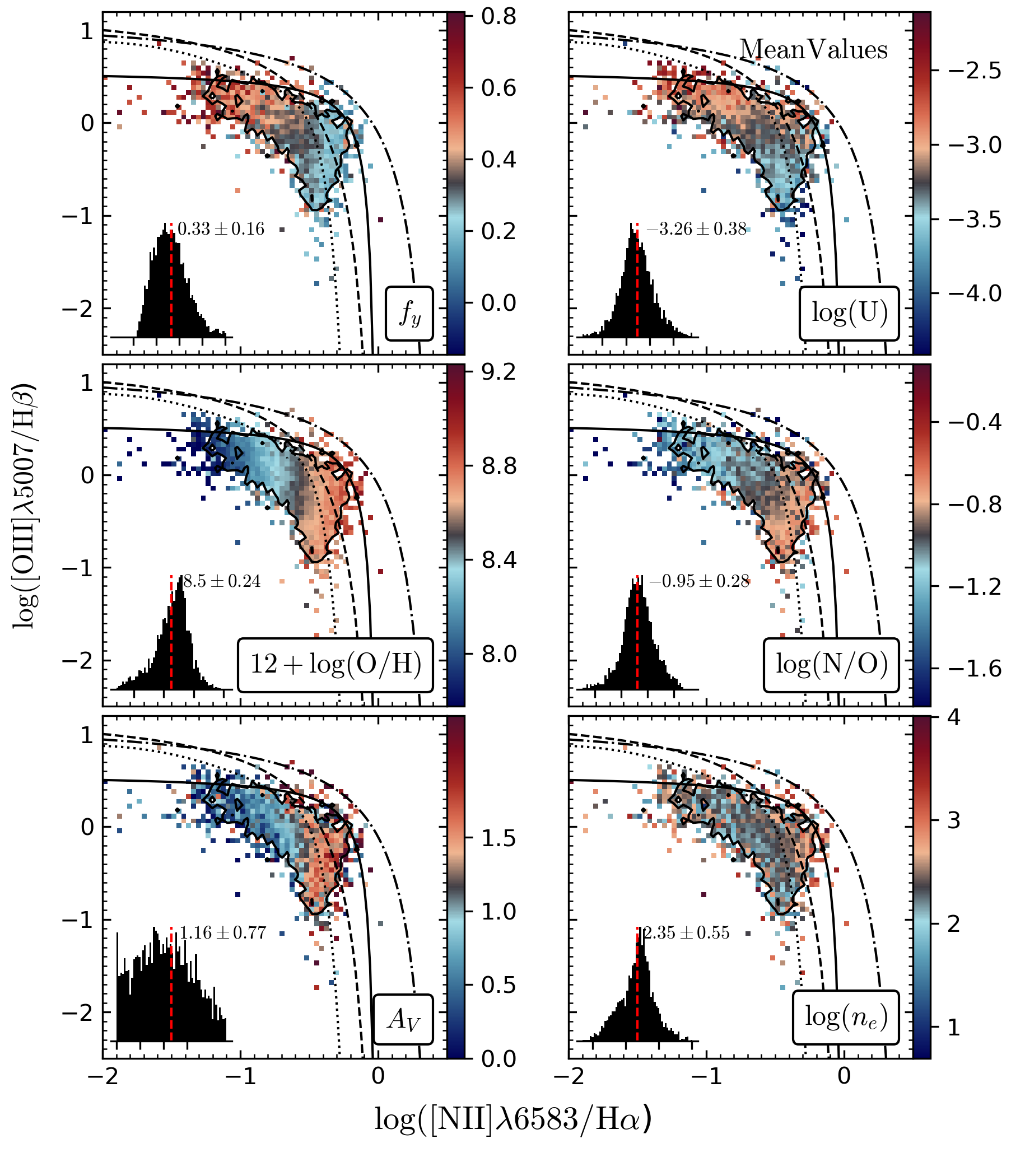}
\caption{Similar plots as those shown in \cref{fig:BPT_O3N2_physical_properties} for the [OIII]$\lambda5007/$H$\beta$ vs. [\ion{N}{ii}]$\lambda6583/$H$\alpha$ diagnostic diagram for the \HII\ regions hosted by edge-on galaxies. }
\label{fig:BPT_O3N2_edge_on_galaxies}
\end{figure}

As we can see in the Figure \ref{fig:BPT_O3N2_edge_on_galaxies}, we find similar trends for all physical properties as those ones found for the non edge-on galaxies, shown in Fig. \ref{fig:BPT_O3N2_physical_properties}. 
The \HII\ regions with high $f_y$ values are found in the upper-left end of the distribution, with values decreasing towards the lower-right zone of the diagram. The trends of log(U), O/H and N/O present the same trends along the Y-axis. The log(U) is higher in those regions on the upper-left of the distribution, and O/H and N/O present a diagonal trend with higher abundances on the upper-left of the distribution and lower abundances on the lower-right of the diagram. Finally, the previously trends for $A_\mathrm{V}$ and $n_e$ are found in this sub-sample of the \HII\ regions. There is a clear diagonal trend for the dust extinction, with higher values at the lower-right zone of the diagram and lower values at the upper-left of the distribution. A weaker trend is observed for the $n_e$, the regions with the highest density are found on the right-hand area of the diagram. The only appreciated difference is that the average dust attenuation is clearly higher for the \HII\ regions residing in edge-on galaxies ($\sim$1.2 mag vs. $\sim$0.9 mag). This is in any case expected just due to inclination effects. In summary the inclusion or exclusion of edge-on galaxies does not affect the main trends described across the diagnostic-diagrams.

\section{Physical Properties along the galactocentric distance: Fitting results}
\label{app:grads}

Tables \ref{tab:fy_coeffs}, \ref{tab:EWHa_coeffs}, \ref{tab:U_coeffs}, \ref{tab:OH_coeffs}, \ref{tab:NO_coeffs}, \ref{tab:ne_coeffs} and \ref{tab:Av_coeffs} show the zero-point ($c_0$) and slopes ($c_1$), and their associated errors, derived from the linear regression to the radial distributions shown in Fig. \ref{fig:DistanceDependenceA}, \ref{fig:DistanceDependenceB} and \ref{fig:DistanceDependenceC} and discussed in  \cref{sec:radialGrads}. In addition we provide with the $rms$ of the residual for each radial distribution once subtracted the best fitted linear relation.

\begin{table*}
  \centering
  \begin{tabular}{ccccccccc}
 $f_{y}$\ \ \ M$_*$ & Parameter & ALL & $\mathrm{E/S0}$ & $\mathrm{Sa}$ & $\mathrm{Sb}$ & $\mathrm{Sbc}$ & $\mathrm{Sc}$ & $\mathrm{Sd/Sm/Irr}$ \\
    \hline
     7.0 -  9.5 & $c_0$ & 0.324 & 0.284 & ... & ... & 0.182 & 0.218 & 0.354 \\
                & $\sigma_{c_0}$ & 0.009 & 0.142 & ... & ... & 0.037 & 0.014 & 0.011 \\
                & $c_1$ & 0.077 & -0.033 & ... & ... & 0.194 & 0.11 & 0.076 \\
                & $\sigma_{c_1}$ & 0.007 & 0.073 & ... & ... & 0.039 & 0.014 & 0.008 \\
                & $rms_r$ & 0.0353 & $\sim$ 0.0 & ... & ... & 0.0138 & 0.0673 & 0.0306 \\
    \hline
    9.5 - 10.5  & $c_0$ & 0.268 & 0.223 & 0.278 & 0.203 & 0.163 & 0.28 & 0.351 \\
                & $\sigma_{c_0}$ & 0.004 & 0.023 & 0.013 & 0.009 & 0.006 & 0.005 & 0.007 \\
                & $c_1$ & 0.049 & -0.022 & -0.035 & 0.038 & 0.066 & 0.068 & 0.051 \\
                & $\sigma_{c_1}$ & 0.003 & 0.023 & 0.008 & 0.007 & 0.004 & 0.004 & 0.004 \\
                & $rms_r$ & 0.003 & 0.0544 & 0.0057 & 0.0263 & 0.0203 & 0.0151 & 0.0186 \\
    \hline
    10.5 - 11.0 & $c_0$ & 0.243 & 0.32 & 0.21 & 0.21 & 0.182 & 0.289 & 0.353 \\
                & $\sigma_{c_0}$ & 0.004 & 0.019 & 0.011 & 0.008 & 0.008 & 0.007 & 0.019 \\
                & $c_1$ & 0.023 & -0.059 & -0.01 & 0.027 & 0.068 & 0.009 & 0.0 \\
                & $\sigma_{c_1}$ & 0.003 & 0.011 & 0.007 & 0.005 & 0.005 & 0.005 & 0.012 \\
                &$rms_r$ & 0.0105 & 0.0275 & 0.0151 & 0.0197 & 0.0192 & 0.0179 & 0.0473 \\
    \hline
    11.0 - 13.0 & $c_0$ & 0.167 & 0.056 & 0.169 & 0.136 & 0.219 & 0.081 & 0.108 \\
                & $\sigma_{c_0}$ & 0.006 & 0.018 & 0.014 & 0.008 & 0.011 & 0.015 & 0.023 \\
                & $c_1$ & 0.037 & 0.039 & -0.002 & 0.056 & 0.04 & 0.13 & 0.159 \\
                & $\sigma_{c_1}$ & 0.004 & 0.012 & 0.007 & 0.005 & 0.007 & 0.012 & 0.019 \\
                & $rms_r$ & 0.0059 & 0.0279 & 0.0218 & 0.0033 & 0.0103 & 0.0334 & 0.0386 \\
    \hline
  \end{tabular}
  \caption{Coefficients of the linear fit ($c_0$ for the zero-point and $c_1$ for the slope) and their associated errors ($\sigma_i$) for the radial distributions of $f_y$ along the galactocentric distance shown in Fig. \ref{fig:DistanceDependenceA}. The \textsc{RMS} of the residuals is listed for each linear regression.}
  \label{tab:fy_coeffs}
\end{table*}

\begin{table*}
  \centering
  \begin{tabular}{ccccccccc}
$\log\mathrm{EW}(\mathrm{H}\alpha)$ M$_*$ &Parameter & ALL & $\mathrm{E/S0}$ & $\mathrm{Sa}$ & $\mathrm{Sb}$ & $\mathrm{Sbc}$ & $\mathrm{Sc}$ & $\mathrm{Sd/Sm/Irr}$ \\
\hline
 7.0 -  9.5 &  $c_0$ &  1.466 & 1.421 & ... & ... & 1.283 & 1.286 & 1.496\\
 &  $\sigma_{c_0}$ &  0.02 & 0.244 & ... & ... & 0.079 & 0.045 & 0.023\\
  &  $c_1$ &  0.086 & -0.095 & ... & ... & 0.269 & 0.169 & 0.084\\
 &  $\sigma_{c_1}$ &  0.014 & 0.121 & ... & ... & 0.073 & 0.032 & 0.016\\
  &  $rms_r$ &  0.0438 & 0.0 & ... & ... & 0.0282 & 0.0954 & 0.0306\\
\hline
 9.5 - 10.5 &  $c_0$ &  1.323 & 1.092 & 1.287 & 1.216 & 1.18 & 1.336 & 1.47\\
 &  $\sigma_{c_0}$ &  0.009 & 0.047 & 0.034 & 0.018 & 0.016 & 0.012 & 0.015\\
  &  $c_1$ &  0.111 & 0.099 & -0.055 & 0.08 & 0.128 & 0.151 & 0.081\\
 &  $\sigma_{c_1}$ &  0.006 & 0.048 & 0.021 & 0.013 & 0.01 & 0.009 & 0.01\\
  &  $rms_r$ &  0.0208 & 0.0633 & 0.0736 & 0.045 & 0.0283 & 0.0328 & 0.023\\
\hline
10.5 - 11.0 &  $c_0$ &  1.269 & 1.165 & 1.211 & 1.153 & 1.169 & 1.406 & 1.51\\
 &  $\sigma_{c_0}$ &  0.009 & 0.033 & 0.026 & 0.015 & 0.017 & 0.015 & 0.034\\
  &  $c_1$ &  0.049 & -0.055 & -0.022 & 0.09 & 0.118 & 0.013 & -0.063\\
 &  $\sigma_{c_1}$ &  0.005 & 0.022 & 0.015 & 0.009 & 0.01 & 0.01 & 0.021\\
  &  $rms_r$ &  0.0117 & 0.076 & 0.0282 & 0.0105 & 0.0319 & 0.0345 & 0.0303\\
\hline
11.0 - 13.0 &  $c_0$ &  1.101 & 1.014 & 0.924 & 1.051 & 1.156 & 1.127 & 1.163\\
 &  $\sigma_{c_0}$ &  0.013 & 0.057 & 0.033 & 0.018 & 0.023 & 0.033 & 0.073\\
  &  $c_1$ &  0.069 & 0.051 & 0.097 & 0.091 & 0.083 & 0.134 & 0.179\\
 &  $\sigma_{c_1}$ &  0.007 & 0.032 & 0.018 & 0.01 & 0.013 & 0.022 & 0.055\\
  &  $rms_r$ &  0.0082 & 0.1037 & 0.0475 & 0.0141 & 0.0302 & 0.0453 & 0.0537\\
\hline
\end{tabular}
  \caption{Coefficients of the linear fit ($c_0$ for the zero-point and $c_1$ for the slope) and their associated errors ($\sigma_i$) for the radial distributions of EW(H$\alpha$) along the galactocentric distance shown in Fig. \ref{fig:DistanceDependenceA}. The \textsc{RMS} of the residuals is listed for each linear regression.}
\label{tab:EWHa_coeffs}
\end{table*}

\begin{table*}
  \centering
  \begin{tabular}{ccccccccc}
$\log(\mathrm{U})$ M$_*$ &Parameter & ALL & $\mathrm{E/S0}$ & $\mathrm{Sa}$ & $\mathrm{Sb}$ & $\mathrm{Sbc}$ & $\mathrm{Sc}$ & $\mathrm{Sd/Sm/Irr}$ \\
\hline
 7.0 -  9.5 &  $c_0$ &  -2.36 & ... & ... & ... & ... & ... & -2.402\\
 &  $\sigma_{c_0}$ &  0.077 & ... & ... & ... & ... & ... & 0.083\\
  &  $c_1$ &  0.006 & ... & ... & ... & ... & ... & 0.032\\
 &  $\sigma_{c_1}$ &  0.066 & ... & ... & ... & ... & ... & 0.073\\
  &  $rms_r$ &  0.0863 & ... & ... & ... & ... & ... & 0.1019\\
\hline
 9.5 - 10.5 &  $c_0$ &  -2.813 & -2.767 & -3.101 & -2.902 & -2.606 & -2.877 & -2.868\\
 &  $\sigma_{c_0}$ &  0.018 & 0.163 & 0.07 & 0.04 & 0.051 & 0.025 & 0.02\\
  &  $c_1$ &  0.035 & 0.342 & 0.046 & 0.08 & -0.031 & 0.051 & 0.105\\
 &  $\sigma_{c_1}$ &  0.012 & 0.325 & 0.091 & 0.032 & 0.029 & 0.016 & 0.015\\
  &  $rms_r$ &  0.0188 & 0.0 & 0.1196 & 0.1547 & 0.0729 & 0.0291 & 0.0178\\
\hline
10.5 - 11.0 &  $c_0$ &  -2.805 & -3.033 & -2.843 & -2.772 & -2.782 & -2.793 & -2.813\\
 &  $\sigma_{c_0}$ &  0.018 & 0.069 & 0.047 & 0.034 & 0.039 & 0.033 & 0.048\\
  &  $c_1$ &  0.0 & 0.015 & 0.024 & 0.01 & 0.028 & -0.087 & -0.096\\
 &  $\sigma_{c_1}$ &  0.011 & 0.073 & 0.036 & 0.021 & 0.021 & 0.021 & 0.036\\
  &  $rms_r$ &  0.0473 & 0.2942 & 0.0858 & 0.0465 & 0.031 & 0.0703 & 0.098\\
\hline
11.0 - 13.0 &  $c_0$ &  -2.919 & -3.035 & -3.04 & -2.596 & -3.122 & -3.175 & -3.515\\
 &  $\sigma_{c_0}$ &  0.038 & 0.179 & 0.11 & 0.059 & 0.055 & 0.118 & 0.229\\
  &  $c_1$ &  -0.031 & -0.095 & 0.071 & -0.147 & 0.03 & -0.053 & 0.274\\
 &  $\sigma_{c_1}$ &  0.022 & 0.107 & 0.058 & 0.035 & 0.035 & 0.074 & 0.166\\
  &  $rms_r$ &  0.0436 & 0.2406 & 0.0286 & 0.0381 & 0.0353 & 0.2389 & 0.0283\\
\hline
\end{tabular}
  \caption{Coefficients of the linear fit ($c_0$ for the zero-point and $c_1$ for the slope) and their associated errors ($\sigma_i$) for the radial distributions of $\log(\mathrm{U})$ along the galactocentric distance shown in Fig. \ref{fig:DistanceDependenceB}. The \textsc{RMS} of the residuals is listed for each linear regression.}
\label{tab:U_coeffs}
\end{table*}

\begin{table*}
  \centering
  \begin{tabular}{ccccccccc}
$12 + \log(\mathrm{O/H})$ M$_*$ &Parameter & ALL & $\mathrm{E/S0}$ & $\mathrm{Sa}$ & $\mathrm{Sb}$ & $\mathrm{Sbc}$ & $\mathrm{Sc}$ & $\mathrm{Sd/Sm/Irr}$ \\
\hline
 7.0 -  9.5 &  $c_0$ &  8.175 & ... & ... & ... & ... & ... & 8.186\\
 &  $\sigma_{c_0}$ &  0.028 & ... & ... & ... & ... & ... & 0.035\\
  &  $c_1$ &  -0.094 & ... & ... & ... & ... & ... & -0.099\\
 &  $\sigma_{c_1}$ &  0.03 & ... & ... & ... & ... & ... & 0.037\\
  &  $rms_r$ &  0.0105 & ... & ... & ... & ... & ... & 0.0143\\
\hline
 9.5 - 10.5 &  $c_0$ &  8.671 & 8.75 & 8.707 & 8.669 & 8.85 & 8.653 & 8.559\\
 &  $\sigma_{c_0}$ &  0.007 & 0.066 & 0.028 & 0.013 & 0.016 & 0.01 & 0.01\\
  &  $c_1$ &  -0.105 & -0.062 & -0.011 & -0.049 & -0.144 & -0.119 & -0.136\\
 &  $\sigma_{c_1}$ &  0.005 & 0.129 & 0.029 & 0.01 & 0.01 & 0.007 & 0.007\\
  &  $rms_r$ &  0.0062 & 0.0 & 0.0382 & 0.0361 & 0.0288 & 0.0139 & 0.0156\\
\hline
10.5 - 11.0 &  $c_0$ &  8.73 & 8.658 & 8.749 & 8.778 & 8.826 & 8.631 & 8.635\\
 &  $\sigma_{c_0}$ &  0.007 & 0.027 & 0.017 & 0.01 & 0.013 & 0.013 & 0.028\\
  &  $c_1$ &  -0.026 & 0.029 & 0.021 & -0.04 & -0.085 & 0.007 & -0.011\\
 &  $\sigma_{c_1}$ &  0.004 & 0.027 & 0.011 & 0.006 & 0.008 & 0.008 & 0.02\\
  &  $rms_r$ &  0.0141 & 0.0307 & 0.0216 & 0.0292 & 0.0282 & 0.0297 & 0.0324\\
\hline
11.0 - 13.0 &  $c_0$ &  8.755 & 8.727 & 8.818 & 8.824 & 8.633 & 8.748 & 8.844\\
 &  $\sigma_{c_0}$ &  0.01 & 0.081 & 0.031 & 0.013 & 0.018 & 0.023 & 0.049\\
  &  $c_1$ &  -0.023 & 0.029 & -0.022 & -0.047 & 0.001 & -0.068 & -0.139\\
 &  $\sigma_{c_1}$ &  0.006 & 0.055 & 0.015 & 0.008 & 0.011 & 0.014 & 0.037\\
  &  $rms_r$ &  0.0097 & 0.0434 & 0.0202 & 0.0044 & 0.0233 & 0.0354 & 0.0191\\
\hline
\end{tabular}
  \caption{Coefficients of the linear fit ($c_0$ for the zero-point and $c_1$ for the slope) and their associated errors ($\sigma_i$) for the radial distributions of 12+$\log(\mathrm{O}/\mathrm{H})$ along the galactocentric distance shown in Fig. \ref{fig:DistanceDependenceB}. The \textsc{RMS} of the residuals is listed for each linear regression.}
  \label{tab:OH_coeffs}
\end{table*}

\begin{table*}
  \centering
  \begin{tabular}{ccccccccc}
$\log(\mathrm{N/O})$ M$_*$ &Parameter & ALL & $\mathrm{E/S0}$ & $\mathrm{Sa}$ & $\mathrm{Sb}$ & $\mathrm{Sbc}$ & $\mathrm{Sc}$ & $\mathrm{Sd/Sm/Irr}$ \\
\hline
 7.0 -  9.5 &  $c_0$ &  -1.102 & ... & ... & ... & ... & ... & -1.105\\
 &  $\sigma_{c_0}$ &  0.056 & ... & ... & ... & ... & ... & 0.071\\
  &  $c_1$ &  -0.119 & ... & ... & ... & ... & ... & -0.117\\
 &  $\sigma_{c_1}$ &  0.051 & ... & ... & ... & ... & ... & 0.062\\
  &  $rms_r$ &  0.1606 & ... & ... & ... & ... & ... & 0.1893\\
\hline
 9.5 - 10.5 &  $c_0$ &  -0.755 & -0.85 & -0.861 & -0.764 & -0.5 & -0.776 & -0.848\\
 &  $\sigma_{c_0}$ &  0.009 & 0.048 & 0.034 & 0.019 & 0.021 & 0.012 & 0.012\\
  &  $c_1$ &  -0.134 & 0.051 & -0.002 & -0.066 & -0.211 & -0.145 & -0.156\\
 &  $\sigma_{c_1}$ &  0.007 & 0.123 & 0.04 & 0.017 & 0.015 & 0.009 & 0.01\\
  &  $rms_r$ &  0.0074 & 0.0 & 0.0566 & 0.1086 & 0.0292 & 0.0301 & 0.0356\\
\hline
10.5 - 11.0 &  $c_0$ &  -0.725 & -0.852 & -0.797 & -0.667 & -0.527 & -0.793 & -0.858\\
 &  $\sigma_{c_0}$ &  0.011 & 0.027 & 0.025 & 0.016 & 0.02 & 0.017 & 0.038\\
  &  $c_1$ &  -0.068 & -0.022 & 0.039 & -0.066 & -0.162 & -0.086 & -0.132\\
 &  $\sigma_{c_1}$ &  0.007 & 0.039 & 0.019 & 0.011 & 0.013 & 0.013 & 0.03\\
  &  $rms_r$ &  0.0558 & 0.213 & 0.0472 & 0.0637 & 0.0837 & 0.0423 & 0.022\\
\hline
11.0 - 13.0 &  $c_0$ &  -0.701 & -0.544 & -0.88 & -0.544 & -0.824 & -0.643 & -0.802\\
 &  $\sigma_{c_0}$ &  0.02 & 0.157 & 0.067 & 0.025 & 0.025 & 0.056 & 0.125\\
  &  $c_1$ &  -0.081 & -0.308 & 0.053 & -0.113 & -0.084 & -0.181 & -0.052\\
 &  $\sigma_{c_1}$ &  0.012 & 0.113 & 0.031 & 0.016 & 0.019 & 0.034 & 0.084\\
  &  $rms_r$ &  0.0129 & 0.08 & 0.0482 & 0.0563 & 0.0375 & 0.1048 & 0.0314\\
\hline
\end{tabular}
  \caption{Coefficients of the linear fit ($c_0$ for the zero-point and $c_1$ for the slope) and their associated errors ($\sigma_i$) for the radial distributions of N/O ratio along the galactocentric distance shown in Fig. \ref{fig:DistanceDependenceB}. The \textsc{RMS} of the residuals is listed for each linear regression.}
    \label{tab:NO_coeffs}
\end{table*}

\begin{table*}
  \centering
  \begin{tabular}{ccccccccc}
$\log(n_e)$ M$_*$ &Parameter & ALL & $\mathrm{E/S0}$ & $\mathrm{Sa}$ & $\mathrm{Sb}$ & $\mathrm{Sbc}$ & $\mathrm{Sc}$ & $\mathrm{Sd/Sm/Irr}$ \\
\hline
 7.0 -  9.5 &  $c_0$ &  2.355 & 3.12 & ... & ... & 2.239 & 2.405 & 2.383\\
 &  $\sigma_{c_0}$ &  0.021 & 0.714 & ... & ... & 0.128 & 0.05 & 0.024\\
  &  $c_1$ &  0.034 & -0.152 & ... & ... & 0.001 & 0.07 & 0.004\\
 &  $\sigma_{c_1}$ &  0.017 & 0.347 & ... & ... & 0.114 & 0.036 & 0.019\\
  &  $rms_r$ &  0.049 & 0.0 & ... & ... & 0.0609 & 0.0431 & 0.041\\
\hline
 9.5 - 10.5 &  $c_0$ &  2.483 & 2.418 & 2.438 & 2.553 & 2.393 & 2.498 & 2.406\\
 &  $\sigma_{c_0}$ &  0.014 & 0.064 & 0.068 & 0.043 & 0.029 & 0.02 & 0.02\\
  &  $c_1$ &  -0.026 & 0.162 & 0.086 & -0.095 & -0.023 & -0.015 & 0.012\\
 &  $\sigma_{c_1}$ &  0.009 & 0.05 & 0.041 & 0.028 & 0.017 & 0.013 & 0.013\\
  &  $rms_r$ &  0.0349 & 0.1272 & 0.0196 & 0.1234 & 0.0301 & 0.0336 & 0.0356\\
\hline
10.5 - 11.0 &  $c_0$ &  2.523 & 2.511 & 2.603 & 2.574 & 2.392 & 2.468 & 2.748\\
 &  $\sigma_{c_0}$ &  0.019 & 0.107 & 0.069 & 0.036 & 0.037 & 0.033 & 0.053\\
  &  $c_1$ &  -0.01 & 0.077 & -0.044 & -0.047 & 0.016 & 0.06 & -0.092\\
 &  $\sigma_{c_1}$ &  0.012 & 0.079 & 0.041 & 0.022 & 0.021 & 0.021 & 0.037\\
  &  $rms_r$ &  0.0645 & 0.1595 & 0.0614 & 0.1008 & 0.0893 & 0.0647 & 0.0431\\
\hline
11.0 - 13.0 &  $c_0$ &  2.604 & 3.163 & 2.756 & 2.572 & 2.34 & 2.906 & 2.294\\
 &  $\sigma_{c_0}$ &  0.034 & 0.175 & 0.082 & 0.05 & 0.064 & 0.112 & 0.221\\
  &  $c_1$ &  -0.072 & -0.222 & -0.107 & -0.052 & -0.025 & -0.206 & 0.37\\
 &  $\sigma_{c_1}$ &  0.019 & 0.095 & 0.042 & 0.03 & 0.038 & 0.071 & 0.173\\
  &  $rms_r$ &  0.0339 & 0.1537 & 0.1078 & 0.0979 & 0.0496 & 0.1494 & 0.0503\\
\hline
\end{tabular}
  \caption{Coefficients of the linear fit ($c_0$ for the zero-point and $c_1$ for the slope) and their associated errors ($\sigma_i$) for the radial distributions of $n_e$ along the galactocentric distance shown in Fig. \ref{fig:DistanceDependenceC}. The \textsc{RMS} of the residuals is listed for each linear regression.}
    \label{tab:ne_coeffs}
\end{table*}

\begin{table*}
  \centering
  \begin{tabular}{ccccccccc}
$\log(A_v)$ M$_*$ &Parameter & ALL & $\mathrm{E/S0}$ & $\mathrm{Sa}$ & $\mathrm{Sb}$ & $\mathrm{Sbc}$ & $\mathrm{Sc}$ & $\mathrm{Sd/Sm/Irr}$ \\
\hline
 7.0 -  9.5 &  $c_0$ &  0.42 & 1.478 & ... & ... & 0.166 & 1.032 & 0.317\\
 &  $\sigma_{c_0}$ &  0.041 & 1.527 & ... & ... & 0.192 & 0.141 & 0.035\\
  &  $c_1$ &  0.172 & -0.188 & ... & ... & 0.313 & -0.048 & 0.198\\
 &  $\sigma_{c_1}$ &  0.03 & 0.739 & ... & ... & 0.183 & 0.09 & 0.028\\
  &  $rms_r$ &  0.0564 & 0.0 & ... & ... & 0.2565 & 0.0925 & 0.0322\\
\hline
 9.5 - 10.5 &  $c_0$ &  1.003 & 0.892 & 1.476 & 0.864 & 1.135 & 1.058 & 0.546\\
 &  $\sigma_{c_0}$ &  0.022 & 0.154 & 0.094 & 0.05 & 0.05 & 0.035 & 0.022\\
  &  $c_1$ &  -0.1 & 0.249 & -0.212 & 0.043 & -0.152 & -0.139 & 0.037\\
 &  $\sigma_{c_1}$ &  0.014 & 0.186 & 0.055 & 0.037 & 0.029 & 0.02 & 0.016\\
  &  $rms_r$ &  0.0477 & 0.5836 & 0.1325 & 0.0979 & 0.0778 & 0.0863 & 0.027\\
\hline
10.5 - 11.0 &  $c_0$ &  1.476 & 1.374 & 1.616 & 1.295 & 1.27 & 1.591 & 1.059\\
 &  $\sigma_{c_0}$ &  0.029 & 0.152 & 0.086 & 0.044 & 0.043 & 0.051 & 0.125\\
  &  $c_1$ &  -0.17 & -0.079 & -0.216 & -0.119 & -0.177 & -0.092 & 0.076\\
 &  $\sigma_{c_1}$ &  0.017 & 0.091 & 0.053 & 0.027 & 0.026 & 0.033 & 0.083\\
  &  $rms_r$ &  0.0635 & 0.2236 & 0.1811 & 0.0464 & 0.1685 & 0.1195 & 0.2395\\
\hline
11.0 - 13.0 &  $c_0$ &  1.685 & 1.312 & 2.03 & 1.548 & 1.674 & 1.629 & 1.3\\
 &  $\sigma_{c_0}$ &  0.049 & 0.256 & 0.161 & 0.077 & 0.093 & 0.106 & 0.155\\
  &  $c_1$ &  -0.129 & 0.312 & -0.291 & -0.088 & -0.165 & -0.071 & -0.146\\
 &  $\sigma_{c_1}$ &  0.029 & 0.151 & 0.083 & 0.045 & 0.056 & 0.065 & 0.122\\
  &  $rms_r$ &  0.026 & 0.1908 & 0.0941 & 0.0601 & 0.0483 & 0.237 & 0.0971\\
\hline
\end{tabular}
  \caption{Coefficients of the linear fit ($c_0$ for the zero-point and $c_1$ for the slope) and their associated errors ($\sigma_i$) for the radial distributions of $A_v$ along the galactocentric distance shown in Fig. \ref{fig:DistanceDependenceC}. The \textsc{RMS} of the residuals is listed for each linear regression.}
    \label{tab:Av_coeffs}
\end{table*}

\begin{table}
  \centering
    \begin{tabular}{ccccccc}
    Parameter & $f_y$ & EW(H$\alpha$) & $\log(U)$ & $\log(\mathrm{N}/\mathrm{O})$ &  $A_v$ & $\log(n_e)$  \\
    \hline
    $c_0$          & 6.78   & 14.16 & 3.81  & -8.85 & -18.2 & -5.79 \\
    $\sigma_{c_0}$ & 1.8    & 3.49  & 2.58  & 1.46  & 5.8   & 3.71   \\
    $c_1$          & -0.76  & -1.48 & -0.83 & 0.93  & 2.21  & 0.98   \\
    $\sigma_{c_1}$ & 0.21   & 0.49  & 0.36  & 0.17  & 0.93  & 0.5    \\
    \end{tabular}
  \caption{Coefficients of the linear fit ($c_0$ for the zero-point and $c_1$ for the slope) and their associated errors ($\sigma_i$) for the distributions shown in Fig. \ref{fig:OHvsPP}, where each of the explored physical properties is fitted as a function of the oxygen abundance (i.e., param=$c_0$+$c_1$(12+log(O/H)).}
    \label{tab:OHvsPP_coeffs}
\end{table}

\begin{table}
  \centering
    \begin{tabular}{cccc}
    Parameter & $\log(\mathrm{Age/yr})$ & $[\mathrm{Z}/\mathrm{H}]$ & $[\mathrm{O}/\mathrm{Z}]$ \\
    \hline
    $c_0$          & -3.11 & -7.27 & -0.21 \\
    $\sigma_{c_0}$ & 1.71  & 2.63  & 0.1 \\
    $c_1$          & 1.4   & 0.82  & -1.22 \\
    $\sigma_{c_1}$ & 0.26  & 0.3   & 0.46 \\
    \end{tabular}
  \caption{Coefficients of the linear fit ($c_0$ for the zero-point and $c_1$ for the slope) and their associated errors ($\sigma_i$) for the distributions shown in Fig. \ref{fig:OHvsStellarProps}, where each of the LW ages, metallicities, and O/Z are fitted as a function of the oxygen abundance (i.e., param=$c_0$+$c_1$(12+log(O/H)).}
    \label{tab:OHvsStellarProps_coeffs}
\end{table}

\begin{table*}
    \centering
    \begin{tabular}{c|c|c|c}
    \hline
    \hline
    ID & Emission Lines & Calibration Type & Reference  \\
    \hline
    OH\textunderscore Kew02\textunderscore N2O2 &  [\ion{N}{II}]$\lambda 6583$/[\ion{O}{II}]$\lambda 3727$ & Theoretical & \cite{kewley02} \\
    OH\textunderscore T04 & [\ion{N}{II}]$\lambda 6548$/H$\alpha$, $R_{23}$ & Empirical & \cite{2004Tremonti_ApJ613} \\
    OH\textunderscore Pet04\textunderscore N2\textunderscore lin &  [\ion{N}{II}]$\lambda 6548$/H$\alpha$ & Empirical & \cite{2004Pettini_MNRAS348} \\
    OH\textunderscore Pet04\textunderscore N2\textunderscore poly &  [\ion{N}{II}]$\lambda 6548$/H$\alpha$ & Empirical & \cite{2004Pettini_MNRAS348} \\
    OH\textunderscore Pet04\textunderscore O3N2 & O3N2 & Empirical & \cite{2004Pettini_MNRAS348} \\
    OH\textunderscore KK04 & $R_{23}$, [\ion{O}{III}]/[\ion{O}{II}] & Theoretical & \cite{2004Kobulnicky_ApJ617} \\
    OH\textunderscore Pil10\textunderscore ONS &  [\ion{N}{II}]$\lambda 6583$/H$\beta$, $R_2$, $R_3$, $P$, \SII & Empirical & \cite{2010Pilyugin_ApJ720} \\
    OH\textunderscore Pil10\textunderscore ON &  [\ion{N}{II}]$\lambda 6583$/H$\beta$, $R_2$, $R_3$, \SII & Empirical & \cite{2010Pilyugin_ApJ720} \\
    OH\textunderscore Pil11\textunderscore NS &  [\ion{N}{II}]$\lambda 6583$/H$\beta$, $R_3$, \SII & Empirical & \cite{2011Pilyugin_MNRAS412} \\
    OH\textunderscore M13\textunderscore N2 &  [\ion{N}{II}]$\lambda 6548$/H$\alpha$ & Empirical & \cite{2013Marino_aa559A} \\
    OH\textunderscore M13\textunderscore O3N2 &  O3N2 & Empirical & \cite{2013Marino_aa559A} \\
    OH\textunderscore Pil16\textunderscore R &  [\ion{N}{II}]$\lambda 6583$/H$\beta$, $R_2$, $R_3$ & Empirical & \cite{2016Pilyugin_MNRAS457} \\
    OH\textunderscore Pil16\textunderscore S &  [\ion{N}{II}]$\lambda 6583$/H$\beta$, $R_3$, \SII & Empirical & \cite{2016Pilyugin_MNRAS457} \\
    OH\textunderscore Cur20\textunderscore R2 & $R_2$  & Empirical & \cite{2020Curti_MNRAS491} \\
    OH\textunderscore Cur20\textunderscore R3 &  $R_3$ & Empirical & \cite{2020Curti_MNRAS491} \\
    OH\textunderscore Cur20\textunderscore R23 &  $R_{23}$ & Empirical & \cite{2020Curti_MNRAS491} \\
    OH\textunderscore Cur20\textunderscore N2 & [\ion{N}{II}]$\lambda 6583$/H$\alpha$  & Empirical & \cite{2020Curti_MNRAS491} \\
    OH\textunderscore Cur20\textunderscore O3N2 & $R_3$, [\ion{N}{II}]$\lambda 6583$/H$\alpha$  & Empirical & \cite{2020Curti_MNRAS491} \\
    OH\textunderscore Cur20\textunderscore O3O2 & [\ion{O}{III}]$\lambda 5007$, [\ion{O}{II}]$\lambda 3727+29$ & Empirical & \cite{2020Curti_MNRAS491} \\
    OH\textunderscore Cur20\textunderscore S2 & \SII  & Empirical & \cite{2020Curti_MNRAS491} \\
    OH\textunderscore Cur20\textunderscore RS32 & $R_3$+\SII/H$\alpha$  & Empirical & \cite{2020Curti_MNRAS491} \\
    OH\textunderscore Cur20\textunderscore O3S2 & $R_3$/\SII/H$\alpha$  & Empirical & \cite{2020Curti_MNRAS491} \\
    OH\textunderscore Ho & $R_2$, $R_3$, [\ion{N}{II}]$\lambda 6583$/H$\beta$, \SII/H$\alpha$  & Empirical & \cite{2019Ho_MNRAS485} \\
    OH\textunderscore NB & \vtop{\hbox{\strut\OII, [\ion{O}{III}]$\lambda 5007$, [\ion{O}{I}]$\lambda 6100$,} \hbox{\strut[\ion{N}{II}]$\lambda 6583$, \SII, H$\alpha$, H$\beta$}} & Theoretical & \cite{2018Thomas_ApJ856} \\
    OH\textunderscore HCm\textunderscore(interp/no\textunderscore interp) & \vtop{\hbox{\strut [\ion{O}{II}]$\lambda 3727$, [\ion{Ne}{III}]$\lambda 3868$, [\ion{O}{III}]$\lambda 4363$,} \hbox{\strut[\ion{O}{III}]$\lambda 4959$, [\ion{O}{III}]$\lambda 5007$, [\ion{N}{II}]$\lambda 6584$, [\ion{S}{II}]$\lambda 6725$ }}& Theoretical & \cite{2014PerezMontero_MNRAS441}\\
    OH\textunderscore IZI\textunderscore value\textunderscore models &
    \vtop{\hbox{\strut [\ion{O}{II}]$\lambda 3727$, [\ion{O}{III}]$\lambda\lambda 4959+5007$,} \hbox{\strut[\ion{N}{II}]$\lambda\lambda 6583,48$,  \SII, H$\alpha$, H$\beta$}}& Theoretical &  \cite{2020Mingozzi_AA636} \\
    \hline
    \end{tabular}
    \caption{List of oxygen abundances calculated for the HII regions sample. }
    \label{tab:OH_calibrators}
\end{table*}

\section{Comparison between several oxygen abundance calibrators}
\label{appx5:compare_oxy}

\begin{figure*}
\centering
\includegraphics[width=\textwidth]{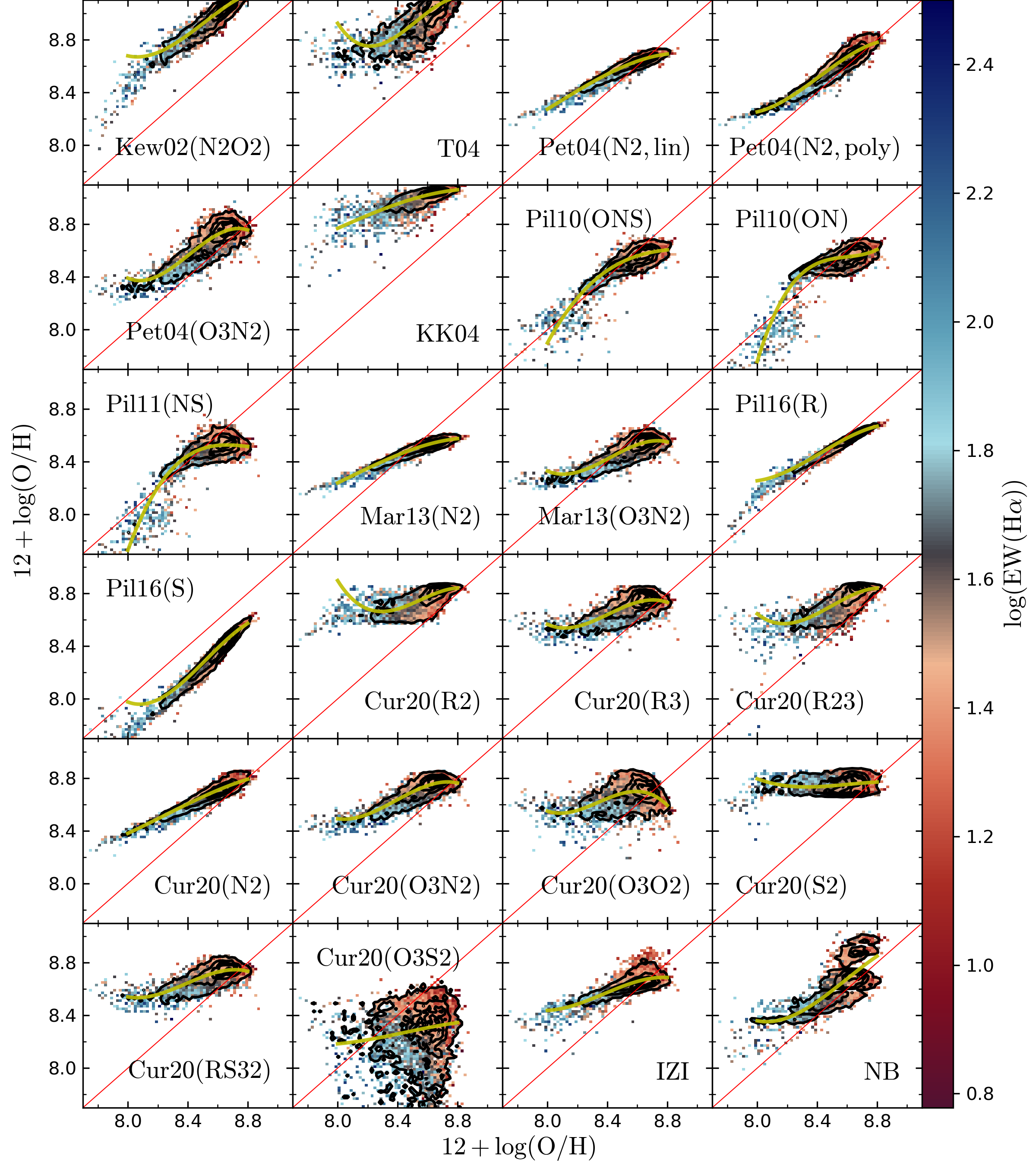}
\caption{Distribution of the oxygen abundance derived using each of the adopted calibrators and codes, one for each panel) as a function of the fiducial one \citet{2019Ho_MNRAS485} for our sample of \HII\ regions. We show the best 3rd order polynomial relation for each calibrator.
In each panel the color code represents the \EWHa\ with, with a scale ranging between $\pm$3$\sigma$ range around its mean value. The contours represent the density distribution, the outermost contour enclosing 95\% of the regions and each consecutive one enclosing 20\% fewer regions. The red-line represents the one-to-one relation and the yellow line corresponds to the fitted 3rd-order polynomial relation between the two parameters.}
\label{fig:OH_comparison}
\end{figure*}

Along this work, we adopt the calibration developed by \cite{2019Ho_MNRAS485} for the oxygen abundance as our fiducial one. However, we calculated the oxygen abundance with several calibrations proposed in the literature. The complete list of calibrators included in the final catalog are listed in  \cref{tab:OH_calibrators}. In this table, we indicate the identification code (ID) adopted to label each calibrator and the emission lines involved. Also, we include if the calibrator is based on observational data (empirical) or photoionization models grid (theoretical). Figure \ref{fig:OH_comparison} shows, for each calibrator, the distribution of oxygen abundances for each \HII\ region as a function of the fiducial one. It is appreciated that there are significant differences in the derived values for each adopted calibrator, although in most of the cases there is monotonic positive trend between the two values. Thus, although many of the results discussed along this article would change quantitatively depending on the adopted O/H calibrator, almost none of them would change from a qualitative point of view. In some cases the correspondance between both estimations of the abundance are well represented by a 1st order polynomial function (e.g., Pil16\textunderscore S), while in other cases a higher order polynomical function is required to transform one estimation to the other (e.g., Cur20(O3N2)). In order to provide with the broadest possible transformation among the different calibrators we fit each distribution using a 1st (linear), 2nd, and 3rd order polynomial function. The results of this analysis are listed in Tables \ref{tab:1st}, \ref{tab:2nd} and \ref{tab:3rd}. We include the values and errors of each coefficient of the adopted polynomial together with the standard deviations of the residuals.


%
\begin{table*}
    \centering
    \begin{tabular}{c|c|c|c|c|c|c|c|c|c|c|c|c|}
    \hline
    \hline
    ID & \multicolumn{2}{c|}{Kew\textunderscore 02\textunderscore N202} & \multicolumn{2}{c|}{T04} & \multicolumn{2}{c|}{Pet04\textunderscore N2\textunderscore lin} & \multicolumn{2}{c|}{Pet04\textunderscore N2\textunderscore poly} & \multicolumn{2}{c|}{Pet04\textunderscore O3N2} & \multicolumn{2}{c|}{KK04} \\
    \hline
$c_{0}$ & 2.203 & 0.194 & 3.085 & 0.385 & 4.6 & 0.254 & 2.574 & 0.16 & 3.678 & 0.319 & 6.432 & 0.155 \\
$c_{1}$ & 0.79 & 0.023 & 0.687 & 0.045 & 0.467 & 0.029 & 0.707 & 0.019 & 0.581 & 0.037 & 0.3 & 0.018 \\
res$_{\mathrm{std}}$ & \multicolumn{2}{c|}{0.052} & \multicolumn{2}{c|}{0.095} & \multicolumn{2}{c|}{0.029} & \multicolumn{2}{c|}{0.044} & \multicolumn{2}{c|}{0.076} & \multicolumn{2}{c|}{0.051} \\

\hline
\hline
    & \multicolumn{2}{c|}{Pil10\textunderscore ONS}& \multicolumn{2}{c|}{Pil10\textunderscore ON}& \multicolumn{2}{c|}{Pil11\textunderscore NS} & \multicolumn{2}{c|}{M13\textunderscore N2} & \multicolumn{2}{c|}{M13\textunderscore O3N2}& \multicolumn{2}{c|}{Pil16\textunderscore R} \\ 
    \hline
$c_{0}$ & 2.784 & 0.954 & 2.321 & 1.419 & 3.77 & 1.858 & 5.168 & 0.096 & 5.068 & 0.19 & 3.567 & 0.204 \\
$c_{1}$ & 0.668 & 0.111 & 0.723 & 0.165 & 0.548 & 0.215 & 0.389 & 0.011 & 0.399 & 0.022 & 0.581 & 0.024 \\
res$_{\mathrm{std}}$ & \multicolumn{2}{c|}{0.069} & \multicolumn{2}{c|}{0.092} & \multicolumn{2}{c|}{0.103} & \multicolumn{2}{c|}{0.023} & \multicolumn{2}{c|}{0.051} & \multicolumn{2}{c|}{0.028} \\
    \hline
    \hline
    & \multicolumn{2}{c|}{Pil16\textunderscore S} & \multicolumn{2}{c|}{Cur20\textunderscore R2} & \multicolumn{2}{c|}{Cur20\textunderscore R3} & \multicolumn{2}{c|}{Cur20\textunderscore R23} & \multicolumn{2}{c|}{Cur20\textunderscore N2} & \multicolumn{2}{c|}{Cur20\textunderscore O3N2} \\  
    \hline
$c_{0}$ & -0.478 & 0.123 & 5.741 & 0.348 & 5.996 & 0.128 & 4.908 & 0.218 & 4.382 & 0.131 & 5.279 & 0.177 \\
$c_{1}$ & 1.029 & 0.014 & 0.352 & 0.04 & 0.314 & 0.015 & 0.447 & 0.025 & 0.502 & 0.015 & 0.399 & 0.021 \\
res$_{\mathrm{std}}$ & \multicolumn{2}{c|}{0.039} & \multicolumn{2}{c|}{0.062} & \multicolumn{2}{c|}{0.062} & \multicolumn{2}{c|}{0.074} & \multicolumn{2}{c|}{0.031} & \multicolumn{2}{c|}{0.05} \\

    \hline
    \hline
    & \multicolumn{2}{c|}{Cur20\textunderscore O3O2} & \multicolumn{2}{c|}{Cur20\textunderscore S2} & \multicolumn{2}{c|}{Cur20\textunderscore RS32} & \multicolumn{2}{c|}{Cur20\textunderscore O3S2} &
    \multicolumn{2}{c|}{IZI} &
    \multicolumn{2}{c|}{NB} \\
    \hline
$c_{0}$ & 7.916 & 0.657 & 8.245 & 0.193 & 6.165 & 0.123 & 6.553 & 0.29 & 5.793 & 0.446 & 2.453 & 0.641 \\
$c_{1}$ & 0.084 & 0.077 & 0.059 & 0.023 & 0.295 & 0.014 & 0.203 & 0.034 & 0.331 & 0.052 & 0.724 & 0.075 \\
res$_{\mathrm{std}}$ & \multicolumn{2}{c|}{0.089} & \multicolumn{2}{c|}{0.048} & \multicolumn{2}{c|}{0.056} & \multicolumn{2}{c|}{0.208} & \multicolumn{2}{c|}{0.053} & \multicolumn{2}{c|}{0.108} \\    \hline
    \end{tabular}
    \caption{Coefficients of the best fitted linear relation between each different calibrator of the oxygen abundance (y) and the fiducial one \citep{2019Ho_MNRAS485} (x): i.e., y=$c_0$+$c_1$x. For each calibrator we include the values of the coefficients and an estimation their errors together with the standard deviation of the residuals (res$_{\rm std}$).}
    \label{tab:1st}
\end{table*}

%

\begin{table*}
    \centering
    \begin{tabular}{c|c|c|c|c|c|c|c|c|c|c|c|c|}
    \hline
    \hline
    ID & \multicolumn{2}{c|}{Kew\textunderscore 02\textunderscore N202} & \multicolumn{2}{c|}{T04} & \multicolumn{2}{c|}{Pet04\textunderscore N2\textunderscore lin} & \multicolumn{2}{c|}{Pet04\textunderscore N2\textunderscore poly} & \multicolumn{2}{c|}{Pet04\textunderscore O3N2} & \multicolumn{2}{c|}{KK04} \\
    \hline
$c_{0}$ & -12.074 & 7.935 & -5.993 & 24.386 & -27.595 & 6.793 & -16.071 & 16.597 & -53.803 & 24.601 & -10.024 & 6.401 \\
$c_{1}$ & 4.134 & 1.859 & 2.775 & 5.681 & 8.044 & 1.584 & 5.076 & 3.874 & 14.068 & 5.74 & 4.162 & 1.494 \\
$c_{2}$ & -0.196 & 0.109 & -0.12 & 0.331 & -0.445 & 0.092 & -0.256 & 0.226 & -0.791 & 0.335 & -0.227 & 0.087 \\
res$_{\mathrm{std}}$ & \multicolumn{2}{c|}{0.051} & \multicolumn{2}{c|}{0.098} & \multicolumn{2}{c|}{0.028} & \multicolumn{2}{c|}{0.047} & \multicolumn{2}{c|}{0.084} & \multicolumn{2}{c|}{0.051} \\
\hline
\hline
    & \multicolumn{2}{c|}{Pil10\textunderscore ONS}& \multicolumn{2}{c|}{Pil10\textunderscore ON}& \multicolumn{2}{c|}{Pil11\textunderscore NS} & \multicolumn{2}{c|}{M13\textunderscore N2} & \multicolumn{2}{c|}{M13\textunderscore O3N2}& \multicolumn{2}{c|}{Pil16\textunderscore R} \\ 
    \hline
$c_{0}$ & -70.712 & 18.553 & -107.157 & 46.237 & -121.66 & 26.435 & -18.786 & 5.615 & -30.287 & 17.76 & -21.729 & 7.248 \\
$c_{1}$ & 18.045 & 4.347 & 26.635 & 10.788 & 30.089 & 6.152 & 6.042 & 1.309 & 8.71 & 4.142 & 6.512 & 1.687 \\
$c_{2}$ & -1.026 & 0.255 & -1.532 & 0.629 & -1.738 & 0.358 & -0.333 & 0.076 & -0.488 & 0.241 & -0.348 & 0.098 \\
res$_{\mathrm{std}}$ & \multicolumn{2}{c|}{0.065} & \multicolumn{2}{c|}{0.072} & \multicolumn{2}{c|}{0.075} & \multicolumn{2}{c|}{0.023} & \multicolumn{2}{c|}{0.055} & \multicolumn{2}{c|}{0.022} \\
\hline
    \hline
    & \multicolumn{2}{c|}{Pil16\textunderscore S} & \multicolumn{2}{c|}{Cur20\textunderscore R2} & \multicolumn{2}{c|}{Cur20\textunderscore R3} & \multicolumn{2}{c|}{Cur20\textunderscore R23} & \multicolumn{2}{c|}{Cur20\textunderscore N2} & \multicolumn{2}{c|}{Cur20\textunderscore O3N2} \\  
    \hline
$c_{0}$ & -16.867 & 11.752 & 12.493 & 13.975 & -35.367 & 21.256 & 8.742 & 13.683 & -8.763 & 5.266 & -39.595 & 16.713 \\
$c_{1}$ & 4.853 & 2.735 & -1.241 & 3.253 & 10.028 & 4.953 & -0.479 & 3.192 & 3.595 & 1.235 & 10.929 & 3.894 \\
$c_{2}$ & -0.223 & 0.159 & 0.094 & 0.189 & -0.57 & 0.288 & 0.056 & 0.186 & -0.182 & 0.072 & -0.617 & 0.227 \\
res$_{\mathrm{std}}$ & \multicolumn{2}{c|}{0.045} & \multicolumn{2}{c|}{0.061} & \multicolumn{2}{c|}{0.068} & \multicolumn{2}{c|}{0.073} & \multicolumn{2}{c|}{0.031} & \multicolumn{2}{c|}{0.056} \\
\hline
    \hline
    & \multicolumn{2}{c|}{Cur20\textunderscore O3O2} & \multicolumn{2}{c|}{Cur20\textunderscore S2} & \multicolumn{2}{c|}{Cur20\textunderscore RS32} & \multicolumn{2}{c|}{Cur20\textunderscore O3S2} &
    \multicolumn{2}{c|}{IZI} &
    \multicolumn{2}{c|}{NB} \\
    \hline
$c_{0}$ & -78.438 & 24.358 & 17.234 & 7.022 & -31.068 & 19.236 & 3.48 & 14.839 & -17.5 & 14.588 & 47.51 & 10.233 \\
$c_{1}$ & 20.416 & 5.665 & -2.062 & 1.654 & 9.039 & 4.488 & 0.928 & 3.473 & 5.789 & 3.441 & -9.96 & 2.423 \\
$c_{2}$ & -1.196 & 0.329 & 0.125 & 0.097 & -0.513 & 0.262 & -0.043 & 0.203 & -0.319 & 0.203 & 0.633 & 0.143 \\
res$_{\mathrm{std}}$ & \multicolumn{2}{c|}{0.093} & \multicolumn{2}{c|}{0.049} & \multicolumn{2}{c|}{0.062} & \multicolumn{2}{c|}{0.208} & \multicolumn{2}{c|}{0.052} & \multicolumn{2}{c|}{0.106} \\
    \hline
    \end{tabular}
    \caption{Coefficients of the best 2nd order polynomial relation between each different calibrator of the oxygen abundance (y) and the fiducial one \citep{2019Ho_MNRAS485} (x): i.e., y=$c_0$+$c_1$x+$c_2$x$^2$. For each calibrator we include the values of the coefficients and an estimation their errors together with the standard deviation of the residuals (res$_{\rm std}$).}
    \label{tab:2nd}
\end{table*}
%
\begin{table*}
    \centering
    \begin{tabular}{c|c|c|c|c|c|c|c|c|c|c|c|c|}
    \hline
    \hline
    ID & \multicolumn{2}{c|}{Kew\textunderscore 02\textunderscore N202} & \multicolumn{2}{c|}{T04} & \multicolumn{2}{c|}{Pet04\textunderscore N2\textunderscore lin} & \multicolumn{2}{c|}{Pet04\textunderscore N2\textunderscore poly} & \multicolumn{2}{c|}{Pet04\textunderscore O3N2} & \multicolumn{2}{c|}{KK04} \\
    \hline
$c_{0}$ & 898.109 & 531.542 & 2007.277 & 623.214 & 172.453 & 148.794 & 666.123 & 164.08 & 1653.529 & 301.103 & 90.595 & 258.552 \\
$c_{1}$ & -315.241 & 186.699 & -705.19 & 218.547 & -62.649 & 52.166 & -236.103 & 57.772 & -588.617 & 106.048 & -32.027 & 90.371 \\
$c_{2}$ & 37.15 & 21.855 & 82.843 & 25.542 & 7.879 & 6.095 & 28.155 & 6.78 & 70.101 & 12.447 & 4.108 & 10.527 \\
$c_{3}$ & -1.455 & 0.853 & -3.24 & 0.995 & -0.327 & 0.237 & -1.115 & 0.265 & -2.779 & 0.487 & -0.173 & 0.409 \\
res$_{\mathrm{std}}$ & \multicolumn{2}{c|}{0.07} & \multicolumn{2}{c|}{0.102} & \multicolumn{2}{c|}{0.026} & \multicolumn{2}{c|}{0.042} & \multicolumn{2}{c|}{0.073} & \multicolumn{2}{c|}{0.051} \\
\hline
\hline
    & \multicolumn{2}{c|}{Pil10\textunderscore ONS}& \multicolumn{2}{c|}{Pil10\textunderscore ON}& \multicolumn{2}{c|}{Pil11\textunderscore NS} & \multicolumn{2}{c|}{M13\textunderscore N2} & \multicolumn{2}{c|}{M13\textunderscore O3N2}& \multicolumn{2}{c|}{Pil16\textunderscore R} \\ 
    \hline
$c_{0}$ & -501.536 & 740.109 & -2163.682 & 1747.281 & -1068.199 & 940.027 & 141.663 & 108.763 & 1220.416 & 268.648 & 569.051 & 171.187 \\
$c_{1}$ & 170.276 & 261.0 & 758.028 & 613.19 & 366.09 & 330.243 & -50.795 & 38.129 & -433.051 & 94.818 & -200.948 & 59.948 \\
$c_{2}$ & -18.95 & 30.671 & -88.189 & 71.714 & -41.476 & 38.663 & 6.375 & 4.455 & 51.505 & 11.153 & 23.93 & 6.996 \\
$c_{3}$ & 0.703 & 1.201 & 3.421 & 2.795 & 1.566 & 1.508 & -0.264 & 0.173 & -2.039 & 0.437 & -0.947 & 0.272 \\
res$_{\mathrm{std}}$ & \multicolumn{2}{c|}{0.071} & \multicolumn{2}{c|}{0.089} & \multicolumn{2}{c|}{0.085} & \multicolumn{2}{c|}{0.021} & \multicolumn{2}{c|}{0.05} & \multicolumn{2}{c|}{0.035} \\
    \hline
    \hline
    & \multicolumn{2}{c|}{Pil16\textunderscore S} & \multicolumn{2}{c|}{Cur20\textunderscore R2} & \multicolumn{2}{c|}{Cur20\textunderscore R3} & \multicolumn{2}{c|}{Cur20\textunderscore R23} & \multicolumn{2}{c|}{Cur20\textunderscore N2} & \multicolumn{2}{c|}{Cur20\textunderscore O3N2} \\  
    \hline
$c_{0}$ & 1199.71 & 498.673 & 1577.692 & 635.709 & 1240.384 & 239.532 & 1272.809 & 770.875 & 136.559 & 128.678 & 1009.504 & 164.634 \\
$c_{1}$ & -421.71 & 174.78 & -551.19 & 222.466 & -440.462 & 84.675 & -447.563 & 270.272 & -47.764 & 45.502 & -359.155 & 58.08 \\
$c_{2}$ & 49.62 & 20.416 & 64.487 & 25.945 & 52.437 & 9.976 & 52.746 & 31.578 & 5.866 & 5.363 & 42.886 & 6.828 \\
$c_{3}$ & -1.941 & 0.795 & -2.513 & 1.008 & -2.078 & 0.392 & -2.069 & 1.23 & -0.237 & 0.211 & -1.704 & 0.268 \\
res$_{\mathrm{std}}$ & \multicolumn{2}{c|}{0.05} & \multicolumn{2}{c|}{0.085} & \multicolumn{2}{c|}{0.059} & \multicolumn{2}{c|}{0.079} & \multicolumn{2}{c|}{0.031} & \multicolumn{2}{c|}{0.047} \\
    \hline
    \hline
    & \multicolumn{2}{c|}{Cur20\textunderscore O3O2} & \multicolumn{2}{c|}{Cur20\textunderscore S2} & \multicolumn{2}{c|}{Cur20\textunderscore RS32} & \multicolumn{2}{c|}{Cur20\textunderscore O3S2} &
    \multicolumn{2}{c|}{IZI} &
    \multicolumn{2}{c|}{NB} \\
    \hline
$c_{0}$ & 1478.095 & 164.607 & 315.066 & 356.235 & 900.092 & 91.215 & 141.086 & 919.541 & 589.36 & 865.24 & 681.624 & 828.078 \\
$c_{1}$ & -529.588 & 58.138 & -107.002 & 125.36 & -319.792 & 32.168 & -47.615 & 324.535 & -208.984 & 305.511 & -236.714 & 292.23 \\
$c_{2}$ & 63.561 & 6.844 & 12.447 & 14.703 & 38.181 & 3.781 & 5.661 & 38.175 & 25.009 & 35.953 & 27.648 & 34.372 \\
$c_{3}$ & -2.541 & 0.269 & -0.482 & 0.575 & -1.517 & 0.148 & -0.223 & 1.497 & -0.995 & 1.41 & -1.072 & 1.347 \\
res$_{\mathrm{std}}$ & \multicolumn{2}{c|}{0.08} & \multicolumn{2}{c|}{0.053} & \multicolumn{2}{c|}{0.053} & \multicolumn{2}{c|}{0.208} & \multicolumn{2}{c|}{0.052} & \multicolumn{2}{c|}{0.106} \\
    \hline
    \end{tabular}
    \caption{Coefficients of the best 3rd order polynomial relation between each different calibrator of the oxygen abundance (y) and the fiducial one \citep{2019Ho_MNRAS485} (x): i.e., y=$c_0$+$c_1$x+$c_2$x$^2$+$c_3$x$^3$. For each calibrator we include the values of the coefficients and an estimation their errors together with the standard deviation of the residuals (res$_{\rm std}$).}
    \label{tab:3rd}
\end{table*}

\section{Oxygen abundance vs ionization parameter}\label{appx6:OHvsUs}

\begin{figure*}
\centering
\includegraphics[width=\textwidth]{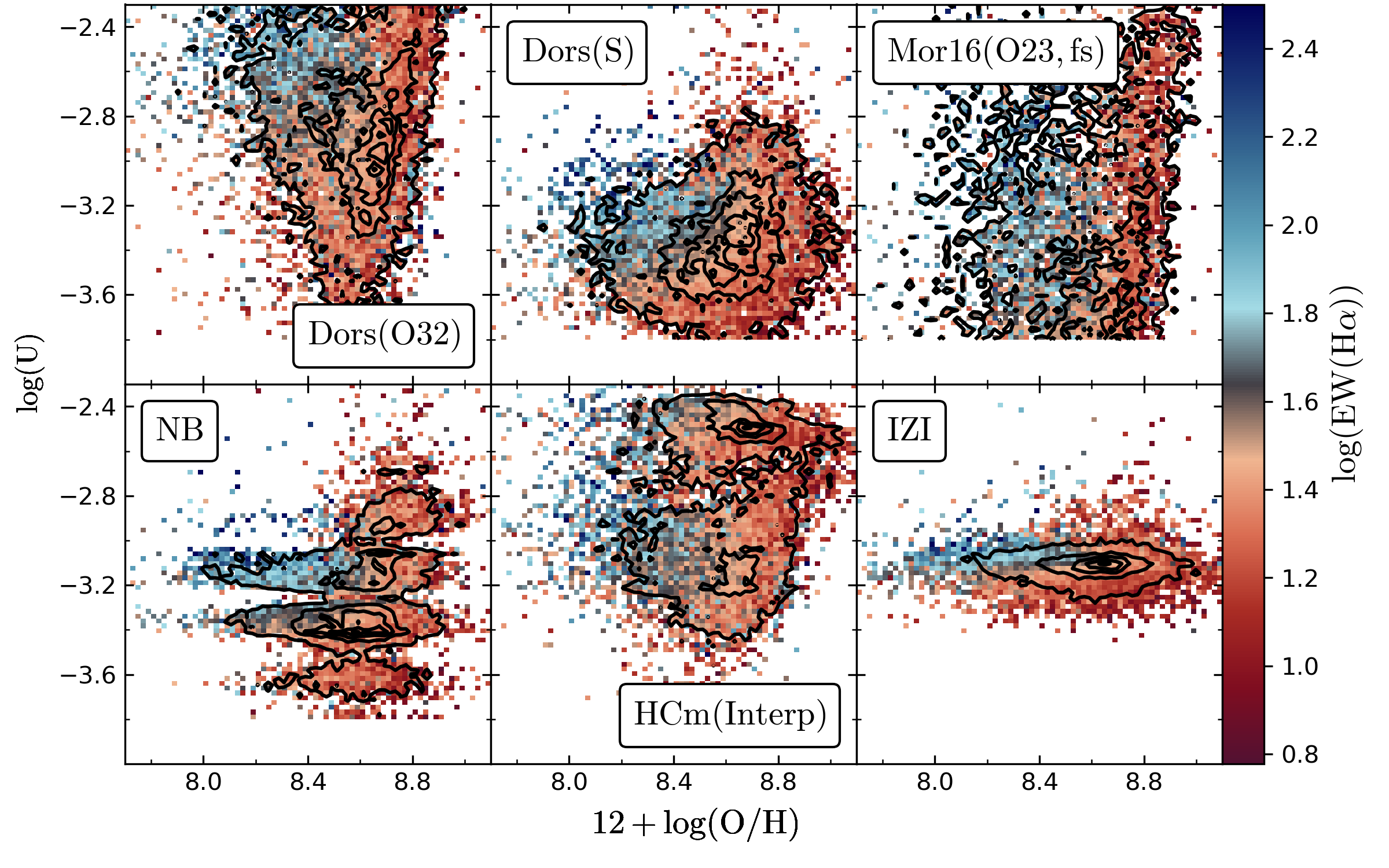}
\caption{Distribution of the ionization parameter, log(U), using each of the adopted calibrators (one for each panel) as a function of the fiducial oxygen abundance \citet{2019Ho_MNRAS485} for our sample of \HII\ regions. In each panel the color code represents the \EWHa\ with, with a scale ranging between $\pm$3$\sigma$ range around its mean value. The contours represent the density distribution, the outermost contour enclosing 95\% of the regions and each consecutive one enclosing 20\% fewer regions. }
\label{fig:oh_vs_u}
\end{figure*}

In \cref{sec:rel_OH} we explore the relation between the oxygen abundances and the ionization parameter using our fiducial calibrators for both parameters. However, the observed trend in \cref{fig:OHvsPP} depends on the calibration used to estimate O/H and log(U). In the case of O/H, based on the results discussed in \cref{appx5:compare_oxy}, the trend would change only quantitatively but not qualitatively. However, this is not the case when adopting a different calibrator of log(U).
In \cref{fig:oh_vs_u} we show the distribution of log(U) as a function of our fiducial O/H for a set of different calibrators for the former parameter: (i) two of the calibrators proposed by \citep[][O32 and S2]{dors11}; (ii) the O32 fill sphere (fs) calibrator proposed by \citet{2016Morisset_aa594A}; (iii) the ionization parameters derived from applying the NebulaBayes, HCm and IZI codes to our dataset \citep[][respectively]{2018Thomas_ApJ856, 2014PerezMontero_MNRAS441, 2020Mingozzi_AA636}. As we can see in the figure, the trend between the two parameters heavely depends on the adopted calibrator. Moreover, the empirical calibrators present a negative trend (weak in all cases), but the theoretical ones (e.g. NebulaBayes and Hcm code) do not have any clear trend, showing a discrete distribution. For HCm code, the distribution covers the same range of parameters as the one shown by our fiducial calibrator. Finally, the IZI code present a flat/constant distribution. It is clear that the theoretical calibrations depends on the adopted grid of photoionization models. It is beyond the scope of the current exploration to determine the nature of this discrepancy, but our comparison undercover a strong difference that may affect the interpretation of any exploration using different calibrators.


\bsp
\label{lastpage}
\end{document}